\documentclass[aps,pra,eqsecnum,superscriptaddress]{revtex4-2}

\usepackage{amssymb,amsmath}
\usepackage{latexsym}

\usepackage{subdepth}
\usepackage{graphicx}
\usepackage{bm}
\usepackage{upgreek}

\usepackage[breaklinks=true]{hyperref}
\hypersetup{
	colorlinks   = true, 
	urlcolor     = blue, 
	linkcolor    = red, 
	citecolor   = green 
}

\DeclareMathOperator{\tr}{tr}

\DeclareMathOperator{\csch}{csch}
\DeclareMathOperator{\spanvec}{span}

\newcommand{\vecJsquared}{\vec J^{\:2}}

\newcommand{\sD}{\mathcal{D}}
\newcommand{\sE}{\mathcal{E}}

\newcommand{\sH}{\mathcal{H}}
\newcommand{\sI}{\mathcal{I}}

\newcommand{\sL}{\mathcal{L}}
\newcommand{\sM}{\mathcal{M}}

\newcommand{\sS}{\mathcal{S}}
\newcommand{\sT}{\mathcal{T}}

\newcommand{\sZ}{\mathcal{Z}}

\newcommand{\ket}[1]{\vert#1\rangle}

\newcommand{\proj}[1]{| #1\rangle\!\langle #1 |}

\newcommand{\smallfrac}[2]{{\textstyle{\frac#1#2}}}

\newcommand{\Rinv}[1]{\underleftarrow{#1}}
\newcommand{\Linv}[1]{\underrightarrow{#1}}

\newcommand{\Odot}{{\mathcal{O}\hspace{-0.575em}\raisebox{0.5pt}{$\boldsymbol{\cdot}$}\hspace{.275em}}}

\newcommand{\ad}{\mathrm{ad}}

\newcommand{\lowerintsub}[1]{{\raisebox{-1.5pt}{$\scriptstyle#1$}}}

\newcommand{\bbZ}{\mathbb Z}
\newcommand{\bbR}{\mathbb R}
\newcommand{\bbC}{\mathbb C}

\newcommand{\SO}{\mathrm{SO}}

\newcommand{\Spin}{\mathrm{Spin}}
\newcommand{\SL}{\mathrm{SL}}
\newcommand{\SU}{\mathrm{SU}}

\newcommand{\GL}{\mathrm{GL}}

\renewcommand{\sl}{\mathfrak{sl}}

\newcommand{\f}{\mathfrak f}
\newcommand{\g}{\mathfrak g}
\newcommand{\gl}{\mathfrak{gl}}
\newcommand{\h}{\mathfrak h}

\renewcommand{\k}{\mathfrak k}

\newcommand{\s}{\mathfrak s}
\newcommand{\cwh}{\mathbb{C}\mathfrak{wh}}

\newcommand{\sfg}{\textsf{g}}
\newcommand{\sfh}{\textsf{h}}
\newcommand{\sfk}{\textsf{k}}

\newcommand{\sfx}{\textsf{x}}

\newcommand{\Go}{G_{\textrm{o}}}
\newcommand{\go}{\g_{\textsf{o}}}

\newcommand{\Ho}{H_{\textrm{o}}}
\newcommand{\gone}{\g_{\textsf{l}}}

\newcommand{\LXdW}{L_X(dW)}
\newcommand{\ZXdt}{\sZ_{X,dt}}
\newcommand{\dZXdW}{d\sZ_X(dW)}
\newcommand{\dZvecXdW}{d\sZ_{\vec X}(d\vec W)}
\newcommand{\LvecXvecdW}{L_{\vec X}(d\vec W)}

\newcommand{\ZvecXdt}{\sZ_{\vec X,dt}}

\usepackage{color}

\newcommand{\subgroupeq}{\leq}

\begin{document}

\title{Simultaneous Measurements of Noncommuting Observables.\\
Positive Transformations and Instrumental Lie Groups}
	
\author{Christopher S.~Jackson}
\email{omgphysics@gmail.com}
\noaffiliation
	
\author{Carlton M.~Caves}
\email{ccaves@unm.edu}
\affiliation{Center for Quantum Information and Control, University of New Mexico,\\Albuquerque, New Mexico 87131-0001, USA}
	
	\date{\today}

\begin{abstract}
	We formulate a general program for describing and analyzing continuous, differential weak, simultaneous measurements of noncommuting observables, which focuses on describing the measuring instrument \emph{autonomously}, without states.
	The Kraus operators of such measuring processes are time-ordered products of fundamental \emph{differential positive transformations}, which generate nonunitary transformation groups that we call \emph{instrumental Lie groups}.
	The temporal evolution of the instrument is equivalent to the diffusion of a \emph{Kraus-operator distribution function} defined relative to the invariant measure of the instrumental Lie group; the diffusion can be analyzed by Wiener path integration, stochastic differential equations, or a Fokker-Planck-Kolmogorov equation.
	This way of considering instrument evolution we call the \textit{Instrument Manifold Program}.
	\hspace{6pt}
	We relate the Instrument Manifold Program to state-based stochastic master equations.
	We then explain how the Instrument Manifold Program can be used to describe instrument evolution in terms of a universal cover we call the universal instrumental Lie group, which is independent not just of states, but also of Hilbert space.
	The universal instrument is generically infinite dimensional, in which situation the instrument's evolution is \emph{chaotic}.
	Special simultaneous measurements have a finite-dimensional universal instrument, in which situation the instrument is considered to be \hbox{\textit{principal\/}} and can be analyzed within the differential geometry of the universal instrumental Lie group.\hspace{6pt}
	Principal instruments belong at the foundation of quantum mechanics.
	We consider the three most fundamental examples:
	measurement of a single observable, of position and momentum,
	and of the three components of angular momentum.
	As these measurements are performed continuously, they limit to strong simultaneous measurements.
	For a single observable, this gives the standard decay of coherence \emph{between\/} inequivalent irreducible representations;
	for the latter two cases, it gives a collapse \emph{within\/} each irreducible representation onto the classical or spherical phase space, locating phase space at the boundary of these instrumental Lie groups.
\end{abstract}

\date{\today}

\maketitle

\vfill\pagebreak

\tableofcontents

\vfill\pagebreak

\section{Introduction}
\label{sec:introduction}

\begin{quote}
{\it``Well, why not say that all the things which should be handled in theory are just
those things which we also can hope to observe somehow.'' \ldots\ I remember that when I
first saw Einstein I had a talk with him about this. \ldots\ [H]e said, ``That may be so,
but still it's the wrong principle in philosophy.''  And he explained that it is the theory
finally which decides what can be observed and what can not and, therefore, one cannot,
before the theory, know what is observable and what not.}\\
\hspace*{\fill}Werner Heisenberg, recalling a conversation with Einstein in 1926,\\
\hspace*{\fill}interviewed by Thomas S.~Kuhn, February~15, 1963~\cite{Heisenberg1963a}
\end{quote}

\begin{quote}
{\it The science of optics, like every other physical science, has two different
directions of progress, which have been called the ascending and the descending scale, the
inductive and the deductive method, the way of analysis and of synthesis. In every physical
science, we must ascend from facts to laws, by the way of induction and analysis; and must
descend from laws to consequences, by the deductive and synthetic way.  We must gather and
group appearances, until the scientific imagination discerns their hidden law, and unity
arises from variety: and then from unity must re-deduce variety, and force the discovered
law to utter its revelations of the future.}\\
\hspace*{\fill}William Rowan Hamilton, 1833~\cite{WRHamilton1833a}
\end{quote}

At the beginning of the emergence of quantum mechanics was Heisenberg's realization that observables have noncommutative algebras (or kinematics), the most fundamental examples being canonical positions and momenta and angular momenta~\cite{vanderWaerden1967}.
This noncommutativity opens up a very deep conversation about the nature of observation and uncertainty.
With Schr\"odinger's wavefunctions~\cite{Schroedinger1982} and Born's interpretation of them~\cite{Born1926a}, observables were then developed within the Dirac-Jordan transformation theory~\cite{Duncan2009,Duncan2013,Oppenheimer1928} and then incorporated into the standard methods and ideas of quantum theory still used today: the inner product and Hilbert space, unitary transformations, and the eigenstate collapse associated with a von Neumann measurement~\cite{vonNeumann1932a,Wheeler1937,Thorne2010,Jammer1966,Wheeler1983}.
The positive transformations of this paper are a development of von Neumann's original ideas about the measuring process~\cite{vonNeumann1932a}, fundamentally changing the perspective on measurement by putting measurement on the same footing as unitary transformations.

Of the three fundamental tools of the standard methodology, von Neumann measurement is the least functional.
After this first generation of quantum theory, the development of radio astronomy and commercialization of radio broadcast, the formulation of stochastic calculus, the development of quantum field theory, and the invention of the laser, the concept of measurement was at last revisited~\cite{Schwinger1959a,Wigner1963a,Feynman1963a}.
Very important measurements such as photodetection, homodyne detection, and heterodyne detection already required a more general understanding than the von Neumann measurement~\cite{Lindblad1976a,Srinivas1981a,Barchielli1991a,Wiseman1993c,Wiseman1993a,Wiseman1993b,Goetsch1994a,Wiseman1994b,Wiseman1996a,CSJackson2022a}.
This generalized measurement theory was accomplished through the introduction of POVMs (positive operator-valued measures), operations, instruments, and Kraus operators~\cite{Jauch1967a,Ludwig1983a,Ludwig1985a,Kraus1983a,Davies1970a,Davies1976a,Peres1993a,Nielsen2000a}.
These tools can be considered an elaboration of another key idea of von Neumann's, the indirect measurement~\cite{vonNeumann1932a}.
More-or-less because of this, this second generation of measurement theory continued to consider the measuring process atemporally---that is, not considering the development over time between when a measurement begins and when it ends.
The positive transformations of this paper offer a comprehensive theory of temporal measuring processes by defining infinitesimal measurements that, we argue, are fundamental.

Now appears to be the end of a third generation of measurement theory, which has focused on continuous (i.e., temporal) measuring processes by incorporating stochastic calculus into the second-generation theory of operations.  These works are usually not about measuring instruments directly, however, but rather about state evolution as described by the stochastic master equation over particular Hilbert spaces~\cite{Carmichael1991,Wiseman1994b,Wiseman1996a,Doherty2000a,Brun2002b,KJacobs2006a,Wiseman2009a,Barchielli2009,Jacobs2014,Chantasri2013a,Albert2016,Karmakar2022}.  As such, though these works are definitely about temporal measurement evolution, these works do not consider the measuring instrument to be what is temporally evolving.
A handful of works have touched on the significance of infinitesimal positive transformations~\cite{Barchielli1982a,Goetsch1994a,Wiseman1996a,KJacobs1998a,KJacobs2006a,Silberfarb2005,LMartin2015a}, but none of these arrive at a clear understanding of simultaneous measurement, which is key to a comprehensive theory of continuous measuring instruments.\\

\begin{figure}[ht!]
	\centering
	\includegraphics[width=6.0in]{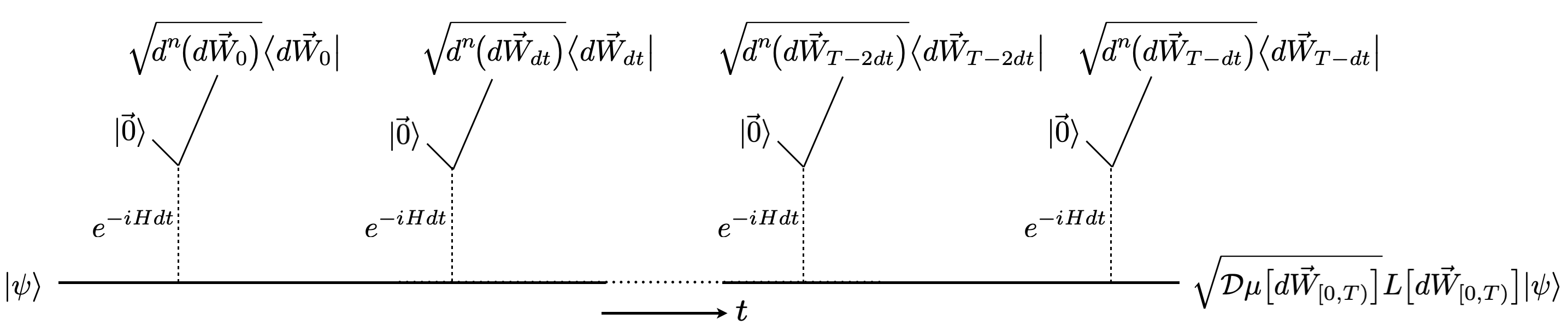}
	\caption{
		Schematic of a sequence of indirect, differential weak measurements; full understanding comes after reading Secs.~\ref{sec:weakmeasurements} and~\ref{sec:continuous}.  A system in a state $|\psi\rangle$ is indirectly measured by a sequence of weak interactions $e^{-iH dt}$, where each set of meters is observed after its interaction; that is, the system is continuously monitored.  The incremental Kraus operator for the measurement at time $t$, given outcomes $d\vec W_t$, is $\sqrt{d^n(d\vec W_t)}\big\langle d\vec W_t\big|e^{-iHdt}\big|\vec 0\,\big\rangle$; under the conditions outlined in Sec.~\ref{sec:weakmeasurements}, this Kraus operator is the differential positive transformation of Eq.~(\ref{fundamentalpositive}), that is, $\sqrt{d\mu(d\vec W_t)}\,L_{\vec X}(d\vec W_t)$, with $L_{\vec X}(d\vec W_t)=e^{-\vec X^2\kappa\,dt+\vec X\cdot\sqrt{\kappa}\,d\vec{W}_t}$.  The incremental Kraus operators ``pile up'' to become, at time $T$, the overall Kraus operator $\sqrt{\sD\mu[d\vec{W}_{[0,T)}]}\,L[d\vec{W}_{[0,T)}]$, which is written as a time-ordered exponential in Eq.~(\ref{toexp}).  The overall Kraus operator gives the unnormalized final state at time $T$, as shown in the figure.  The collection of Kraus operators at time $T$, for all Wiener outcome paths $d\vec W_{[0,T)}$, defines an \emph{instrument}, which can be analyzed on its own, independent of system states---simply omit $\ket\psi$ from the figure---a style of analysis we call \textit{instrument autonomy}.  The Kraus operators move across the manifold of an \emph{instrumental Lie group}, which is generated by the measured observables; placing the instrument within its instrumental Lie group and analyzing its evolution there we call the \emph{Instrument Manifold Program}.
	}\label{fig:sequential}
\end{figure}

In this paper, we formulate a program for directly analyzing continuous measuring instruments, which we call the \emph{Instrument Manifold Program}.
Similar to how (time-dependent) Hamiltonians generate unitary transformation groups, continuous measuring instruments also generate transformation groups, which we call \emph{instrumental Lie groups.}
Continuous measuring instruments consist of Kraus operators generated by incremental (i.e., infinitesimally generated) \emph{differential positive transformations\/} of the form
\begin{equation}\label{fundamentalpositive}
	\sqrt{d\mu(d\vec{W}_t)}\,L_{\vec X}(d\vec{W}_t) = \sqrt{d(dW^1_t)\cdots d(dW^n_t)\frac{e^{-d\vec{W}_t\cdot d\vec{W}_t/2dt}}{(2\pi dt)^{n/2}}\,}\,
e^{-\vec X^2\kappa\,dt+\vec X\cdot\sqrt{\kappa}\,d\vec{W}_t}\,.
\end{equation}
Here $\vec{X}=(X_1,\ldots,X_n)$ is an $n$-tuple of dimensionless observables being weakly measured simultaneously at a time $t$ with rate $\kappa$, $\vec X^2=\vec X\!\cdot\!\vec X$, and $d\vec{W}_t = (dW^1_t,\ldots,dW^n_t)$ is the conjugate $n$-tuple of Wiener outcome increments that are registered by weak measurements.
These differential positive transformations ``pile up'' as successive measurements are performed; at time $T$ the instrument is the collection of Kraus operators,
\begin{align}\label{toexp}
	\left\{L\big[d\vec{W}_{[0,T)}\big]=\sT\exp\!\bigg(\int_0^{T-dt}\hspace{-6pt}-\vec X^2\kappa\,dt+\vec X\cdot\sqrt{\kappa}\,d\vec W_t\bigg)\,:\;d\vec{W}_{[0,T)}\text{ is a Wiener path}\right\}\,,
\end{align}
where $\sT$ denotes a time-ordered exponential.  This scenario of piling up incremental Kraus operators is sketched in Fig.~\ref{fig:sequential}. These instruments are contained in the Lie group $G$ infinitesimally generated by the measured observables, $\{X_1,\ldots,X_n\}$, and the quadratic term $\vec X^2$.  We call $G$ the \emph{instrumental Lie group}.  At every time $T$, the instrument~(\ref{toexp}) is equivalent to a \emph{Kraus-operator distribution function},
\begin{align}
	D_T(L)\equiv\int\sD\mu[d\vec{W}_{[0,T)}]\;\delta\big(L,L\big[d\vec{W}_{[0,T)}\big]\big)\,,
\end{align}
where $\sD\mu[d\vec{W}_{[0,T)}]$ is the Wiener path measure and $\delta\big(L,L\big[d\vec{W}_{[0,T)}\big]\big)$ is a Dirac $\delta$-function with respect to the left-invariant measure of $G$.
The Kraus-operator distribution function describes how the instrument is distributed in the instrumental Lie group.  The Markovianity or group property of the instrument,
\begin{equation}\label{Markov}
	L[d\vec{W}_{[0,t+dt)}]=L(d\vec{W}_t)L[d\vec{W}_{[0,t)}]\,,
\end{equation}
means that the Kraus-operator distribution function evolves according to a Fokker-Planck-Kolmogorov equation,
\begin{equation}
	\frac{1}{\kappa}\frac{\partial}{\partial t}D_t(L) = \Bigg(\Rinv{\vec{X}^2}+\frac12\sum_i\Rinv{X_\mu}\Rinv{X_\mu}\Bigg)[D_T](L)\,,
\end{equation}
where $\Rinv{X}$ denotes a right-invariant derivative,
\begin{equation}
	\Rinv{X}[f](L) \equiv \lim_{h\to0}\frac{f(e^{hX}L)-f(L)}{h}\,.
\end{equation}
Therefore the instrument can be considered to evolve within the manifold that is the instrumental Lie group $G$.
These are the topics of Secs.~\ref{sec:weakmeasurements} and~\ref{sec:continuous}.
Section~\ref{sec:stateevolution} applies the objects of the Instrument Manifold Program to state evolution for the purpose of connecting to conventional works on continuous measurement.

As a manifold, the instrumental Lie group $G$ can be considered either within a matrix representation or universally; that is, the time-ordered exponentials of Eq.~(\ref{toexp}) can be processed either with a matrix algebra or with abstract Lie brackets.  The corresponding instruments we will distinguish by the names \emph{quantum instrument\/} and \emph{universal instrument}.  For special choices of observables, the universal instrument is finite-dimensional, in which case we will call it a \emph{principal instrument}; otherwise, the universal instrument evolves chaotically, and we will call it a \emph{chaotic instrument}.
The details of this are the subject of Sec.~\ref{sec:detaching}.

In Sec.~\ref{sec:1-2-3} we apply the Instrument Manifold Program to the three most fundamental principal instruments: Sec.~\ref{sec:measX} discusses the measurement of a single observable; Sec.~\ref{sec:measQP} discusses the simultaneous momentum and position measurement (SPQM)~\cite{CSJackson2023b}; and Sec.~\ref{sec:meas3J} discusses simultaneous measurement of the three components of angular momentum, a.k.a.~the isotropic spin measurement (ISM)~\cite{CSJackson2021a}.  The second and third of these measurements have a very different character from the first.  While the first instrument evolves in a 2-dimensional abelian Lie group, the second and third evolve in 7-dimensional nonabelian Lie groups.  While the first measurement collapses onto the von Neumann POVM, the second and third measurements collapse onto the canonical coherent POVM and the spin-coherent POVM~\cite{Shojaee2018a,CSJackson2021a,CSJackson2023b}.
The key to analyzing the properties of these last two instruments is in establishing a co\"ordinate system on the universal instrumental Lie group, and for that purpose the Cartan decomposition is just the ticket.\\

The main purpose of the name Instrument Manifold Program is to bring attention to the fact that this work consists of mathematical techniques from the theory of transformation groups as they apply to the theory of measurement: universal covers~\cite{Poincare2010a, Weyl1913a, Bourbaki1989a, Knapp1986a, Knapp2002a}, Haar measures~\cite{Haar1933a, vonNeumann1999a, Pontrjagin1946a, Bourbaki2004a, Nachbin1965a, Montgomery1974a, Barut1980a}, the Maurer-Cartan form~\cite{Chern1952a, Knapp1988a, Borel2001, Chirikjian2009a, Frankel2012a}, and Cartan decompositions~\cite{Chern1952a, Helgason1978a, Barut1980a, Borel1998a, Knapp2002a}.
While the theories of transformation groups and quantum mechanics essentially developed simultaneously, they barely came into contact and basically could not come together until stochastic calculus~\cite{Chirikjian2009a, Einstein1905a, Lemons1997, Uhlenbeck1930a, Feller1949, Stratonovich1963, Stratonovich1967, Ito1950a, Ito1996a, Gardiner2009a, Gardiner2021a} became established in measurement theory.
The history of these mathematical techniques and the theory of measurement and the extent to which they coexisted and influenced each other is complicated and fascinating, and perhaps we will write about them in the future.
For now, it suffices to affirm that we believe this work is the first to demonstrate that measurement can be considered a theory of positive transformations, putting it on the same footing as unitary transformations.
By doing so, two quite important connections have so far been realized: the connection between simultaneous measurements and phase-space POVMs (both standard and spin) and a surprising connection of simultaneous measurements to chaos, which promises a way forward on the problem of quantum chaos and dynamical complexity.

An understanding of the Instrument Manifold Program can be broken into three important steps or ``perspectival shifts,'' which are pointed out as the paper moves along:
\begin{enumerate}
	\item The first shift, in Sec.~\ref{sec:SDEsub}, is about considering infinitesimally generated positive transformations as the fundamental measuring processes, similar to how infinitesimally generated unitary transformations are considered fundamental dynamical processes.
	\item The next shift, in Sec.~\ref{sec:FPKEsKOD}, is about how such instruments can therefore be understood as evolutions on an autonomous instrument manifold, relying not on states for their existence, but rather finding their home in an abstract instrumental Lie group.
	\item The final shift, in Sec.~\ref{sec:detaching}, takes this new autonomy of the instrument a step further by pointing out that the definition of such instruments with instrumental Lie groups can be considered universally, independent even of the matrix representation of the observables and therefore not relying even on the specific Hilbert space.
\end{enumerate}
We now invite the reader to embark on the journey of understanding and appreciating these three perspectival shifts.

\section{Continuous, differential weak measurements of noncommuting observables}
\label{sec:continuousweak}

\subsection{Differential weak measurements and incremental Kraus operators}
\label{sec:weakmeasurements}

A differential weak measurement of multiple observables is made by doing a sequence of indirect weak measurements of the several observables; these indirect measurements are implemented by coupling independent Gaussian meters to the system, one for each observable.  We call this a ``differential weak measurement'' because the Kraus operators are differentially close to the identity; these incremental Kraus operators can then be regarded as fundamental, infinitesimally generated \emph{differential positive transformations} of a differentiable manifold.  Although a differential measurement is definitely weak, there are measurements that are generally construed as weak but that have some Kraus operators that are not close to the identity (e.g., jump processes).  The ``weak'' in differential weak measurement is thus both insufficient by itself and unnecessary when preceded by ``differential''; it is included to throw a lifeline to conventional usage.

The key accomplishment of this section is to show that at the level of differential weak measurements, the commutators of the observables can be ignored, so there is no temporal order to the measurements of the several observables and these measurement can be regarded as occurring \emph{simultaneously.}

\subsubsection{Differential weak measurement of a single observable}
\label{sec:single}

We start by considering differential weak measurement of a single observable (Hermitian operator) $X$ of a system~$\sS$, described in a Hilbert space $\sH$, during an increment of time $dt$.  The system is coupled to a canonical (essentially classical) position-momentum ($Q$-$P$) meter~$\sM$.  The interaction Hamiltonian $H$ acting over time $dt$,
\begin{align}\label{eq:HdtX}
H\,dt=2\sqrt{\kappa\,dt}\,X\otimes\sigma P\,,
\end{align}
generates a controlled displacement of the meter.  The meter begins in a state $\ket0$, which is assumed to have a Gaussian wave function,
\begin{align}\label{eq:Gaussianmeter}
\langle q|0\rangle=\sqrt{\frac{e^{-q^2/2\sigma^2}}{\sqrt{2\pi\sigma^2}}}\,.
\end{align}
The Kraus operator for the differential weak measurement of~$X$ with outcome~$q$ within~$dq$ is~\cite{Kraus1983a,Davies1976a,Nielsen2000a}
\begin{align}\label{eq:Kraussingle}
\begin{split}
\sqrt{dq}\,\langle q|e^{-iH\,dt/\hbar}|0\rangle
&=\sqrt{dq}\,\langle q|e^{-i2\sqrt{\kappa\,dt}\,X\otimes\,\sigma P/\hbar}|0\rangle\\
&=\sqrt{dq}\,e^{-2\sqrt{\kappa\,dt}\,\sigma Xd/dq}\langle q|0\rangle\\
&=\sqrt{dq}\,\big\langle q-2\sqrt{\kappa\,dt}\,\sigma X\big|0\big\rangle\\
&=\sqrt{\frac{dq}{\sqrt{2\pi\sigma^2}}}\,
\exp\!\bigg(\!{-}\frac{(q-2\sqrt{\kappa\,dt}\,\sigma X)^2}{4\sigma^2}\bigg)\,.
\end{split}
\end{align}
Here we deliberately do not set $\hbar=1$, thus making clear that $X$ is a ``dimensionless'' system observable.  Processing the incremental Kraus operator further, we have \begin{align}\label{eq:Kraussingle2}
\begin{split}
\sqrt{dq}\,\langle q|e^{-iH\,dt/\hbar}|0\rangle
&=\sqrt{dq\,\frac{e^{-q^2/2\sigma^2}}{\sqrt{2\pi\sigma^2}}}\,
e^{Xq\sqrt{\kappa\,dt}/\sigma-\kappa\,dt\,X^2}\\
&=\sqrt{d(dW)\,\frac{e^{-dW^2/2 dt}}{\sqrt{2\pi dt}}}\,
e^{X\sqrt\kappa\,dW-X^2\kappa\,dt}\,.
\end{split}
\end{align}
In the final form, the outcome is rescaled to be
\begin{align}
dW=\frac{q}{\sigma}\sqrt{dt}\,,
\end{align}
which is a standard Wiener increment---we call $dW$ a Wiener outcome increment---with Gaussian probability measure,
\begin{align}\label{eq:Wienerdmu}
	d\mu(dW)\equiv d(dW)\,\frac{e^{-dW^2/2\,dt}}{\sqrt{2\pi\,dt}}\,.
\end{align}
Readers uncomfortable with the notations $d\mu(dW)$ and $d(dW)$ should pause for a moment to get comfortable by reading the last sentence again.

The reader should appreciate that the $1/\sqrt{dt}$ scaling of the controlled displacement~(\ref{eq:HdtX}),
\begin{align}
H=2\sqrt{\frac{\kappa}{dt}}\,X\otimes\sigma P\,,
\end{align}
anticipates that as $dt$ goes to zero, the interaction strength $\sqrt{\kappa/dt}$ must go to infinity as $1/\sqrt{dt}$.  This is so that the allegedly differentiable process associated with a Hamiltonian, which is conjugate to time, becomes a diffusive process associated with a positive Kraus operator.  The incremental Kraus operator~(\ref{eq:Kraussingle2}) has a term linear in $X$ that is conjugate to a Wiener outcome increment $dW$, stochastically of order $\sqrt{dt}$, and a term quadratic in $X$ that is conjugate to~$dt$.

Defining a Kraus operator with the Wiener measure omitted,
\begin{align}\label{eq:LX}
	\LXdW\equiv e^{X\sqrt\kappa\,dW-X^2\kappa\,dt}\,,
\end{align}
brings the Kraus operator~(\ref{eq:Kraussingle2}) into the form
\begin{align}\label{eq:qtomuL}
\sqrt{dq}\,\langle q|e^{-iH\,dt/\hbar}|0\rangle
=\sqrt{d\mu(dW)}\,\LXdW\,.
\end{align}
The (completely positive) superoperator for outcome $dW$,
\begin{align}\label{eq:instrumentelX}
	\dZXdW = d\mu(dW)\,\LXdW\!\odot\!\LXdW^\dag\,,
\end{align}
we call an \textit{instrument element}.  We stress that the outcome increment $dW$ is literally the outcome of the measurement, scaled to have a variance $dt$.  We also note that the exponential expressions here are exact in the sense that they hold even when $dt$ is not infinitesimal.  The set of instrument elements corresponding to all outcomes is the \textit{instrument}~\cite{Davies1970a,Davies1976a,Dressel2013}.   Here we also introduce the ``odot'' ($\odot$) notation~\cite{Caves1999c,Rungta2001b,Menicucci2005a} for a superoperator, defined by
\begin{align}
A\odot B^\dagger(C)=ACB^\dagger\,.
\end{align}
The $\odot$ is literally a tensor product, but if one doesn't want to think about that, one can think of the $\odot$ as just a placeholder for an operator on which the superoperator acts.  We say a bit more about the odot notation below.

Integrating the instrument elements over outcomes gives the (unconditional) \textit{quantum operation\/} associated with the instrument,
\begin{align}\label{eq:unconQOX}
\begin{split}
\ZXdt&\equiv\int\dZXdW\\
&=\int d\mu(dW)\,\LXdW\!\odot\!\LXdW^\dag\\
&=e^{-\kappa\,dt(X^2\odot 1+1\odot X^2)}\int d\mu(dW)\,e^{X\cdot\sqrt\kappa\,dW}\!\odot\!e^{X\cdot\sqrt\kappa\,dW}\\
&=e^{-\kappa\,dt(X^2\odot 1+1\odot X^2)}\int d(dW)\,\frac{e^{-dW^2/2 dt}}{\sqrt{2\pi dt}}e^{\sqrt\kappa\,dW(X\odot 1+1\odot X)}\\
&=e^{-\kappa\,dt(X^2\odot 1+1\odot X^2)}e^{(\kappa\,dt/2)(X\odot 1+1\odot X)^2}\\
&=e^{{-}(\kappa\,dt/2)(X\odot 1-1\odot X)^2}\\
&=e^{{-}(\kappa\,dt/2)\ad_X^2}\,,
\end{split}
\end{align}
which is a trace-preserving, completely positive superoperator.  The last line introduces the adjoint, defined by $\ad_X(A)=[X,A]$, as the superoperator
\begin{align}
\ad_X=X\odot 1-1\odot X\,.
\end{align}
The instrument is said to \textit{unravel\/} the quantum operation~\cite{Carmichael1991}.  Unravelings are not unique: the Kraus operators~(\ref{eq:LX}) are the particular unraveling of $\ZXdt$ that is a differential weak measurement of $X$.  We say that $\ZXdt$ is \textit{woven\/} from these differential instrument elements.

A brief digression on terminology is in order~\cite{Nielsen2000a}. The term ``instrument'' originated with Davies and Lewis~\cite{Davies1970a,Davies1976a,Dressel2013}.  We adopt it, as opposed to other possible terminology, because it evokes the notion of an autonomous physical device or sense organ that is independent of the state of the system, a device ready to be stimulated or ``played'' by an input system state in the manner described below.
The style of our analysis, wholly in terms of instrument elements (or Kraus operators) and bereft of quantum states, we refer to as~\textit{instrument autonomy\/} (we sometimes think of this as more than just a style and elevate it to the \emph{Principle of Instrument Autonomy}~\cite{CSJackson2022a}); for continuous measurements, this style of analysis emerged from the work of Shojaee~\textit{et al.}\ on continuous isotropic measurements of the three components of angular momentum~\cite{Shojaee2018a,CSJackson2021a}.  We reserve the term ``quantum operation'' for a \textit{trace-preserving\/} completely positive superoperator, often distinguished as an ``unconditional quantum operation.''  In place of the ``unconditional'' in unconditional quantum operation, we could use ``total'' or ``nonselective.''   The term quantum operation often also includes trace-decreasing completely positive maps, like our instrument elements, and these are sometimes distinguished as ``selective quantum operations.''  The unraveling of a quantum operation into an instrument is often called a Kraus decomposition or an operator-sum decomposition~\cite{Kraus1983a,Schumacher1996a,Nielsen2000a}.  The instrument elements of an unraveling are taken apart into Kraus operators; we often slough over the distinction between a Kraus operator $L$ and the corresponding instrument element $L\!\odot\!L^\dag$.

The only aspect of the odot notation used here, but not presented in the previous literature\cite{Caves1999c,Rungta2001b,Menicucci2005a} is a faux bra-ket notation that writes the matrix elements of a superoperator $\sS$ as $(B)\sS(A)=\tr\!\big(B^\dagger\sS(A)\big)$.  Since $\sS(A)$ is read ``$\sS$ \textit{of} $A$,'' we like to read  $(B)\sS$ as ``$B$ \textit{faux} $\sS$.''  The faux bra-ket notation will be developed in detail elsewhere.  The only features we need for the present are the following: (i)~a trace-preserving superoperator satisfies $\tr(A)=\tr\!\big(\sS(A)\big)=(1)\sS(A)$ for all operators $A$, and thus trace preservation is expressed by $(1)\sS=1$; (ii)~$(1)A\!\odot\!A^\dagger(B)=\tr(ABA^\dagger)=\tr(A^\dagger AB)$, which implies that
\begin{align}
(1)A\!\odot\!A^\dagger=A^\dagger A
\end{align}
is the projection that maps $A$ to $A^\dagger A$.  In the absence of a complete understanding of or interest in the faux bra-ket notation, one can regard these two features as notational conveniences.

An instrument is a refinement of two fundamental state-independent objects.  The first is the unconditional quantum operation~\cite{Ludwig1983a,Ludwig1985a,Kraus1983a,Nielsen2000a}, as in Eq.~(\ref{eq:unconQOX}).  The second, the \textit{positive-operator-valued measure\/} (POVM)~\cite{Jauch1967a,Kraus1983a,Peres1993a,Nielsen2000a}, is made up of the operators
\begin{align}\label{eq:SPOVMelement}
 dE_X(dW)\equiv(1)\dZXdW=d\mu(dW)\,\LXdW^\dag\LXdW\,,
\end{align}
each of which is called a \textit{POVM element\/}; often, just as for Kraus operators, one omits the measure when talking about POVM elements.  An immediate consequence of Eq.~(\ref{eq:qtomuL}) is that the POVM satisfies a \textit{completeness relation\/}: the POVM elements integrate over outcomes to the identity operator,
\begin{align}\label{eq:POVMcompleteness}
1=(1)\ZXdt=\int dE_X(dW)=\int d\mu(dW)\,\LXdW^\dag\LXdW\,.
\end{align}
Equivalent to $\ZXdt$ being trace preserving, this completeness relation can be regarded in the case at hand as a trivial consequence of the last two forms in Eq.~(\ref{eq:unconQOX}) because $(1)\ad_X=11X-X11=0$.\\

So far there has been no mention of quantum states---instrument autonomy!---but it is useful to review, before the notation makes it hard to discern the forest for the trees, how operations, instruments, Kraus operators, and POVMs emerged from state-dependent consideration of indirect measurements~\cite{Ludwig1983a,Ludwig1985a,Kraus1983a,Davies1970a,Davies1976a,Nielsen2000a}.   Given initial system state $\rho$, the probability for outcome $dW=(q/\sigma)\sqrt{dt}$,
\begin{align}
\begin{split}
d(dW)\,P(dW|\rho)
&=dq\,\tr_\sS\!\Big(\langle q|e^{-iH\,dt/\hbar}\rho\otimes\proj0 e^{iH\,dt/\hbar}|q\rangle\Big)\\
&=d\mu(dW)\,\tr\!\big(\LXdW\rho\LXdW^\dag\big)\\
&=(1)\dZXdW(\rho)\,,
\end{split}
\end{align}
is a matrix element of the instrument element $\dZXdW$; this can be converted to the POVM element~(\ref{eq:SPOVMelement}),
\begin{align}\label{eq:POVM}
\begin{split}
d(dW)\,P(dW|\rho)
&=d\mu(dW)\,\tr\!\big(\LXdW^\dag\LXdW\rho\big)\\
&=\tr\!\big(dE_X(dW)\rho\big)\,.
\end{split}
\end{align}
The completeness relation~(\ref{eq:POVMcompleteness}) expresses the normalization of this probability for all normalized input states~$\rho$.  The normalized state of the system after a measurement with outcome $dW$ is
\begin{align}\label{eq:postmeasurementstate}
\begin{split}
\rho(dW|\rho)
&=\frac{dq\,\langle q|e^{-iH\,dt/\hbar}\rho\otimes\proj0 e^{-iH\,dt/\hbar}|q\rangle}
{dq\,\tr_S\!\Big(\langle q|e^{-iH\,dt/\hbar}\rho\otimes\proj0 e^{-iH\,dt/\hbar}|q\rangle\Big)}\\
&=\frac{\LXdW\rho\LXdW^\dag}{\tr\!\big(\LXdW\rho\LXdW^\dag\big)}\\
&=\frac{\dZXdW(\rho)}{d(dW)\,P(dW|\rho)}\,.
\end{split}
\end{align}
In the second line of Eq.~(\ref{eq:POVM}), the POVM element combines with the initial system state $\rho$ to give an outcome probability,
and in the final form of Eq.~(\ref{eq:postmeasurementstate}), the instrument element $\dZXdW$ maps the initial state to the unnormalized post-measurement state, conditioned on outcome $dW$, with the normalization given by the outcome probability.  If one ignores the outcome, the post-measurement state is given by the unconditional quantum operation,
\begin{align}
\int d(dW)\,P(dW|\rho)\rho(dW|\rho)=\ZXdt(\rho)= e^{{-}(\kappa\,dt/2)\ad_X^2}(\rho)\,.
\end{align}

\subsubsection{Differential weak measurements of multiple observables simultaneously}
\label{sec:multiple}

Suppose now that one measures several, generally noncommuting observables, $\{\,X_\mu\mid\mu=1,\ldots,n\,\}\equiv\vec X$, during an increment $dt$.  Initially (but only temporarily), we think of the $n$ measurements as occurring sequentially during $dt$, each taking up an increment $dt/n$.  There is a meter for each observable.  The meter wave functions are assumed to be identical Gaussians, that of Eq.~(\ref{eq:Gaussianmeter}); the interaction strengths are adjusted so that the interaction Hamiltonians, each acting over a time $dt/n$,~are
\begin{align}\label{eq:HdtXmu}
H_\mu dt/n=2\sqrt{\kappa\,dt}\,X_\mu\otimes\sigma P_\mu\,,
\end{align}
thus giving a Kraus operator of the form~(\ref{eq:qtomuL}) for each of the observables.  The Kraus operator for all $n$ measurements~is
\begin{align}
\begin{split}
\sqrt{dq_n}\cdots\sqrt{dq_1}\,&\langle q_n,\ldots,q_1|e^{-iH_n dt/n\hbar}\cdots e^{-iH_1 dt/n\hbar}|0_n,\ldots,0_1\rangle\\
&=\sqrt{dq_n}\,\langle q_n|e^{-iH_n dt/n\hbar}|0_n\rangle\cdots\sqrt{dq_1}\,\langle q_1|e^{-iH_1 dt/n\hbar}|0_1\rangle\\\
&=\sqrt{d\mu(dW^n)\cdots d\mu(dW^1)}\,L_{X_n}(dW^n)\cdots L_{X_1}(dW^1)\,.\\
&=\sqrt{d\mu(d\vec W)}\,\LvecXvecdW\,.
\end{split}
\end{align}
where
\begin{align}\label{eq:Ljoint1}
\LvecXvecdW\equiv L_{X_n}(dW^n)\cdots L_{X_1}(dW^1)
\end{align}
and the measure for the $n$ independent outcome increments is given by the isotropic Gaussian
\begin{align}\label{eq:tinyiso}
d\mu(d\vec{W})
\equiv d\mu(dW^n)\cdots d\mu(dW^1)
=\frac{d(dW^n)\cdots d(dW^1)}{(2\pi\,dt)^{n/2}}\exp\!\bigg({-}\frac{d\vec W\cdot d\vec W}{2 dt}\bigg)\,.
\end{align}
Here $d\vec W\equiv\{dW^1,\ldots,dW^n\}$ and $d\vec W\cdot d\vec W=\sum_{\mu=1}^n(dW^\mu)^2$.  The $n$ outcome increments are uncorrelated, zero-mean Wiener increments, having variances keyed to the measurement time $dt$; the outcome increments thus satisfy the It\^o rule
\begin{align}\label{eq:tinyIto}
dW^\mu dW^\nu=\delta^{\mu\nu} dt\,.
\end{align}
The instrument element for the $n$ measurements during the increment $dt$ comes from composing the instrument elements for the $n$ observables,
\begin{align}\label{eq:dZvecXinstrumentel}
\dZvecXdW&=d\sZ_{X_n}(dW^n)\circ\cdots\circ d\sZ_{X_1}(dW^1)=d\mu(d\vec{W})\,\LvecXvecdW\!\odot\!\LvecXvecdW^\dag\,.
\end{align}
It is important to appreciate that the Kraus operators for the individual measurements pile up as a linear product to make the incremental Kraus operator for all $n$ measurements.  In the instrument elements this becomes composition of the individual superoperators.

The Kraus operator~(\ref{eq:Ljoint1}) can be manipulated in the following ways,
\begin{align}\label{eq:Ljoint2}
\begin{split}
\LvecXvecdW&=e^{X_n\sqrt{\kappa}\,dW^n-X_n^2\kappa\,dt}
\cdots
e^{X_1\sqrt{\kappa}\,dW^1-X_1^2\kappa\,dt}\\
&=e^{\vec{X}\cdot\sqrt{\kappa}\,d\vec{W}-\vec{X}^{2}\kappa\,dt}\\
&=e^{-\vec{X}^{2}\kappa\,dt}e^{\vec{X}\cdot\sqrt{\kappa}\,d\vec{W}}\\
&=1-\vec{X}\cdot\sqrt{\kappa}\,d\vec{W}-\frac12\vec{X}^{2}\kappa\,dt\,,
\end{split}
\end{align}
where $\vec X\cdot d\vec W=X_\mu dW^\mu$ (the Einstein summation convention is used to sum on matched lower and upper indices)~and
\begin{align}
\vec{X}^{2}=\vec X\cdot\vec X= \sum_{\mu=1}^n X_\mu^2\,.
\end{align}
The key to these manipulations is this: because the Wiener increments $dW^\mu$ are independent, the It\^o rule~(\ref{eq:tinyIto}) sets to zero all the outcome-increment cross terms that arise in expanding $\LvecXvecdW$ to order $dt$, regardless of whether the observables commute, thus making the temporal ordering of the $n$ differential weak measurements irrelevant and allowing us to combine the Kraus operators for the individual measurements into the forms on the last three lines of Eq.~(\ref{eq:Ljoint2})~\cite{Barchielli1982a,CSJackson2021a,Karmakar2022,CSJackson2023b}.   This means that as opposed to the serial measurements of the $n$ observables contemplated initially, we can think of $\LvecXvecdW$ as coming from \textit{simultaneous\/} measurement of the $n$ observables over the entire increment $dt$, with each observable using the interaction Hamiltonian~(\ref{eq:HdtX}) with the standard interaction strength.  It also means that whereas the exponential expressions for a single observable are exact, those for multiple, noncommuting observables are good only to order $dt$, as in the last line in Eq.~(\ref{eq:Ljoint2}); that being sufficient, we can move forward with the exponential expressions with confidence.

The incremental Kraus operator for the $n$ measurements,
\begin{align}\label{eq:Ldtupdelta}
	\boxed{
		\vphantom{\Bigg(}
		\hspace{15pt}
L_{dt}=\LvecXvecdW=e^\updelta\,,\qquad\updelta\equiv\vec{X}\cdot\sqrt{\kappa}\,d\vec{W}-\vec{X}^{2}\kappa\,dt\,,
\hspace{10pt}
\vphantom{\Bigg)}
}
\end{align}
generates the stochastic evolution produced by the measurement; $L_{dt}$ is a \emph{differential positive transformation}.  The logarithm, $\updelta=\ln L_{dt}$, is the key object in the theory; we refer $\updelta$ as the \emph{forward generator}.

The unconditional quantum operation is obtained by integrating over the outcomes~$d\vec W$,
\begin{align}\label{eq:unconqovecX}
\begin{split}
\ZvecXdt&=\int\dZvecXdW=\prod_{\mu=1}^n\int d\sZ_{X_\mu}(dW^\mu)=e^{-(\kappa\,dt/2)(\vec X\odot 1-1\odot\vec X)^2}\,,
\end{split}
\end{align}
an expression good to order $dt$.  The notation is perhaps a shorthand too far, so we spell out that
\begin{align}\label{eq:LindbladXmu}
\begin{split}
-\frac12\big(\vec X\odot 1-1\odot\vec X\big)^2
&=-\frac12\sum_\mu (X_\mu\odot 1-1\odot X_\mu)\circ(X_\mu\odot 1-1\odot X_\mu)\\
&=-\frac12\sum_\mu\ad_{X_\mu}\circ\ad_{X_\mu}\\
&=\sum_\mu X_\mu\odot X_\mu-\frac12\big(X_\mu^2\odot 1+1\odot X_\mu^2\big)\\
&=-\frac12\Big(\vec X^2\odot 1+1\odot\vec X^2\Big)+\sum_\mu X_\mu\odot X_\mu\,,
\end{split}
\end{align}
which is the Lindbladian for the master equation whose Lindblad operators are the measured observables $X_\mu$.\\

We stress that the Lindbladian---put differently, the quantum operation---is determined by the interaction with the meters and the quantum state of the meters and is independent of how the meter is read out.  Weakly measuring the Lindblad operators~$X_\mu$ unravels the quantum operation (or the Lindbladian) into instrument elements whose Kraus operators are constructed from the measured observables as in Eq.~(\ref{eq:Ldtupdelta}).  Other unravelings arise from making different measurements on the meter.  For example, the incremental quantum operation~(\ref{eq:unconqovecX}) can be unraveled into Kraus operators that are differential stochastic-unitary transformations,
\begin{align}\label{eq:stochasticunitary}
\sqrt{d\mu(d\vec W)}\,e^{-i\vec X\cdot\sqrt{\kappa}\,d\vec W}\,,
\end{align}
where
\begin{align}\label{eq:stochasticunitary2}
e^{-i\vec X\cdot\sqrt{\kappa}\,d\vec W}=1-i\vec X\cdot\sqrt\kappa\,d\vec W-\frac12\vec X^2\kappa\,dt\,,
\end{align}
Integrating, one finds that the unconditional quantum operation is indeed still $\ZvecXdt$,
\begin{align}
\begin{split}
\int d\mu(d\vec W)\,e^{-i\vec X\cdot\sqrt{\kappa}\,d\vec W}\!\odot\!e^{i\vec X\cdot\sqrt{\kappa}\,d\vec W}
&=\int d\mu(d\vec W)\,e^{-i\sqrt\kappa\,d\vec W\cdot(\vec X\odot 1-1\odot\vec X)}\\
&=e^{-(\kappa\,dt/2)(\vec X\odot 1-1\odot\vec X)^2}=\ZvecXdt\,.
\end{split}
\end{align}
As we show in App.~\ref{app:unravelings}, the differential unitary transformations~(\ref{eq:stochasticunitary}) arise from the same meter model that gives the incremental Kraus operators~(\ref{eq:Ldtupdelta}), but with registration of the meter momenta, instead of the meter positions. Comparing the last form of $\LvecXvecdW$ in Eq.~(\ref{eq:Ljoint2}) with the stochastic unitary~(\ref{eq:stochasticunitary2}), one sees that the Lindblad operators change according to $X_\mu\rightarrow-iX_\mu$.  This is an example of a symmetry of the general Lindbladian,
\begin{align}
	\sL = \sum_j A_j \odot A_j^\dag - \frac12\Big({A_j^\dag A_j\odot 1+1\odot A_j^\dag A_j}\Big)\,,
\end{align}
which is that the Lindbladian remains unchanged under unitary transformations of the Lindblad operators, $A_j\rightarrow A_k {U^k}_j$.

Another unraveling of the Lindbladian~(\ref{eq:LindbladXmu}) is the ``jump unraveling'' into discrete Kraus operators,
\begin{align}
\mbox{no jump:}\quad&K_0=e^{-(\kappa\,dt/2)\vec X^2}=1-\frac12\kappa\,dt\,\vec X^2\,,\label{eq:nojump}\\
\mbox{jump:}\quad&K_\mu=\sqrt{\kappa\,dt}\,X_\mu\,,\qquad\mu=1,\ldots,n\,.\label{eq:jump}
\end{align}
Obvious this is because
\begin{align}
K_0\!\odot\!K_0^\dagger+\sum_\mu K_\mu\!\odot\!K_\mu^\dagger=1+\kappa\,dt\bigg(\!{-}\frac12\Big(\vec X^2\odot 1+1\odot\vec X^2\Big)+\sum_\mu X_\mu\odot X_\mu\bigg)\,.
\end{align}
We show in App.~\ref{app:unravelings} how this jump unraveling follows from the same Gaussian meter model, but with registration of the meter in its number basis, instead of registration of position or momentum.

We stress that the incremental Kraus operators~(\ref{eq:Ldtupdelta}) and the stochastic-unitary Kraus operators~(\ref{eq:stochasticunitary2}) are both differential, i.e., close to the identity.  In contrast, the jump Kraus operators~(\ref{eq:jump}) are not close to the identity; thus the jump unraveling is not suitable for formulating an instrumental Lie-group manifold---or really any group at all, because the jump operators generally do not have an inverse.

The differential weak measurements of noncommuting observables that we consider in this paper are of the sort first considered by Barchielli for the case of simultaneous measurements of position and momentum~\cite{Barchielli1982a,Barchielli2009,Karmakar2022,CSJackson2023b} and by Jackson {\it et.~al.} for the case of angular-momentum components~\cite{Shojaee2018a,CSJackson2021a}.  These measurements give rise to all Lindbladians that have Hermitian Lindblad operators.  What happens with nonHermitian Lindblad operators was the focus of work in quantum optics in the 1990s and 2000s.  This work was pioneered by Wiseman and Milburn~\cite{Wiseman1993c,Wiseman1993a,Wiseman1993b}, pushed forward by Goetsch and Graham~\cite{Goetsch1994a}, and brought to perfection in Wiseman's PhD dissertation~\cite{Wiseman1994b} and a subsequent publication~\cite{Wiseman1996a}; it started from the standard quantum-optical master equation, which describes an optical mode decaying to vacuum, physically by leaking out of an optical cavity and mathematically via a Lindblad equation whose Lindblad operator is the mode's (nonHermitian) annihilation operator.  These researchers unraveled this Lindblad master equation in terms of the standard measurements of quantum optics---photon counting, homodyne detection, and heterodyne detection---by considering measurements on the field leaking from the cavity (this is indirect measurement of the cavity mode).  The resulting theory serves as the basis for quantum feedback and control~\cite{Doherty2000a,Wiseman2009a}.  Jackson~\cite{CSJackson2022a} has recently developed the group-theoretic aspects of the photodetector and the heterodyne instrument, with emphasis on their autonomy.   It is important to appreciate that in the current paper we are considering only Hermitian Lindblad operators, which arise from the controlled-displacement system-meter interaction of Eq.~(\ref{eq:HdtX}); nonHermitian Lindblad operators emerge from a different system-meter interaction.  The general interaction that gives rise to all Hermitian and nonHermitian Lindblad operators is not tied to quantum optics and thus is richer than the leaky-cavity quantum-optical master equation.  We have explored these general interactions and the measurements that unravel them and will provide an account of that work in future~papers.

\subsection{Continuous measurements of noncommuting observables.  Piling up incremental Kraus operators}
\label{sec:continuous}

In this section we pile up the incremental Kraus operators $\LvecXvecdW$ as a time-ordered product and thus develop a description of a continuous measurement of the generally noncommuting observables $\vec X$.  We deliberately do not include any unitary system dynamics, because we want to focus on the evolution of the measurement itself; this means that we are assuming that any dynamical time scales of the measured system are long compared to $1/\kappa$.

We formulate the description in terms of the three faces of the stochastic trinity: a Wiener-like path integral, stochastic differential equations (SDEs), and a Fokker-Planck-Kolmogorov (diffusion) equation (FPKE) for an evolving Kraus-operator distribution function.  The three faces of the trinity describe motion of the Kraus operators within a manifold that we call the \emph{instrumental Lie group\/}; this section is thus the essential start of our development of the Instrument Manifold Program.

\subsubsection{Stochastic differential equations and path integrals}
\label{sec:SDEsub}
Suppose one performs a continuous sequence of differential weak, simultaneous measurements, starting at $t=0$ and ending at $t=T$ (the last set of measurements commences at $T-dt$).  The defining mathematical object is the instrument element for an outcome sequence $d\vec W_{[0,T)}\equiv\{d\vec W_{0dt},d\vec W_{1dt},\cdots,d\vec W_{T-dt}\}$,
\begin{align}\label{eq:contmeasinstrumentel}
\begin{split}
\sD\sZ[d\vec W_{[0,T)}]
&= d\sZ_{\vec X}\big(d\vec{W}_{T-dt}\big)\circ \cdots\circ\,d\sZ_{\vec X}\big(d\vec{W}_{1dt}\big)\circ d\sZ_{\vec X}\big(d\vec{W}_{0dt}\big)\\
&\equiv\sD\mu[d\vec{W}_{[0,T)}]\;L[d\vec{W}_{[0,T)}]\!\odot\!L[d\vec{W}_{[0,T)}]^\dag\,.
\end{split}
\end{align}
Here
\begin{align}\label{eq:IsoWiener}
\begin{split}
\sD\mu[d\vec{W}_{[0,T)}]
&\equiv d\mu(d\vec W_{T-dt})\cdots d\mu(d\vec W_{1dt})\,d\mu(d\vec W_{0dt})\\
&=\left(\prod_{k=0}^{T/dt-1} d^n\!\big(d\vec{W}_{kdt}\big)\right)
\left(\frac{1}{2\pi dt}\right)^{nT/2\,dt}\exp\!\left(-\int_0^{T_-}\frac{d\vec{W}_t\cdot d\vec W_t}{2\,dt}\,\right)
\end{split}
\end{align}
is the isotropic Wiener measure.  The open parenthesis in the outcome sequence $d\vec W_{[0,T)}$ reminds us that the last vector of outcomes in the sequence is $d\vec W_{T-dt}$; likewise, the minus subscript on the upper integration limit, $T_-\equiv T-dt$, indicates that the integral does \textit{not\/} include the outcome increment $d\vec W_T$. The overall Kraus operator is
\begin{align}\label{eq:overallKraus}
\begin{split}
L_T\equiv L[d\vec{W}_{[0,T)}]&\equiv L_{\vec X}(d\vec W_{T-dt})\cdots L_{\vec X}(d\vec W_{1dt})L_{\vec X}(d\vec W_{0dt})\vphantom{\bigg)}\\
&=e^{\vec{X}\cdot\sqrt\kappa\,d\vec{W}_{T-dt}-\vec X^2\kappa\,dt}\cdots
e^{\vec{X}\cdot\sqrt\kappa\,d\vec{W}_{1dt}-\vec X^2\kappa\,dt}
e^{\vec{X}\cdot\sqrt\kappa\,d\vec{W}_{0dt}-\vec X^2\kappa\,dt}\\
&=\sT\,\prod_{t=0}^{T_-}\;\exp\!\Big(\vec{X}\cdot\sqrt\kappa\,d\vec{W}_t-\vec X^2\kappa\,dt\Big)\\
&=\sT\,\exp\!\bigg(\int_0^{T_-}\vec{X}\cdot\sqrt\kappa\,d\vec{W}_t-\vec X^2\kappa\,dt\bigg)\,;
\end{split}
\end{align}
the last two lines use $\sT$ to denote the time-ordered product and the time-ordered exponential.
In brief, the simultaneous measurement of a possibly noncommuting set of observables $\vec{X}=\{X_1,\ldots,X_n\}$ defines an instrument which registers simultaneous Wiener paths $d\vec{W}_{[0,T)}=\{dW_{[0,T)}^1,\ldots,dW_{[0,T)}^n\}$ with Kraus operators
\begin{align}\label{fundamentalTexp}
	\boxed{
		\vphantom{\Bigg(}
		\hspace{15pt}
		L[d\vec{W}_{[0,T)}]=\sT\,\exp\!\bigg(\int_0^{T_-}\vec{X}\cdot\sqrt\kappa\,d\vec{W}_t-\vec X^2\kappa\,dt\bigg)\,.
		\hspace{10pt}
		\vphantom{\Bigg)}
	}
\end{align}
The first place in the literature where we have seen this time-ordered product---the ``piling up''---of incremental Kraus operators written out explicitly is in a paper by Jacobs and Knight~\cite{KJacobs1998a}, albeit for a single measured observable that is mixed up with system dynamics.
Less explicitly and with different Kraus operators, similar time-ordered products appear in papers by Srinivas and Davies~\cite{Srinivas1981a} and by Goetsch and Graham~\cite{Goetsch1994a}.

The successive Kraus operators contributing to $L_T$ in Eq.~(\ref{eq:overallKraus}) must be time ordered whenever the measured observables do not commute.  Please appreciate that for any finite number of increments $dt$, the commutators can be ignored, temporal ordering is unnecessary, and the finite number of increments can simply be regarded as a ``bigger'' infinitesimal increment.
Once one proceeds to a finite time $T$, time ordering must be respected.  Being able to amalgamate any finite number of infinitesimal increments allows one to start with nonGaussian outcome increments, with the Gaussian behavior emerging from a kind of central-limit theorem over a bigger infinitesimal increment.  This freedom was used by Gross \textit{et al.}~\cite{JGross2018a}, who replaced Gaussian meters with qubit meters in a state-based formulation of continuous measurements.  The conditions for the emergence of Gaussian behavior should rightly be the subject of further investigation.

The incremental Kraus operators~(\ref{eq:Ldtupdelta}) and the overall Kraus operators~(\ref{eq:overallKraus}) were derived above from a meter model in which a measurement of position, a continuous variable, is made on each of the meters; von Neumann essentially introduced this meter model and called it an indirect measurement~\cite{vonNeumann1932a}.  We ask the reader now to join us in a {\bf shift in perspective}, the {\bf first} of three: \emph{regard the incremental Kraus operators for simultaneous measurements of noncommuting observables, $L_{dt}=e^\updelta$ of~(\ref{eq:Ldtupdelta}), not as derived objects, but as the fundamental differential positive transformations, more fundamental in quantum measurement theory than von Neumann projectors.  The forward generator $\updelta$ plays the role for positive transformations that anti-Hermitian Hamiltonian generators, $-iH\,dt$, play in generating unitary transformations.}  Continuously measuring commuting observables leads, over time, to von Neumann's original conception of eigenstates of Hermitian operators as measurement outcomes.  The perspectival shift is that Hermitian operators now play the more important role of generating positive transformations, acting via exponentiation of the forward generator~$\updelta$ to produce the incremental Kraus operators.  For noncommuting observables, these incremental Kraus operators, piled up over time, lead to $\dots$---\ well, that is the subject of the rest of this paper.

Although several researchers have hinted at or touched on the significance of positive transformations~\cite{Barchielli1982a,KJacobs2006a,Silberfarb2005}, especially those who work or comment on linear quantum trajectories~\cite{Goetsch1994a,Wiseman1996a,KJacobs1998a,LMartin2015a}, none has had a complete understanding of how differential weak, simultaneous measurements lead to the differential positive transformations, $L_{dt}=e^\updelta$ of~(\ref{eq:Ldtupdelta}), nor of how these transformations pile up to construct instrument manifolds.

The overall Kraus operator~(\ref{eq:overallKraus}) is the solution to the SDE
\begin{align}\label{eq:LSDE}
\begin{split}
dL_t\,L_t^{-1}&=L_{t+dt}L_t^{-1}-1\\
&=L_{\vec X}(d\vec W_t)-1\\
&=\updelta_t+\frac12\updelta_t^2\\
&=\vec{X}\!\cdot\!\sqrt{\kappa}\,d\vec{W}_t - \frac{1}{2}\vec{X}^2\kappa\,dt\,,
\end{split}
\end{align}
with initial condition $L_0=1$.  The left side of the SDE, called the \textit{Maurer-Cartan form}, is processed by expanding the exponential $L_{\vec X}(d\vec W_t)=L_{dt}=e^{\updelta_t}$ and applying the It\^o rule~(\ref{eq:tinyIto}); this obscures the role of the quadratic drift term $-\vec X^2\kappa\,dt$ in the exponential.  To respect the exponentials, Jackson and Caves introduced the \textit{modified Maurer-Cartan stochastic differential\/} (MMCSD) of $L_t$, which satisfies
\begin{align}\label{eq:LMMCSD}
	\boxed{
		\vphantom{\Bigg(}
		\hspace{15pt}
dL_t\,L_t^{-1} - \frac{1}{2}(dL_t\,L_t^{-1})^2 = \vec{X}\!\cdot\!\sqrt{\kappa}\,d\vec{W}_t-\vec{X}^2\kappa\,dt=\updelta_t\,.
\hspace{10pt}
\vphantom{\Bigg)}
}
\end{align}
This result comes from expanding the exponential in $L_{\vec X}(d\vec W_t)$ to second order, but does not rely on the It\^o rule~(\ref{eq:tinyIto}).  The MMCSD form of the SDE respects the exponential form of the incremental Kraus operators in Eq.~(\ref{eq:overallKraus}), which means that the MMCSD is equal to the forward generator $\updelta_t$; the quadratic term is unavoidable and traces back to the displacement of the Gaussian meter wave functions.

Equations~(\ref{eq:LSDE}) and~(\ref{eq:LMMCSD}) are It\^o-form SDEs, a fact recognized by noting that the ``coefficient'' of the increment $dL_t$, in this case $L_t^{-1}$, is evaluated at the beginning of the increment.  The equivalent Stratonovich-form SDE uses mid-point evaluation in the Maurer-Cartan form,
\begin{align}\label{eq:LSDEStratonovich}
	\boxed{
		\vphantom{\Bigg(}
		\hspace{15pt}
dL_t\,L_{t+dt/2}^{-1}=\delta_t\,.
\hspace{10pt}
\vphantom{\Bigg)}
}
\end{align}
Those who object that midpoint evaluation does not exist in the stochastic calculus should regard it as defined by $a_{t+dt/2}=\frac12(a_t+a_{t+dt})=a_t+\frac12 da_t$, which is precisely what one would write down for midpoint evaluation without thinking about this technicality.  One sees the equivalence to the It\^o-form SDE by finding the It\^o correction~\cite{Gardiner2009a,Gardiner2021a},
\begin{align}
\begin{split}
dL_t\,L_{t+dt/2}^{-1}
&=dL_t\bigg(L_t^{-1}+\frac12 dL_t^{-1}\bigg)\\
&=dL_t\bigg(L_t^{-1}-\frac12 L_t^{-1}dL_t\,L_t^{-1}\bigg)\\
&=dL_t\,L_t^{-1}-\frac12(dL_t\,L_t^{-1})^2\,,
\end{split}
\end{align}
which shows that the Stratonovich version of the Maurer-Cartan form, $dL_t\,L_{t+dt/2}^{-1}$, is a sort of shorthand for the It\^o-form~\hbox{MMCSD}.

The unconditional quantum operation~$\sZ_T$ is woven from the instrument elements~$\sD\sZ[d\vec W_{[0,T)}]$, the weaving expressed as a Wiener-like path integral of the measurement record,
\begin{align}\label{eq:contmeasqo1}
\sZ_T\equiv\int\sD\sZ[d\vec W_{[0,T)}]
=\int\sD\mu[d\vec{W}_{[0,T)}]\;L[d\vec{W}_{[0,T)}]\!\odot\!L[d\vec{W}_{[0,T)}]^\dag\,.
\end{align}
The ``-like'' indicates, first, that the functional integral sums over superoperators, not just c-numbers and, second, that there is no restriction on the endpoint of the Wiener paths.  This unraveling of $\sZ_T$ we call the \textit{Wiener differential unraveling}.  It is easy to integrate $\sZ_T$ because it is the composition of the incremental quantum operations~$\ZvecXdt$ of Eq.~(\ref{eq:unconqovecX}),
\begin{align}\label{eq:contmeasqo2}
\sZ_T
=\underbrace{\ZvecXdt\circ\cdots\ZvecXdt\circ\ZvecXdt}_{\mbox{$T/dt$ terms}}
=e^{-(\kappa T/2)(\vec X\odot 1-1\odot\vec X)^2}\,.
\end{align}
The integrated form is familiar to anyone who works with Lindblad master equations: it is the exponential of the Lindbladian~(\ref{eq:LindbladXmu}).  Since the incremental quantum operations are trace preserving, so is the composite quantum operation,
\begin{align}
1=(1)\sZ_T=\int\sD\mu[d\vec{W}_{[0,T)}]\;L[d\vec{W}_{[0,T)}]^\dag L[d\vec{W}_{[0,T)}]\,,
\end{align}
and this is equivalent to saying that the corresponding POVM, consisting of POVM elements
\begin{align}
\sD E[d\vec W_{[0,T)}]=(1)\sD\sZ[d\vec W_{[0,T)}]=\sD\mu[d\vec{W}_{[0,T)}]\,L[d\vec{W}_{[0,T)}]^\dagger L[d\vec{W}_{[0,T)}]\,,
\end{align}
is complete.

\subsubsection{The Kraus-operator distribution function and subsequent Fokker-Planck-Kolmogorov equation}
\label{sec:FPKEsKOD}

The third element of the stochastic trinity, FPKEs, involves introduction of a new mathematical object, the \emph{Kraus-operator distribution function}, and a new mathematical tool, \textit{right-invariant derivatives}, which appear naturally in the FPKE that evolves the Kraus-operator distribution function.  The authors introduce these two mathematical objects with some trepidation, because unlike path integrals and SDEs, they require most physics readers to appreciate and to understand new concepts.  Still, appreciate and understand the reader must, because these two objects are at the heart of the Instrument Manifold Program.  so we take the plunge and introduce these new concepts in this section.  There will be a crash of cymbals just below to indicate when the reader needs to wake up and pay special attention.

To get started, we need to think of the Kraus operators as occupying some ``space.''  Provisionally, we can think of the space of Kraus operators as being the general linear group on $\sH$, the Lie group $\GL(\sH,\bbC)$.  We can and must refine this provisional conception of the Kraus-operator space, a task that we take up in Secs.~\ref{sec:detaching} and~\ref{sec:1-2-3}, but for the present this is all we need.  We assume that there is a right- and left-invariant measure $d\mu(L)$ on the space of Kraus operators, and again provisionally, $d\mu(L)$ can be taken to be the Haar measure for $\GL(\sH,\bbC)$.  The invariance properties of the measure are
\begin{align}\label{eq:Haarleftright}
d\mu(L'L)=d\mu(L)=d\mu(LL')\,,
\end{align}
with the left equality expressing left invariance and the right equality expressing right invariance.  It is useful to note that
\begin{align}\label{eq:Haarinverse}
d\mu(L)=d\mu(1)=d\mu(L^{-1})\,,
\end{align}
which follows from left and right invariance.

The $\delta$-function that is conjugate to this measure, $\delta(L,L')$, satisfies the reproduction property~\cite{Brif1999a,STAli1999a},
\begin{align}
\int d\mu(L)\,f(L)\,\delta(L,L')=f(L')\,,
\end{align}
for any function $f$ on the space of Kraus operators.  We have
\begin{align}
\begin{split}
\int d\mu(L_1)\,f(L_1)\,\delta(LL_1,LL_2)
&=\int d\mu(L^{-1}L')\,f(L^{-1}L')\,\delta(L',LL_2)\\
&=\int d\mu(L')\,f(L^{-1}L')\,\delta(L',LL_2)\\
&=f(L_2)\,.
\end{split}
\end{align}
where the second step uses the left invariance of the measure.  This result implies that
\begin{align}\label{eq:deltaleft}
\delta(L_1,L_2)=\delta(LL_1,LL_2)=\delta(1,L_1^{-1}L_2)=\delta(L_2^{-1}L_1,1)\,,
\end{align}
which can be regarded as expressing the consequences of left invariance for the $\delta$-function.  Proceeding in the same way, one finds that the consequences of right invariance for the $\delta$-function are
\begin{align}\label{eq:deltaright}
\delta(L_1,L_2)=\delta(L_1 L,L_2 L)=\delta(1,L_2L_1^{-1})=\delta(L_1L_2^{-1},1)\,.
\end{align}
Finally, we have
\begin{align}
\begin{split}
\int d\mu(L_1)\,f(L_1)\,\delta(L_1^{-1},L_2^{-1})
&=\int d\mu(L_1^{-1})\,f(L_1)\,\delta(L_1^{-1},L_2^{-1})\\
&=\int d\mu(L)\,f(L^{-1})\,\delta(L,L_2^{-1})\\
&=f(L_2)\,,
\end{split}
\end{align}
where the first step uses the property~(\ref{eq:Haarinverse}).   This final property implies that $\delta(L_1,L_2)=\delta(L_1^{-1},L_2^{-1})$.  Applying Eqs.~(\ref{eq:deltaleft}) and~(\ref{eq:deltaright}) gives
\begin{align}\label{eq:deltaexchange}
\delta(L_1,L_2)=\delta(L_2,L_1)\,,
\end{align}
a property that requires both left and right invariance of the measure.  The relations~(\ref{eq:deltaleft}), (\ref{eq:deltaright}), and~(\ref{eq:deltaexchange}), which might mistakenly be thought of as trivially equivalent ways of requiring that $L_1=L_2$ in an integral, have content because the $\delta$-function must pay attention to how the measure changes from point to point in the group manifold.  That there are no position-dependent multipliers in these relations comes from the way right and left invariance relate the measure at different points in the manifold.

We can now partition the measurement-record paths into sets, all paths of which lead to a particular Kraus operator~$L$, and we use the $\delta$-function to add up all the Wiener-measure probability for a set into a \textit{Kraus-operator distribution function\/},
\begin{align}\label{eq:DTL}
	\boxed{
		\vphantom{\Bigg(}
		\hspace{15pt}
D_T(L)\equiv\int\sD\mu[d\vec{W}_{[0,T)}]\;\delta\big(L,L[d\vec{W}_{[0,T)}]\big)\,.
\hspace{10pt}
\vphantom{\Bigg)}
}
\end{align}
This functional integral over the Wiener measure involves only c-numbers and is constrained by a path-end $\delta$-function and thus is what is usually called a Wiener path integral~\cite{Wiener1921b,Wiener1921a,Wiener1924a,Kac1947a,Kac1959a,Chaichian2001a}.   We use ``distribution function,'' ``distribution,'' and ``density'' interchangeably, despite subtle differences some might attribute to these usages, and we abbreviate Kraus-operator distribution function as \hbox{KOD} to invite the reader to use whichever of these terms makes the reader happy.

The KOD is trivially normalized to unity because the Wiener measure is normalized to unity:
\begin{align}
\int d\mu(L)\,D_T(L)=\int\sD\mu[d\vec{W}_{[0,T)}]=1\,.
\end{align}
The unconditional quantum operation at time $T$, given by Eq.~(\ref{eq:contmeasqo1}), can be unraveled in terms of this distribution,
\begin{align}\label{eq:DTLunraveling}
	\boxed{
		\vphantom{\Bigg(}
		\hspace{15pt}
\sZ_T=\int d\mu(L)\,D_T(L)\,L\!\odot\!L^\dagger\,,
\hspace{10pt}
\vphantom{\Bigg)}
}
\end{align}
an unraveling we call the \textit{KOD unraveling}.  This was called the semisimple unraveling by Jackson and Caves~\cite{CSJackson2021a} in the context of $\SL(2,\bbC)$ and semisimple Lie groups, but it is more general than that context, so we give it a more general name here.  That $\sZ_T$ is trace preserving implies that
\begin{align}
1=(1)\sZ_T=\int d\mu(L)\,D_T(L)\,L^\dagger L\,.
\end{align}

In terms of the differential positive transformation~(\ref{eq:Ldtupdelta}),
\begin{align}\label{eq:Ldtupdelta2}
L_{dt}=L_{\vec X}(d\vec W_t)=e^\updelta=e^{-\vec{X}^{2}\kappa\,dt+\vec{X}\cdot\sqrt{\kappa}\,d\vec{W}_t}\,,
\end{align}
the KOD satisfies an incremental Chapman-Kolomogorov equation,
\begin{align}\label{eq:CKlike}
\begin{split}
D_{t+dt}(L)
&=\int \sD\mu[d\vec W_{[0,t+dt)}]\;\delta\big(L,L[d\vec{W}_{[0,t+dt)}]\big)\\
&=\int d\mu(d\vec W_t)\,\sD\mu[d\vec W_{[0,t)}]\;\delta\big(L,L_{dt}L[d\vec{W}_{[0,t)}]\big)\\
&=\int d\mu(d\vec W_t)\int\sD\mu[d\vec W_{[0,t)}]\;\delta\big(L_{dt}^{-1}L,L[d\vec{W}_{[0,t)}]\big)\\
&=\int d\mu(d\vec W_t)\,D_t\big(L_{dt}^{-1}L\big)\,.
\end{split}
\end{align}
In words, the value of the KOD at $L$ at time $t+dt$ is the value to be at a precursor point $L_{dt}^{-1}L$ multiplied by the Wiener probability $d\mu(d\vec W_t)$ to transition from $L_{dt}^{-1}L$ to $L$, with this product then averaged over the precursor points.  The reader should appreciate that the incremental Chapman-Kolmogorov equation only requires the $\delta$-function relations~(\ref{eq:deltaleft}) and thus relies only on left invariance of $d\mu(L)$.  As is generally the case, the incremental Chapman-Kolmogorov equation is the basis for developing a diffusion equation.

To do that development, we introduce the \textit{right-invariant derivative}~\cite{Knapp1988a,Frankel2012a,Kitaev2018a,CSJackson2021a} of a function $f$ along a path $e^{hX}L$ leading from~$L$:
\begin{align}\label{eq:RinvXf}
	\boxed{
		\vphantom{\Bigg(}
		\hspace{15pt}
		\Rinv{X}[f](L)\equiv\frac{d}{dh}f(e^{hX}L)\bigg|_{h=0}
		=\lim_{h\rightarrow0}\frac{f\big(e^{hX}L\big)-f(L)}{h}=\lim_{h\rightarrow0}\frac{f(L+hXL)-f(L)}{h}\,.
		\hspace{10pt}
		\vphantom{\Bigg)}
	}
\end{align}
The underarrow points to the left because the path $e^{hX}$ is applied on the left side of $L$.  The derivative is called right-invariant because the derivative of a right-displaced function $g(L)=f(LL')$ is also right-displaced, that is, $\Rinv{X}[g](L)=\Rinv{X}[f](LL')$.  The right-invariant derivative is trivially $\bbR$-linear,
\begin{align}
\Rinv{aX+bY}=a\Rinv{X}+b\Rinv{Y}\,, \quad a,b\in\bbR\,.
\end{align}
The definition of the right-invariant derivative means that it generates a Taylor expansion along the path $e^{hX}L$,
\begin{align}
f\big(e^{hX}L\big)
=e^{h\Rinv{X}}[f](L)
=f(L)+h\Rinv{X}[f](L)+\frac12 h^2\Rinv{X}\big[\Rinv{X}[f]\big](L)+\cdots\,.
\end{align}
Now notice that right-invariant derivatives are not $\bbC$-linear.  Indeed, for a complex number $z$,
\begin{align}
\Rinv{zX}\ne z\Rinv{X}\,;
\end{align}
in particular,
\begin{align}\label{eq:noCauchyRiemann}
\Rinv{\,iX}\ne i\,\Rinv{X}\,.
\end{align}

Crash of cymbals!  The reader should be wide awake and asking who ordered a right-invariant derivative for the anti-Hermitian generator $iX$.  Any Kraus operator, that is, any element of $\GL(\sH,\bbC)$, has a polar decomposition,
\begin{align}\label{eq:polardecomposition}
L=W\sqrt E\,,\qquad\qquad E=L^\dagger L\,,
\end{align}
where $W$ is a unitary operator and $E$ is a positive operator, indeed (within a constant) a POVM element.  The unitary operators are a subgroup of $\GL(\sH,\bbC)$, but the positive operators are not.  Our incremental Kraus operators, generated by Hermitian operators, are differential positive operators, but when one starts piling up these incremental Kraus operators, the overall Kraus operator acquires a unitary piece in the polar decomposition.  This means that Instrument Manifold Program must be able to deal with Hermitian and anti-Hermitian generators.

The {\bf second shift in perspective}, which we now ask the reader to contemplate, is this:
\emph{instrument evolution is the stochastic motion of Kraus operators across a manifold, specifically the manifold of a complex Lie group.}  This is the essence of the Instrument Manifold Program.  The incremental Kraus operators $L_{dt}=e^\updelta$ generate this motion; this can be seen quite clearly in the Wiener path integrals and the SDEs.
In an FPKE, the motion is smooth and right-invariant motion such as Eq.~(\ref{fundamentalTexp}) is most naturally described by right-invariant derivatives acting on the \hbox{KOD}.  The right-invariant derivatives are expressions of motion or flow, with $\Rinv{X}$ describing flow locally, at each point in the manifold.  For Hermitian $X$, $\Rinv{X}$ describes flow in the direction of positive operators; $\Rinv{iX}$ similarly describes flow in the direction of unitary operators.  Such derivatives are vector fields---a vector at all points in the group manifold---and they sit in the (real) tangent bundle to the group manifold.  The vector fields $\Rinv{X}$ and $\Rinv{iX}$, which describe quite distinct flows on the group manifold, are $\bbR$-linearly independent.  Holomorphic functions would be those $f$ satisfying $\Rinv{iX}[f](L)=i\Rinv{X}[f](L)$.  The KOD is not holomorphic precisely because of the difference between the motion associated with unitary and positive transformations.

We now call out explicitly the terminology we have been using: The emphasis on motion and on the transformation groups whose elements generate the motion means that we generally refer to unitary operators as unitary transformations, something most physicists are perfectly happy with, but we also designate positive operators as positive transformations, in recognition of their role as transformations in instrument evolution.

To make the right-invariant derivative explicit, one can consider $f$ to be a function of the matrix elements $L_{jk}$.  The preceding discussion makes clear that we should regard $L_{jk}$ and $L_{jk}^*$ as independent variables. Noting that to order~$h$,
\begin{align}
f(L+hXL)=f(L)+h\bigg((XL)_{jk}\frac{\partial f}{\partial L_{jk}}+(XL)^*_{jk}\frac{\partial f}{\partial L^*_{jk}}\bigg)\,,
\end{align}
where we use the summation convention on the indices of the matrix elements, we have a chain rule,
\begin{align}
\Rinv{X}[f](L)=\bigg((XL)_{jk}\frac{\partial}{\partial L_{jk}}+(XL)^*_{jk}\frac{\partial}{\partial L^*_{jk}}\bigg)f\,.
\end{align}
This licenses us to regard $\Rinv{X}$ as acting directly on $L$ and $L^\dagger$---this is the action on linear functions---according to
\begin{align}\label{eq:RinvXL}
\Rinv{X}[L]=XL\qquad
\textrm{and}
\qquad\Rinv{X}[L^\dagger]=\Rinv{X}[L]^\dagger=L^\dagger X^\dagger\,.
\end{align}
It should be noted that $L$ is a holomorphic function and $L^\dagger$ is an anti-holomorphic function.  Not every function is one of these, which is why the chain rule is in terms of both.

The right-invariant derivatives do not commute---they are not co\"ordinate derivatives---but their commutators are vector fields, as we see from
\begin{align}\label{eq:ricomm1}
\begin{split}
\Rinv{X}\big[\Rinv{Y}[f]\big]
&=\bigg((XL)_{jk}\frac{\partial}{\partial L_{jk}}+(XL)^*_{jk}\frac{\partial}{\partial L^*_{jk}}\bigg)
\bigg((YL)_{lm}\frac{\partial f}{\partial L_{lm}}+(YL)^*_{lm}\frac{\partial f}{\partial L^*_{lm}}\bigg)\\
&=(YXL)_{jk}\frac{\partial f}{\partial L_{jk}}+(YXL)^*_{jk}\frac{\partial f}{\partial L^*_{jk}}\\
&\qquad+(XL)_{jk}(YL)_{lm}\frac{\partial^2 f}{\partial L_{jk}\partial L_{lm}}+(XL)^*_{jk}(YL)^*_{lm}\frac{\partial^2 f}{\partial L^*_{jk}\partial L^*_{lm}}\\
&\qquad+(XL)^*_{jk}(YL)_{lm}\frac{\partial^2 f}{\partial L^*_{jk}\partial L_{lm}}+(XL)_{jk}(YL)^*_{lm}\frac{\partial^2 f}{\partial L_{jk}\partial L^*_{lm}}\,,
\end{split}
\end{align}
which implies that
\begin{align}\label{eq:ricomm2}
\big[\Rinv{X},\Rinv{Y}\big][f]
=\bigg(([Y,X]L)_{jk}\frac{\partial}{\partial L_{jk}}+([Y,X]L)^*_{jk}\frac{\partial}{\partial L^*_{jk}}\bigg)f
=\Rinv{[Y,X]}[f]\,,
\end{align}
thus giving a commutator antihomomorphism,
\begin{align}\label{eq:RinvLiebracket}
\big[\Rinv{X},\Rinv{Y}\big]=-\Rinv{[X,Y]}\,.
\end{align}
The right-invariant derivatives inherit the commutators of the path generators $X$ and $Y$, with a minus sign coming from the right invariance.  Although Eqs.~(\ref{eq:ricomm1})--(\ref{eq:RinvLiebracket}) are instructive in showing how the commutators emerge as vector fields from their action on an arbitrary function, the relation~(\ref{eq:RinvLiebracket}) follows immediately from letting the derivatives act on linear functions, as in Eq.~(\ref{eq:RinvXL}).

It is useful to appreciate that for left-invariant derivatives, defined by
\begin{align}\label{eq:LinvXf}
\Linv{X}[f](L)\equiv\frac{df(Le^{hX})}{dh}\bigg|_{h=0}
=\lim_{h\rightarrow0}\frac{f\big(Le^{hX}\big)-f(L)}{h}\,,
\end{align}
we have
\begin{align}\label{eq:LinvXL}
\Linv{X}[L]=LX\qquad\textrm{and}\qquad\Linv{X}[L^\dagger]=X^\dagger L^\dagger=\Linv{X}[L]^\dagger\,,
\end{align}
which implies that
\begin{align}\label{eq:LinvLiebracket}
\big[\Linv{X},\Linv{Y}\big]=\Linv{[X,Y]}\,.
\end{align}
It is also trivial to see that right-invariant derivatives commute with left-invariant derivatives,
\begin{align}
\big[\Rinv{X},\Linv{Y}\big]=0\,.
\end{align}

Returning now to the incremental Chapman-Kolmogorov equation, notice that we can write things in terms of right-invariant derivatives,
\begin{align}
D_t(L_{dt}^{-1}L)=D_t(e^{-\updelta_t}L)=e^{-\Rinv{\updelta_t}}D_t(L)\,,\qquad\Rinv{\updelta_t}=-\Rinv{\vec{X}^{2}}\kappa\,dt+\Rinv{\vec{X}}\cdot\sqrt{\kappa}\,d\vec{W}_t\,,
\end{align}
so the incremental Chapman-Kolmogorov equation~(\ref{eq:CKlike}) becomes
\begin{align}\label{eq:CKlike2}
D_{t+dt}(L)=\int d\mu(d\vec W_t)\,e^{-\Rinv{\updelta_t}}D_t(L)\,.
\end{align}
Expanding to order $dt$,
\begin{align}\label{eq:TaylorDt}
\begin{split}
e^{-\Rinv{\updelta}}D_t(L)
&=D_t(L)-\Rinv{\updelta}[D_t](L)+\frac12\Rinv{\updelta}\big[\Rinv{\updelta}[D_t]\big](L)+\cdots\\
&=D_t(L)
+\kappa\,dt\,\Rinv{\vec X^2}[D_t](L)-\sqrt\kappa\,dW^\mu\Rinv{X_\mu}[D_t](L)
+\frac12\kappa\,dW^\mu dW^\nu\Rinv{X_\mu}\big[\Rinv{X_\nu}[D_t]\big](L)+\cdots\,,
\end{split}
\end{align}
and plugging this into the incremental Chapman-Kolmogorov equation~(\ref{eq:CKlike2}) gives the FPKE for the KOD,
\begin{align}\label{eq:DTFPKE}
	\boxed{
		\vphantom{\Bigg(}
		\hspace{15pt}
\frac{1}{\kappa}\frac{\partial D_t(L)}{\partial t}=\Delta[D_t](L)\,,
\hspace{10pt}
\vphantom{\Bigg)}
}
\end{align}
where introduced is the \emph{Kolmogorov forward generator},
\begin{align}\label{eq:Kolmogorovforward}
	\boxed{
		\vphantom{\Bigg(}
		\hspace{15pt}
\Delta\equiv\Rinv{\vec X^2}+\frac12\nabla^2\,,
\hspace{10pt}
\vphantom{\Bigg)}
}
\end{align}
which has a Laplacian diffusion operator in the positive directions,
\begin{align}\label{eq:Laplacian}
	\boxed{
		\vphantom{\Bigg(}
		\hspace{15pt}
\nabla^2\equiv\sum_\mu\Rinv{X_\mu}\Rinv{X_\mu}\,.
\hspace{10pt}
\vphantom{\Bigg)}
}
\end{align}
The initial condition corresponding to the path integral~(\ref{eq:DTL}) is $D_0(L)=\delta(L,1)$.

Rudely awakened by the crash of cymbals, the reader is urged now to savor the reward: pause and contemplate the really quite blissful~\hbox{FPKE} and its Kolmogorov forward generator~$\Delta$.  Meanwhile, it is important to recognize that the crucial step in the Chapman-Kolmogorov derivation occurs when $L_{dt}^{-1}=e^{-\updelta}$ is pulled out of the argument of $D_t$ to become an exponential of right invariant derivatives, $e^{-\Rinv{\updelta}}$.  With this step one goes from thinking about points---Kraus operators---moving stochastically through the group manifold to motion on the manifold described by vector fields.  Accompanying this perspective is the appreciation that, in Eqs.~(\ref{eq:TaylorDt})--(\ref{eq:Laplacian}), the deterministic and stochastic parts of the forward-generator vector field $\Rinv{\updelta}$ contribute differently to the Kolomogorov forward generator $\Delta$: the right-invariant derivative coming from the deterministic quadratic term, $-\Rinv{\vec X^2}\kappa\,dt$, gives the first-derivative term in $\Delta$; the right-invariant derivatives coming from the measured observables, $\Rinv{X_\mu}\sqrt\kappa\,dW^\mu_t$, don't know, in some sense, which direction to point because of the Wiener increment, and so give rise to the second-derivative, diffusion terms in the Laplacian.
Our Wiener path integrals are right-invariant versions of what are often called Feynman-Kac formulas for the solution of a diffusion equation.  It appears that Kac, inspired by Feynman's translation of the Schr\"odinger equation to Feynman path integrals~\cite{Feynman1942a,Feynman1948a,Feynman2010a}, pioneered the translation of Wiener path integrals---then referred to as averages over continuous functions---to diffusion equations~\cite{Kac1947a,Kac1959a,Chaichian2001a}.

With the initial condition $D_0(L)=\delta(L,1)$ comes a tale that needs to be told.  Suppose the initial Kraus operator is $L_0$ instead of the identity.  This situation is described by replacing the unconditional quantum operation~(\ref{eq:contmeasqo1}) with
\begin{align}
\sZ_{T|L_0}
=\int\sD\mu[d\vec{W}_{[0,T)}]\;L[d\vec{W}_{[0,T)}]L_0\!\odot\!L_0^\dagger L[d\vec{W}_{[0,T)}]^\dag
=\sZ_T\circ L_0\!\odot\!L_0^\dagger\,,
\end{align}
which can be unraveled as
\begin{align}\label{eq:DTLL0unraveling}
\sZ_{T|L_0}=\int d\mu(L)\,D_T(L|L_0)\,L\!\odot\!L^\dagger\,,
\end{align}
where the KOD is defined by the Wiener path integral,
\begin{align}\label{eq:DTLL0}
D_T(L|L_0)=\int\sD\mu[d\vec{W}_{[0,T)}]\;\delta\big(L,L[d\vec{W}_{[0,T)}]L_0\big)\,.
\end{align}
Because the SDE and the FPKE involve only what is happening at the end of the chain of incremental Kraus operators, $L_t=L[d\vec{W}_{[0,t)}]L_0$ obeys the SDE~(\ref{eq:LSDE}), with initial condition $L_0$, and $D_t(L|L_0)$ obeys the FPKE~(\ref{eq:DTFPKE}), with initial condition~$D_0(L|L_0)=\delta(L,L_0)$.  The path-integral definition of $D_T(L|L_0)$ in Eq.~(\ref{eq:DTLL0}) implies that the unit normalization of $D_t(L|L_0)$ is preserved by the FPKE~(\ref{eq:DTFPKE}): thus appreciate that $D_t(L|L_0)$ is the normalized Green function of the FPKE.  Yet $D_T(L|L_0)$ does not weave a \textit{trace-preserving\/} superoperator except when $L_0^\dagger L_0=1$---thus $L_0$ is a unitary operator---as one sees trivially from
\begin{align}
(1)\sZ_{T|L_0}=(1)\sZ_T\circ\,L_0\odot L_0^\dagger=(1)L_0\odot L_0^\dagger=L_0^\dagger L_0\,.
\end{align}
Obvious, yes, yet useful it is to stress what this means.  The initial condition $L_0$ can be thought of in the following way: precede the string of differential weak measurements by a two-outcome measurement whose Kraus operators are $L_0$ and $\sqrt{1-L_0^\dagger L_0}$, and keep only the result $L_0$.  This necessarily discards probability corresponding to the second result, thus rendering $\sZ_{T|L_0}$ trace decreasing, unless the prior measurement is completely uninformative, having just one result $L_0$, necessarily satisfying $L_0^\dagger L_0=1$, which makes $L_0$ a unitary operator.

Actually to use the continuous-measurement SDE~(\ref{eq:LMMCSD}) and FPKE~(\ref{eq:DTFPKE}) requires knowing more about the space occupied by the Kraus operators.  In particular, one needs to characterize---ultimately this means to co\"ordinate---the space of Kraus operators so that the SDE can be converted to SDEs for the co\"ordinates and the FPKE can be written in terms of co\"ordinate derivatives.  (The use of the matrix elements of $L$ in considering right-invariant derivatives is a mindless way of co\"ordinating the Kraus operators when no other structure has been recognized.)  The further characterization of the space of Kraus operators is the task of placing the instrument in a more refined Lie-group manifold, a task we take up in Secs.~\ref{sec:detaching} and~\ref{sec:1-2-3}.  Here we digress to show how our instrument-autonomous approach is related to conventional accounts of continuous measurements in terms of a quantum state evolving via a stochastic master equation.

\subsection{Stepping back into state evolution}
\label{sec:stateevolution}

Methods for quantum state evolution, such as Lindblad master equations, quantum trajectories, and stochastic master equations, are far more popular than the methods just introduced for analyzing measuring instruments autonomously.  This section makes contact between the two methods.  There is an important difference in the philosophy of the two methods concerning the nature of outcomes: in instrument-autonomous evolution, the outcomes are Wiener distributed; in state evolution, the outcomes are Born-rule distributed.

This section only exists to comfort the reader who feels bereft without the presence of a state.  Those who are perfectly fine with the state-independent instrument formalism can safely skip this section.  We do point out, however, just this once, that this section provides a very neat formulation of how quantum states evolve in the presence of continuous, differential weak measurements.\\

To convert our instrument-autonomous description to state evolution, we begin by noting that the quantum state at time $T$, given an initial state $\rho_0$ and outcomes $d\vec W_{[0,T)}$, is obtained by applying the instrument element~(\ref{eq:contmeasinstrumentel}) to $\rho_0$ and then normalizing,
\begin{align}\label{eq:rhoTW}
\begin{split}
\rho\big[d\vec W_{[0,T)}\big|\rho_0\big]
&\equiv\frac{\sD\sZ[d\vec W_{[0,T)}](\rho_0)}{\tr\!\big(\sD\sZ[d\vec W_{[0,T)}](\rho_0)\big)}\\
&=\frac{\sD\sZ[d\vec W_{[0,T)}](\rho_0)}
{\sD p\big[d\vec W_{[0,T)}\big|\rho_0\big]}\\
&=\frac{d\sZ_{\vec X}\big(d\vec{W}_{T-dt}\big)\circ \cdots\circ\,d\sZ_{\vec X}\big(d\vec{W}_{1dt}\big)\circ d\sZ_{\vec X}\big(d\vec{W}_{0dt}\big)(\rho_0)}
{\sD p\big[d\vec W_{[0,T)}\big|\rho_0\big]}\\
&=\frac{\sD\mu[d\vec{W}_{[0,T)}]}{\sD p\big[d\vec W_{[0,T)}\big|\rho_0\big]}\;L[d\vec{W}_{[0,T)}]\rho_0L[d\vec{W}_{[0,T)}]^\dag\\
&=\frac{L[d\vec{W}_{[0,T)}]\rho_0L[d\vec{W}_{[0,T)}]^\dag}{\tr\!\big(L[d\vec{W}_{[0,T)}]\rho_0 L[d\vec{W}_{[0,T)}]^\dag\big)}\,.
\end{split}
\end{align}
Here
\begin{align}\label{eq:dpTW}
\sD p\big[d\vec W_{[0,T)}\big|\rho_0\big]
\equiv\tr\!\big(\sD\sZ[d\vec W_{[0,T)}](\rho_0)\big)
=\sD\mu\big[d\vec{W}_{[0,T)}\big]\tr\!\big(L[d\vec{W}_{[0,T)}]^\dag L[d\vec{W}_{[0,T)}]\rho_0\big)
\end{align}
is the Born-rule probability for the outcome sequence $d\vec W_{[0,T)}$ given the initial state $\rho_0$.  ``Given the initial state $\rho_0$'' is the reason for the conditional notation in $\rho\big[d\vec W_{[0,T)}\big|\rho_0\big]$ and $\sD p\big[d\vec W_{[0,T)}\big|\rho_0\big]$.   The probability~(\ref{eq:dpTW}), constructed in the standard way from the initial state and the POVM element for the outcome sequence, we call the \textit{Born-rule measure}.  The density operator~$\rho\big[d\vec W_{[0,T)}\big|\rho_0\big]$ is often called the \emph{quantum trajectory\/} associated with the outcome path $d\vec W_{[0,T)}$~\cite{Carmichael1991,Doherty2000a,Brun2002b,KJacobs2006a}.

Another way to handle quantum states is to work with an unnormalized density operator that depends on the outcome record and on the initial state,
\begin{align}\label{eq:tilderhodW}
\begin{split}
\tilde\rho\big[d\vec W_{[0,T)}\big|\rho_0\big]
&\equiv\frac{\sD\sZ[d\vec W_{[0,T)}](\rho_0)}{\sD\mu\big[d\vec{W}_{[0,T)}\big]}\\
&=\frac{d\sZ_{\vec X}\big(d\vec{W}_{T-dt}\big)\circ \cdots\circ\,d\sZ_{\vec X}\big(d\vec{W}_{1dt}\big)\circ d\sZ_{\vec X}\big(d\vec{W}_{0dt}\big)(\rho_0)}
{\sD\mu\big[d\vec{W}_{[0,T)}\big]}\\
&=L[d\vec{W}_{[0,T)}]\rho_0L[d\vec{W}_{[0,T)}]^\dag\,.
\end{split}
\end{align}
This unnormalized density operator comes from applying the piled-up incremental Kraus operators in the overall Kraus operator~(\ref{eq:overallKraus}) to the input quantum state.  The product of incremental Kraus operators~\cite{Wiseman1996a,KJacobs1998a,KJacobs2006a} was named a \textit{linear quantum trajectory\/} by Wiseman~\cite{Wiseman1996a}; the associated SDE, derived below, was developed by Goetsch and Graham~\cite{Goetsch1994a} and called by them a \emph{linear}~\hbox{SDE}.

It is useful to record and to contemplate three interconnected relations among the quantities introduced here:
between the Born-rule measure and the Wiener measure,
\begin{align}\label{eq:dpTW2}
\sD p\big[d\vec W_{[0,T)}\big|\rho_0\big]
=\sD\mu\big[d\vec{W}_{[0,T)}\big]\tr\!\big(\tilde\rho\big[d\vec W_{[0,T)}\big|\rho_0\big]\big)\,,
\end{align}
between the normalized trajectories and the linear trajectories,
\begin{align}
\rho\big[d\vec W_{[0,T)}\big|\rho_0\big]
=\frac{\tilde\rho\big[d\vec W_{[0,T)}\big|\rho_0\big]}{\tr\!\big(\tilde\rho\big[d\vec W_{[0,T)}\big|\rho_0\big]\big)}\,,
\end{align}
and the invariance of their product,
\begin{align}\label{eq:twounravelings}
\sD p\big[d\vec W_{[0,T)}\big|\rho_0\big]\,\rho\big[d\vec W_{[0,T)}\big|\rho_0\big]
=\sD\mu[d\vec{W}_{[0,T)}]\,\tilde\rho\big[d\vec W_{[0,T)}\big|\rho_0\big]\,.
\end{align}
The last of these associates the two density operators with their conjugate measures, which is key to the unravelings we turn to now.

The Wiener differential unraveling~(\ref{eq:contmeasqo1}) and the KOD unraveling~(\ref{eq:DTLunraveling}) are state-independent unravelings of the unconditional quantum operation $\sZ_T$.  State-based unravelings start by applying $\sZ_T$, in the form of these two unravelings, to the initial state to get an unconditional, normalized final state $\sZ_T(\rho_0)$.  For each unraveling, there are two ways to proceed, by using unnormalized or normalized states and their conjugate distributions.  The result is four state-based unravelings:
\begin{align}
\sZ_T(\rho_0)
&=\int\sD\mu[d\vec{W}_{[0,T)}]\;L[d\vec{W}_{[0,T)}]\rho_0 L[d\vec{W}_{[0,T)}]^\dag=\int\sD\mu[d\vec{W}_{[0,T)}]\;\tilde\rho\big[d\vec W_{[0,T)}\big|\rho_0\big]\,,\label{eq:Wienerdiffunravel}\\
\sZ_T(\rho_0)
&=\int\sD p\big[d\vec W_{[0,T)}\big|\rho_0\big]\;\rho\big[d\vec W_{[0,T)}\big|\rho_0\big]\,,\label{eq:Borndiffunravel}\\[5pt]
\sZ_T(\rho_0)
&=\int d\mu(L)\,D_T(L)\,L\rho_0 L^\dagger=\int d\mu(L)\,D_T(L)\,\tilde\rho(L|\rho_0)\,,\label{eq:Krausunravel}\\
\sZ_T(\rho_0)
&=\int d\mu(L)\,D_T(L)\,\tr(L^\dagger L\rho_0)\,\rho(L|\rho_0)=\int dp_T(L|\rho_0)\,\rho(L|\rho_0)\,.\label{eq:Bornunravel}
\end{align}
The first two of these unravelings are differential and thus serve as the basis for developing SDEs for an evolving quantum state, a development we take up below. The first is a state-based version of the Wiener differential unraveling~(\ref{eq:contmeasqo1})---just put $\rho_0$ in place of the $\odot$; it gives rise to linear quantum trajectories and a linear \hbox{SDE}.  The second unravels $\sZ_T(\rho_0)$ into normalized states and thus leads to stochastic master equations; notable is that to get to the stochastic master equation, one must decompose into incremental time steps both the Born-rule measure~$\sD p\big[d\vec W_{[0,T)}\big|\rho_0\big]$ and the normalized state $\rho\big[d\vec W_{[0,T)}\big|\rho_0\big]$.  We call this second unraveling, that of Eq.~(\ref{eq:Borndiffunravel}), the \textit{Born-rule differential unraveling}.

The third and fourth unravelings are based on the KOD unraveling~(\ref{eq:DTLunraveling}).  The third is a direct expression of the KOD unraveling---just put $\rho_0$ in place of the $\odot$. It introduces an overall unnormalized linear state,
\begin{align}\label{eq:tilderhoL}
\tilde\rho(L|\rho_0)=L\rho_0 L^\dagger\,,
\end{align}
which has the path-integral expression
\begin{align}\label{eq:tilderhopathintegral}
D_T(L)\,\tilde\rho(L|\rho_0)=\int\sD\mu[d\vec{W}_{[0,T)}]\,\tilde\rho\big[d\vec W_{[0,T)}\big|\rho_0\big]\;\delta\big(L,L[d\vec{W}_{[0,T)}]\big)\,.
\end{align}
This follows immediately from the path-integral formula~(\ref{eq:DTL}) for $D_T(L)$ and exhibits the importance of the KOD in the context of the linear states.  The fourth unraveling is similar to the third, but unravels into normalized overall states,
\begin{align}\label{eq:rhoL1}
\rho(L|\rho_0)
=\frac{L\rho_0 L^\dagger}{\tr(L^\dagger L\rho_0)}
=\frac{\tilde\rho(L|\rho_0)}{\tr\!\big(\tilde\rho(L|\rho_0)\big)}\,.
\end{align}

This fourth unraveling deserves extra attention.  We call it the \textit{Born-rule unraveling}, because if one thinks of the overall Kraus operators $L$ as outcomes, $\rho(L|\rho_0)$ is the normalized overall state conditioned on outcome $L$ and given input state $\rho_0$:
\begin{align}\label{eq:rhoL2}
\rho(L|\rho_0)=\frac{d\mu(L)\,D_T(L)\,L\rho_0 L^\dagger}{dp_T(L|\rho_0)}\,.
\end{align}
Defined here is the Born-rule probability,
\begin{align}\label{eq:probL}
	\boxed{
		\vphantom{\Bigg(}
		\hspace{15pt}
dp_T(L|\rho_0)=d\mu(L)\,D_T(L)\tr(L^\dagger L\rho_0)=d\mu(L)\,D_T(L)\tr\!\big(\tilde\rho(L|\rho_0)\big)\,,
\hspace{10pt}
\vphantom{\Bigg)}
}
\end{align}
for transition~$L$ within the infinitesimal volume $d\mu(L)$, given initial state $\rho_0$.  This probability can be interpreted as the probability to transition from initial state $\rho_0$ to final state $\rho(L|\rho_0)$ at time~$T$.  Just as in Eq.~(\ref{eq:twounravelings}), it is worth explicitly associating measures with the states,
\begin{align}
d\mu(L)\,D_T(L)\,\tilde\rho(L|\rho_0)=dp_T(L|\rho_0)\,\rho(L|\rho_0)\,.
\end{align}

The Born-rule probability $dp_T(L|\rho_0)$ splits into three factors: the invariant measure $d\mu(L)$, the KOD~$D_T(L)$, and the state-dependent factor $\tr(L^\dagger L\rho_0)$.  From a state-based perspective, one might be tempted to shunt part or all of the KOD into the measure or into the Kraus operators themselves; the extremes are to incorporate the KOD wholly into the measure or wholly into renormalized Kraus operators $\sqrt{D_T(L)}\,L$.  The reason not to do any shunting in an instrument-autonomous approach is that the KOD has a real meaning: it expresses how the Kraus operators become concentrated at different points in the manifold of Kraus operators relative to the measure $d\mu(L)$, which itself defines what is meant by a uniform distribution.  This justification becomes even stronger in the context of the instrument's natural Lie group, where the group's Haar measure provides the dominant measure.  We take up this group-theoretic perspective in Sec.~\ref{sec:detaching} in order to detach the instrument from Hilbert space.  The Born-rule-based approaches, with their state-dependent probabilities, cannot be detached from Hilbert space.

Taking the trace of Eq.~(\ref{eq:tilderhopathintegral}) gives Wiener path-integral expressions for the transition-probability distribution function,
\begin{align}\label{eq:dpTLrho0paths}
\begin{split}
\frac{dp_T(L|\rho_0)}{d\mu(L)}
&=\int\sD\mu[d\vec{W}_{[0,T)}]\tr\!\Big(\tilde\rho\big[d\vec W_{[0,T)}\big|\rho_0\big]\Big)\;\delta\big(L,L[d\vec{W}_{[0,T)}]\big)\\
&=\int\sD p\big[d\vec W_{[0,T)}\big|\rho_0\big]\;\delta\big(L,L[d\vec{W}_{[0,T)}]\big)\,.
\end{split}
\end{align}
This is the Wiener-path-integral solution, that is, the Feynman-Kac solution, of a diffusion equation for the distribution function $dp_T(L|\rho_0)/d\mu(L)$ (not given here, nor anywhere else, as far we can tell).  It is also the point of departure for the state-based path-integral description of continuous, differential weak measurements developed in a sequence of three papers by Chantasri~\textit{et al.}~\cite{Chantasri2013a,Chantasri2015a,Chantasri2018a}; Chantasri~\textit{et al.}'s path-integral formulation is considered in some detail in {App.~\ref{app:Chantasri}}.  The path integral~(\ref{eq:tilderhopathintegral}) for $\tilde\rho(L|\rho_0)$ is equivalent to path integrals for~$\rho(L|\rho_0)$,
\begin{align}
\begin{split}
D_T(L)\tr(L^\dagger L\rho_0)\,\rho(L|\rho_0)
&=\int\sD\mu[d\vec{W}_{[0,T)}]\,\tilde\rho\big[d\vec W_{[0,T)}\big|\rho_0\big]\;\delta\big(L,L[d\vec{W}_{[0,T)}]\big)\\
&=\int\sD p[d\vec{W}_{[0,T)}]\,\rho\big[d\vec W_{[0,T)}\big|\rho_0\big]\;\delta\big(L,L[d\vec{W}_{[0,T)}]\big)\,.
\end{split}
\end{align}

Now let's turn attention to the SDEs for the state evolution described by the above path integrals.  For this purpose, we strip down our state notation, leaving implicit the outcomes in the conditional density operators, thus writing
\begin{align}
\rho_t=\rho\big[d\vec W_{[0,t)}\big|\rho_0\big]\qquad\textrm{and}\qquad
\tilde\rho_t=\tilde\rho\big[d\vec W_{[0,t)}\big|\rho_0\big]\,.
\end{align}
We do this partly because everyone else does it and partly because retaining all the dependences so clutters up the equations that they can hardly be read.  Nonetheless, we do it reluctantly because failure to appreciate all the dependences can lead to confusion and even serious misconceptions.  The SDE for the linear, unnormalized state $\tilde\rho_t$ comes trivially from applying the differential positive transformation~(\ref{eq:Ldtupdelta2}) to update $\tilde\rho_t$ according to
\begin{align}
\tilde\rho_{t+dt}=L_{dt}\tilde\rho_t L_{dt}^\dagger\,,
\end{align}
with the result that
\begin{align}\label{eq:dtilderhot}
d\tilde\rho_t
=\tilde\rho_{t+dt}-\tilde\rho_t
=\sqrt\kappa\,d\vec W_t\cdot\big(\vec X\tilde\rho_t+\tilde\rho_t\vec X\big)
+\kappa\,dt\bigg({-}\frac12\big(\vec X^2\tilde\rho_t+\tilde\rho_t\vec X^2\big)+\sum_\mu X_\mu\tilde\rho_t X_\mu\bigg)
\,.
\end{align}
This SDE for $\tilde\rho_t$ is called a \textit{linear\/}~SDE~\cite{Goetsch1994a}.  The first term represents conditioning on the outcome increments~$d\vec W_t$; integrating over these increments leaves the second term, which describes the Lindblad evolution of an unconditional density operator under the Lindbladian~(\ref{eq:LindbladXmu}).  The linear SDE follows directly from piling up incremental Kraus operators in Eq.~(\ref{eq:overallKraus}); the outcome increments are drawn from the Wiener measure.  Wiseman, in a careful analysis of stochastic state evolution, classifies this way of handling state evolution as Method~C and calls the Wiener-measure probabilities ``ostensible,'' because they are not the probabilities for sampling from an evolving quantum state~\cite{Wiseman1996a}.

Conventional stochastic master equations describe the evolution of the normalized state~(\ref{eq:rhoTW}) and thus use incremental probabilities that are determined by the evolving state---Wiseman calls this Method~{A}~\cite{Wiseman1996a}.  The place to begin is by unraveling the Born-rule measure $\sD p\big[d\vec W_{[0,T)}\big|\rho_0\big]$ of Eq.~(\ref{eq:dpTW2}) into a product of incremental probabilities; our treatment here can be traced back to the analysis of Goetsch and Graham~\cite{Goetsch1994a}.  Updating by one increment gives
\begin{align}
\begin{split}
\sD p\big[d\vec W_{[0,t+dt)}\big|\rho_0\big]
&=\sD\mu[d\vec{W}_{[0,t+dt)}]\,\tr(\tilde\rho_{t+dt})\\
&=d\mu(d\vec W_t)\sD\mu[d\vec{W}_{[0,t)}]\frac{\tr(\tilde\rho_{t+dt})}{\tr(\tilde\rho_t)}\tr(\tilde\rho_t)\\
&=d\mu(d\vec W_t)\frac{\tr(\tilde\rho_{t+dt})}{\tr(\tilde\rho_t)}\sD\mu[d\vec{W}_{[0,t)}]\tr(\tilde\rho_t)\\
&=dp(d\vec W_t|\rho_t)\sD p\big[d\vec W_{[0,t)]}\big|\rho_0\big]\,,
\end{split}
\end{align}
where the Born-rule incremental measure is
\begin{align}\label{eq:dpdWgivenrho}
\begin{split}
dp(d\vec W_t|\rho_t)
&=d\mu(d\vec W_t)\frac{\tr(\tilde\rho_{t+dt})}{\tr(\tilde\rho_t)}\\
&=d\mu(d\vec W_t)\frac{\tr(L_{dt}\tilde\rho_{t}L_{dt}^\dag)}{\tr(\tilde\rho_t)}\\
&=d\mu(d\vec W_t)\tr\!\big(L_{dt}\rho_t L_{dt}^\dagger\big)\\
&=d\mu(d\vec W_t)\tr\!\Big(L_{\vec X}(d\vec W_t)\rho\big[d\vec W_{[0,t)}\big|\rho_0\big]L_{\vec X}(d\vec W_t)^\dagger\Big)\,.
\end{split}
\end{align}
The last form restores all the decorations, to remind the reader why we should and should not include them.  The Born-rule incremental measure is the probability for outcomes $d\vec W_t$, given state $\rho_t=\rho\big[d\vec W_{[0,t)}\big|\rho_0\big]$ at time $t$.  We can now factor the Born-rule measure~(\ref{eq:dpTW}) into a product of incremental probabilities,
\begin{align}
\sD p\big[d\vec W_{[0,T)}\big|\rho_0\big]=dp(d\vec W_{T-dt}|\rho_{T-dt})\cdots dp(d\vec W_{1dt}|\rho_{1dt})\,dp(d\vec W_{0dt}|\rho_0)\,.
\end{align}
Notice that if the first form of the Born-rule incremental measure in Eq.~(\ref{eq:dpdWgivenrho}) is substituted into this product, the denominator of each ratio $\tr(\tilde\rho_{t+dt})/\tr(\tilde\rho_t)$ cancels with the numerator of the next term in the product, leaving the Born-rule measure in the form~(\ref{eq:dpTW2}).

We need one more piece of information, $\tr\!\big(L_{dt}\rho_t L_{dt}^\dagger\big)$.  Since $L_{dt}\rho_t L_{dt}^\dagger$ looks just like the update of~$\tilde\rho_t$,
\begin{align}
L_{dt}\rho_t L_{dt}^\dagger
=\rho_t+\sqrt\kappa\,d\vec W_t\cdot\big(\vec X\rho_t+\rho_t\vec X\big)+\kappa dt\bigg({-}\frac12\big(\vec X^2\rho_t+\rho_t\vec X^2\big)+\sum_\mu X_\mu\rho_t X_\mu\bigg)\,,
\end{align}
we have
\begin{align}\label{eq:Borndp}
dp(d\vec W_t|\rho_t)
&=d\mu(d\vec W_t)\,\big(1+2\sqrt\kappa\,d\vec W_t\cdot\langle\vec X\rangle_{\rho_t}\big)\,.
\end{align}
The Born-rule incremental measure is normalized, and relative to it the means and second-moment matrix of the Wiener outcome increments are
\begin{align}\label{eq:Wieneroutcomemeandp}
\langle dW_t^\mu\rangle_{\rho_t}&=\int dp(d\vec W|\rho_t)\,dW^\mu=2\sqrt\kappa\langle X_\mu\rangle_{\rho_t}\,dt\,,\\
\langle dW_t^\mu dW_t^\nu\rangle_{\rho_t}&=\int dp(d\vec W|\rho_t)\,dW^\mu dW^\nu=\delta^{\mu\nu}dt\,.
\label{eq:Wieneroutcomematrixdp}
\end{align}
One should appreciate that relative to the Born-rule incremental measure, each Wiener outcome increment $dW_t^\mu$ acquires a mean value proportional to the expected value of its observable $X_\mu$ and also proportional to $dt$.

The covariance matrix of the Wiener increments relative to the Born-rule incremental measure is
\begin{align}
\big\langle\big(dW_t^\mu-2\sqrt\kappa\,dt\langle X_\mu\rangle_{\rho_t}\big)\big(dW_t^\nu-2\sqrt\kappa\,dt\langle X_\nu\rangle_{\rho_t}\big)\big\rangle_{\rho_t}
=\delta^{\mu\nu}dt-4\kappa\,dt^2\,\langle X_\mu\rangle_{\rho_t}\langle X_\nu\rangle_{\rho_t}=\delta^{\mu\nu}dt\,.
\end{align}
Dropping the mean-product term, on the grounds that being proportional to $dt^2$ makes it zero in the stochastic calculus, is crucial to further developments.  Indeed, with this omission in mind, we can process the Born-rule incremental measure~(\ref{eq:Borndp}) to a new form,
\begin{align}\label{eq:Borndp2}
\begin{split}
dp(d\vec W_t|\rho_t)
&=d\mu(d\vec W_t)\,\exp\!\Big(2\sqrt\kappa\,d\vec W_t\cdot\langle\vec X\rangle_{\rho_t}-2\kappa\,dt\,\langle\vec X\rangle_{\rho_t}^2\Big)\\
&=\frac{d(dW_t^n)\cdots d(dW_t^1)}{(2\pi\,dt)^{n/2}}\exp\!\bigg({-}\frac{\big(d\vec W_t-2\sqrt\kappa\,dt\langle\vec X\rangle_{\rho_t}\big)^2}{2 dt}\bigg)\,,
\end{split}
\end{align}
which is consistent with the second-moment matrix of the Wiener outcome increments in Eq.~(\ref{eq:Wieneroutcomematrixdp}) only if we drop mean-product terms proportional to $dt^2$ when calculating second moments.

Conventional at this point is to introduce outcome- and state-dependent \textit{innovations}~\cite{Kailath1974a,Doherty2000a},
\begin{align}
d\vec\sI_t\equiv d\vec W_t-2\sqrt\kappa\langle\vec X\rangle_{\rho_t}\,dt\,,
\end{align}
which have the means of the Born-rule-distributed outcome increments removed.  The Born-rule incremental measure is a Wiener measure in the innovations,
\begin{align}
dp(d\vec{W}_t|\rho_t)
=\frac{d(d\sI_t^n)\cdots d(d\sI_t^1)}{(2\pi\,dt)^{n/2}}\exp\!\bigg({-}\frac{d\vec\sI_t^2}{2 dt}\bigg)
=d\mu(d\vec\sI_t)\,.
\end{align}
It is critical to appreciate that the outcome increments and the innovations satisfy the It\^o rule, $dW_t^\mu dW_t^\nu=\delta^{\mu\nu}dt$ and $d\sI_t^\mu d\sI_t^\nu=\delta^{\mu\nu}dt$, regardless of whether they are drawn from the Wiener measure or the Born-rule measure.  The reason is that the difference between the outcome increments and the innovations is proportional to $dt$ and thus makes a vanishing correction to the It\^o rule.  This is equivalent to dropping the $dt^2$ terms from second-moment calculations in the above.

Advancing the normalized density operator by one increment, one finds
\begin{align}\label{eq:rhotplusdt}
\begin{split}
\rho_{t+dt}
&=\frac{\tilde\rho_{t+dt}}{\tr(\tilde\rho_{t+dt})}\\
&=\frac{L_{dt}\rho_t L_{dt}^\dagger}{\tr(L_{dt}\rho_t L_{dt}^\dagger)}\\
&=\Bigg(\rho_t+\sqrt\kappa\,d\vec W_t\cdot\big(\vec X\rho_t+\rho_t\vec X\big)+\kappa dt\bigg({-}\frac12\big(\vec X^2\rho_t+\rho_t\vec X^2\big)+\sum_\mu X_\mu\rho_t X_\mu\bigg)\Bigg)\\
&\qquad\times\big(1-2\sqrt\kappa\,d\vec W_t\cdot\langle\vec X\rangle_{\rho_t}
+4\kappa\,dt\,\langle\vec X\rangle_{\rho_t}^2\big)\\
&=\rho_t
+\sqrt\kappa\big(d\vec W_t-2\sqrt\kappa\langle\vec X\rangle_{\rho_t}\,dt\big)\cdot\big(\vec X\rho_t+\rho_t\vec X-2\langle\vec X\rangle_{\rho_t}\rho_t\big)
+\kappa dt\bigg({-}\frac12\big(\vec X^2\rho_t+\rho_t\vec X^2\big)+\sum_\mu X_\mu\rho_t X_\mu\bigg)\,,
\end{split}
\end{align}
and this gives the conventional \textit{stochastic master equation}~\cite{Carmichael1991,Wiseman1994b,Wiseman1996a,Doherty2000a,Brun2002b,KJacobs2006a},
\begin{align}
d\rho_t=\rho_{t+dt}-\rho_t
=\sqrt\kappa\,d\vec\sI_t\cdot\big(\vec X\rho_t+\rho_t\vec X-2\langle\vec X\rangle_{\rho_t}\rho_t\big)
+\kappa dt\bigg(\sum_\mu X_\mu\rho_t X_\mu-\frac12\big(\vec X^2\rho_t+\rho_t\vec X^2\big)\bigg)\,,
\end{align}
which is written in terms of the innovations.  It is instructive to write the first term as
\begin{align}
\sqrt\kappa\,d\vec\sI_t\cdot\Big(\big(\vec X-\langle\vec X\rangle_{\rho_t}\big)\rho_t+\rho_t\big(\vec X-\langle\vec X\rangle_{\rho_t}\big)\Big)\,.
\end{align}
This term, nonlinear in $\rho_t$ because of the expectation value $\langle\vec X\rangle_{\rho_t}$, describes how the outcomes $d\vec W_t$ affect the evolving quantum state: each innovation $d\sI_t^\mu$, drawn from the Born-rule measure, is conjugate to the deviation of the corresponding observable $X_\mu$ from its expected value.  It is trivial to see, first, that the stochastic master equation is trace preserving and, second, that averaging over the innovations gives the Lindblad master equation for the unconditional density operator.

Wiseman~\cite{Wiseman1996a} explains that the (Method~A) stochastic master equation is better suited to simulations of state evolution than the (Method~C) linear SDE, because the outcome paths are guided by the quantum states that one is trying to simulate, whereas the outcome paths in linear SDEs are free to wander over the entire manifold of possible Kraus operators, not guided by state-based probabilities.  Instrument autonomy, in contrast, is all about letting the Kraus operators go where they may within the instrument manifold, thereby revealing the structure of the instrument, a structure defined purely by the measured observables.  The next step, which we turn to now, is to identify more precisely the instrument manifold and thus to detach the instrument not just from quantum states, but from Hilbert space itself.

\subsection{Getting out of Hilbert space. Universal instruments, towers of chaos, and principal instruments}
\label{sec:detaching}

This section is about how continuous, differential weak measurements are all about the time-ordered exponential~(\ref{eq:overallKraus}) for the overall Kraus operator.  The instruments defined by these measurements are thus not just autonomous, but precede Hilbert space.  This means that the methods of Sec.~\ref{sec:continuous}, culminating in the Instrument Manifold Program, can be considered universally, independent of Hilbert space.

The Hilbert space $\sH$ came up twice in the preceding general discussion of simultaneous measurements of noncommuting observables.  The first time was right at the beginning of Sec.~\ref{sec:single}, where $\sH$ is mentioned once in setting up the problem of differential weak measurement.  The second, more substantial, was in Sec.~\ref{sec:FPKEsKOD}, where the space of Kraus operators was identified provisionally as the Lie-group manifold $\GL(\sH,\bbC)$; this was done in order to formulate the Kraus-operator distribution as a function of $\GL(\sH,\bbC)$ and to find its diffusion equation in terms of right-invariant derivatives acting as vector fields tangent to $\GL(\sH,\bbC)$.  Readers reconciled with the idea of a group manifold to house the Kraus operators and right-invariant derivatives acting on functions of that manifold---those readers are ready for the third, final, and most important change in perspective: detaching the instrument from Hilbert space.  Here we outline the procedure for constructing the Lie group generated by the measured observables $\vec X$ and the quadratic term $\vec X^2$.  We call this group the \textit{instrumental Lie group}, and it is the proper home of the Kraus operators.  The instruments we have been considering assumed a Hilbert space $\sH$; for different Hilbert spaces, the instrumental group can look quite different, and we call these $\sH$-specific groups \emph{quantum instrumental Lie groups}.  This section shows that there is a \emph{universal instrumental Lie group}, in the usual concept of a universal covering group~\cite{Poincare2010a,Weyl1913a,Bourbaki1989a,Pontrjagin1946a,Knapp2002a}, which is Hilbert-space independent and unifies all the quantum instrumental Lie groups.\\

As we embark on this adventure, it is important to appreciate that any Lie algebra considered in quantum theory is embedded in an associative algebra that has the operations of a complex matrix algebra: commutative addition, associative multiplication, scalar multiplication by complex numbers, and Hermitian conjugation.  An associative algebra is a complex vector space under addition and scalar multiplication.  The Lie algebra inherits the vector-space property, but as already explained, we regard it as a real vector space in which Hermitian and anti-Hermitian elements are $\bbR$-linearly independent.  A quantum-physicist reader has been assuming all along that we have been operating in the associative algebra $A_\sH=\gl(\sH,\bbC)$ of operators (or matrices) on $\sH$.   Detaching from Hilbert space might be thought of as the adventure of getting out of~$A_\sH$.

The incremental Kraus operator or differential positive transformation, $L_{dt}=e^\updelta$ of Eq.~(\ref{eq:Ldtupdelta}), is differentially close to the identity and is generated by the measured observables, $\vec X=\{X_1,\ldots,X_n\}$, and the quadratic, completeness-preserving term~$\vec X^2$.  Piling up incremental Kraus operators leads to the overall Kraus operator $L_T=L[d\vec{W}_{[0,T)}]$, which is written as a time-ordered product in Eq.~(\ref{eq:overallKraus}).  This product can be reduced to a product of finitely many exponential factors, each of which, by the Magnus expansion~\cite{Magnus1954a,Hollander2017a}, has an argument given by a series of integrals of the operators $\{X_1,\ldots,X_n,\vec X^2\}=\{\vec X,\vec X^2\}$ and their successive commutators.  This is to say that the overall Kraus operator is an element of the instrumental Lie group $G=e^{\g}$, where $\g$ is the Lie algebra generated by the set $\{\vec X,\vec X^2\}$.   Below we get to the instrumental groups, both universal and $\sH$-specific (or quantum), in two steps, which highlight the difference between the measured observables and the quadratic generator.

First, however, we bring forward some general properties, which are based on the fact that the real vector space $\g$ is the direct sum of a subspace $\gone$ of Hermitian generators and a subspace $\go$ of anti-Hermitian generators:
\begin{align}\label{eq:gdirectsum}
\g=\go\oplus\gone\,.
\end{align}
The two subspaces satisfy
\begin{align}
[\go,\go]&\subseteq\go\,,\label{eq:kk}\\
[\go,\gone]&\subseteq\gone\,,\label{eq:kl}\\
[\gone,\gone]&\subseteq\go\label{eq:ll}\,,
\end{align}
thus identifying $\go\subset\g$ as a Cartan pair.  Equation~(\ref{eq:kk}) implies that $\go$ is a Lie subalgebra, which generates the subgroup $\Go=e^{\go}$ of unitary transformations within $G$.  In contrast, Eq.~(\ref{eq:ll}) indicates that the Hermitian subspace~$\gone$ is not a subalgebra; $\gone$ generates the positive transformations, which are not a subgroup of $G$, but should be thought of as a base manifold $\sE$ within $G$.  The incremental Kraus operator $L_{dt}=e^\updelta$ is a differential positive transformation, and the forward generator $\updelta$ is an element of $\gone$.  Equation~(\ref{eq:kl}) says that unitary conjugation of a positive transformation gives another positive transformation; conjugation of the base manifold $\sE$ by an element of the unitary subgroup $\Go$ is a rotation of the base manifold.

The Kraus operators are points in the group manifold $G$ and, at the same time, in the way of groups, they are also transformations of $G$.  Any Kraus operator has a unique group-theoretic polar decomposition $L=W\sqrt E$, as in Eq.~(\ref{eq:polardecomposition}), where $W$ is an element of $\Go$ and $\sqrt E=\sqrt{L^\dagger L}$ is a positive transformation and thus in $\sE$; the group-theoretic polar decomposition is a consequence of the Lie-algebraic direct sum~$\g=\go\oplus\gone$ the commutation relations implied by Eqs.~(\ref{eq:kk})--(\ref{eq:ll}).   The stochastic motion of the Kraus operators traces a path on $G$.  These stochastic paths are described by the SDE~(\ref{eq:LSDE}); the outcome-increment paths are the domain of integration in the path integral~(\ref{eq:contmeasqo1}) for the unconditional quantum operation.  The KOD, a function of $G$, is a distribution relative to the Haar measure of $G$, describing how the outcome paths accumulate on different Kraus operators.  The right-invariant derivatives are vector fields that describe flow on $G$.  The instrument assumes a shape within~$G$.  The unitary subgroup $\Go$ is called the \textit{structure group}, the base manifold $\sE$ is the space of POVM elements, and $G$ is a principal bundle~\cite{Sharpe1997}.  One way of thinking is that $\Go$ is a fiber of unitary transformations at each point in the base manifold $\sE$ and $G$ is the principal fiber bundle.

Let's proceed now to the two-step process for generating $\g$.   In the first step we find the Lie algebra $\f$ generated by the measured observables, which we call the \textit{observable Lie algebra}.  Starting with the subspace spanned by the measured observables,
\begin{align}
\Gamma^{(0)}=\spanvec\{X_1,\ldots,X_n\}\,,
\end{align}
we generate the Lie algebra through successive commutators:
\begin{align}\label{eq:iteration1}
\begin{split}
\Gamma^{(1)}&=\Gamma^{(0)}\oplus\big[\Gamma^{(0)},\Gamma^{(0)}\big]\\
\Gamma^{(2)}&=\Gamma^{(1)}\oplus\big[\Gamma^{(1)},\Gamma^{(1)}\big]\\
&\;\;\vdots\\
\Gamma^{(j)}&=\Gamma^{(j-1)}\oplus\big[\Gamma^{(j-1)},\Gamma^{(j-1)}\big]\\
&\;\;\vdots
\end{split}\quad.
\end{align}
This iterative process continues till it closes, say, after $N$ iterations, that is, $\Gamma^{(N+1)}=\Gamma^{(N)}$, in which case the observable Lie algebra is $\f=\Gamma^{(N)}$.  The corresponding Lie group $F=e^{\f}$ we call the \textit{observable Lie group}.  We need to think a bit about wherefrom comes the commutator algebra to generate $\f$.  If the measured observables are selected from the Hermitian generators of a Lie algebra $\s$ that is represented in the Hilbert space $\sH$, then one already knows the commutators, independent of $\sH$.  If the measured observables are arbitrary Hermitian operators on $\sH$, one resorts to the associative algebra $A_\sH$ to evaluate the commutators, which is equivalent to saying that $\s=\gl(\sH,\bbC)$.  It is trivial to see that $\f\subseteq\gl(\sH,\bbC)$ and thus that $F\subgroupeq\GL(\sH,\bbC)$.  Letting $d=\dim_{\bbC}{\sH}$, we have $\dim_{\bbR}\!\big(\gl(\sH,\bbC)\big)=2d^2$, so $\dim_{\bbR}(\f)\le2d^2$.  Thus if $\sH$ is finite-dimensional, so is $\f$, implying that the iterative process~(\ref{eq:iteration1}) closes after a finite number of iterations.  Once we have constructed $\f$, we can regard it and $F$ abstractly, that is, as detached from $\sH$ and its associative operator algebra $A_\sH$.

Now for the second step, which is to add the quadratic term to the observable Lie algebra $\f$.  Starting with the subspace
\begin{align}
\Delta^{(0)}=\f\oplus\vec X^2\,,
\end{align}
we generate the Lie algebra $\g$ through successive commutators:
\begin{align}\label{eq:iteration2}
\begin{split}
\Delta^{(1)}&=\Delta^{(0)}\oplus\big[\Delta^{(0)},\Delta^{(0)}\big]=\f\oplus\vec X^2\oplus\big[\f,\vec X^2\big]\\
\Delta^{(2)}&=\Delta^{(1)}\oplus\big[\Delta^{(1)},\Delta^{(1)}\big]
=\Delta^{(1)}\oplus\big[\f,\big[\f,\vec X^2\big]\big]\oplus\big[\vec X^2,\big[\f,\vec X^2\big]\big]\oplus\big[\big[\f,\vec X^2\big],\big[\f,\vec X^2\big]\big]\\
&\;\;\vdots\\
\Delta^{(j)}&=\Delta^{(j-1)}\oplus\big[\Delta^{(j-1)},\Delta^{(j-1)}\big]\\
&\;\;\vdots
\end{split}\quad.
\end{align}
The Lie algebra this iterative process generates we call the \textit{instrumental Lie algebra}.  The definite article here is misleading, however, because there are now two genuinely different ways to evaluate the commutators, corresponding to different choices of the associative operator algebra that is associated with $\f$.  The first way is to work within $A_\sH$; this uses the $d\times d$ matrix representations, starting with $\f$ and $\vec X^2$, to evaluate the commutators.  The iterative process~(\ref{eq:iteration2}) necessarily closes at a Lie algebra $\h\subseteq\gl(\sH,\bbC)$, the $\sH$-specific instrumental Lie algebra, whose corresponding $\sH$-specific Lie group, $e^\h\subgroupeq\GL(\sH,\bbC)$, we call a \emph{quantum instrumental Lie group}.  If $\sH$ is finite-dimensional, closure occurs after a finite number $N$ of iterations, so $\h=\Delta^{(N)}$.  The second way to evaluate the commutators in the iterative process~(\ref{eq:iteration2}) is within the \textit{universal enveloping algebra\/} $U_\f$ of the observable Lie algebra $\f$~\cite{Bourbaki1989a,Dixmier1996a}; this is the associative algebra that is free of constraints, \textit{except\/} for the commutators coming from $\f$.  In general, when one works in the universal enveloping algebra, the iterations~(\ref{eq:iteration2}) do not close, so $\g=\Delta^{(\infty)}$ is an infinite-dimensional Lie algebra, and the corresponding Lie group $G=e^\g$ is also infinite dimensional.  We call $G$ the \emph{universal instrumental Lie group}.

Working within $A_\sH$ yields a $\sH$-specific instrumental Lie algebra~$\h$ and an $\sH$-specific quantum instrumental group $e^\h$, whereas working within the universal enveloping algebra $U_\f$ gives the Hilbert-space-independent Lie algebra~$\g$ and the universal instrumental group $G=e^\g$. It is instructive to consider the difference between $\h$ and $\g$.  The quadratic term is quadratic in the ``linear'' measured observables, and its matrix commutators generally generate higher and higher powers of the elements of $\f$.  When working with matrices on a finite-dimensional $\sH$, sufficiently high powers are constrained to be related to lower powers by the dimensionality of~$\sH$, so the iterative process~(\ref{eq:iteration2}) closes after a finite number of steps.  This is particularly obvious in the extreme case that $\f=\gl(\sH,\bbC)$; then $\vec X^2$ is already in $\f$, so the iterative process goes nowhere and $\h=\f=\gl(\sH,\bbC)$.  In contrast, when working in the universal enveloping algebra $U_\f$, where the associative algebra is constrained only by the commutators coming from $\f$, high powers of elements of $\f$ are not constrained to be related to lower powers, so the iterative process defining $\g$ can and generally does go on forever.  This universal iterative process yields the universal instrumental Lie algebra~$\g$ and the corresponding Lie group, the universal instrumental group $G=e^\g$, which is a kind of universal covering group that unifies all the $\sH$-specific quantum instrumental groups.  We summarize this as the {\bf third perspectival shift}: \emph{detach the instrument from Hilbert space and place it in its proper home, the universal instrumental Lie group, where the three faces of the stochastic trinity can be applied universally.}

Only very special instruments have a finite-dimensional universal instrumental group; we call these \textit{principal\/} (universal) instruments.  These are pre-quantum~\cite{STAli2005a}, Hilbert-space-independent objects that structure any Hilbert space in which they reside.  The cases 1-2-3 of Sec.~\ref{sec:1-2-3} are all principal instruments.  Universal instruments that are not principal instruments we call \textit{chaotic\/} (universal) instruments.

We need to examine more carefully the relation between the Lie algebras and the Lie groups and between the ($\sH$-specific) quantum and universal realizations.  There is an associative-algebra homomorphism $\hat\pi\,:\,U_\f\rightarrow A_\sH$, meaning that the map respects the algebraic properties:
\begin{align}
\hat\pi(z_1\sfx_1+z_2\sfx_2)&=z_1\hat\pi(\sfx_1)+z_2\hat\pi(\sfx_2)\,,\\
\hat\pi(\sfx_1\sfx_2)&=\hat\pi(\sfx_1)\hat\pi(\sfx_2)\,,\\
\hat\pi(\sfx^\dagger)&=\hat\pi(\sfx)^\dagger\,,
\end{align}
for any $\sfx_1,\sfx_2\in U_\f$ and $z_1,z_2\in\bbC$.  Restricting the domain of this map to the universal instrumental Lie algebra $\g$ gives a Lie-algebra homomorphism $\pi\,:\,\g\rightarrow\h$ that projects the universal instrumental Lie algebra $\g$ onto the $\sH$-specific instrumental Lie algebra $\h$.  The kernel of this projection map,
\begin{align}
\ker\pi=\pi^{-1}(0)=\{\sfx\in\g\mid\pi(\sfx)=0\}\equiv\k\,,
\end{align}
is an ideal of $\g$, since $[\sfk,\sfg]\in\k$ for any $\sfk\in\k$ and $\sfg\in\g$.  The Lie group $e^\k$ is a normal subgroup of $G=e^\g$.  The quotient group $G/e^{\k}$ is not, however, $e^\h$ because $e^\h$ knows that elements of $\h$ other than~0 exponentiate to the identity.

To go further, we extend $\pi$ to a group projection map $\Pi\,:\,G\rightarrow e^\h$, defined by $\Pi(e^\sfg)=e^{\pi(\sfg)}$ for any $\sfg\in\g$.  It is important to realize that $\Pi$ is the associative-algebra projection map $\hat\pi$ restricted to $G$:
\begin{align}\label{eq:Pihatpi}
\Pi(e^\sfg)=e^{\pi(\sfg)}=e^{\hat\pi(\sfg)}=\hat\pi(e^{\sfg})\,.
\end{align}
The kernel of this map,
\begin{align}
\ker\Pi=\Pi^{-1}(1)=\{\sfg\in G\mid\Pi(\sfg)=1\}\equiv K\,,
\end{align}
is a normal subgroup of $G$, as one sees easily by applying the projection map~(\ref{eq:Pihatpi}).  Moreover, it is also easy to see that the quotient group, $H=G/K$, is isomorphic to $\Pi(G)=e^\h$,
\begin{align}
H=G/K\cong\Pi(G)=e^\h\,.
\end{align}
This is the sense in which the universal instrumental Lie group $G=e^\g$ is a universal cover: for every Hilbert space in which the instrument is represented, there is a subgroup of $G$, the kernel $K$, such that the quantum instrumental Lie group $e^\sfh$ is isomorphic to $H=G/K$.

To make this more concrete, we consider briefly two examples.  The first is case~3, simultaneous measurements of the three components of angular momentum, $J_x$, $J_y$, and $J_z$.  Such measurements have been considered for spin-$\frac12$ (qubits)~\cite{HWei2008a,Ruskov2010} and in great detail in an instrument-autonomous, universal fashion (there called Kraus-operator-centric and representation-independent) by Jackson and Caves~\cite{CSJackson2021a}.  In this situation, the observable Lie algebra closes after just one step: $\f=\Gamma^{(1)}=\spanvec\{-iJ_x,-iJ_y,-iJ_z,J_x,J_y,J_z\}=\sl(2,\bbC)$, with corresponding Lie group $F=e^\f=\SL(2,\bbC)$.  The quadratic term, $\vecJsquared=J_x^2+J_y^2+J_z^2$, is the Casimir invariant: in a spin-$j$ representation, $\vecJsquared=j(j+1)1_j$, and in the universal enveloping algebra $U_\f$, $\vecJsquared$ commutes with all the elements of $\f$.  Thus the iterative process~(\ref{eq:iteration2}) ends before it begins, at the 7-dimensional Lie algebra $\g=\Delta^{(0)}=\spanvec\{\vecJsquared,-iJ_x,-iJ_y,-iJ_z,J_x,J_y,J_z,\vecJsquared\}=\bbR\oplus\sl(2,\bbC)$ and the 7-dimensional universal instrumental Lie group $G=e^{\bbR}\times\SL(2,\bbC)$.  The same thing happens in any spin-$j$ representation, with the result that $\h_j\cong\g$ and the quantum instrumental Lie groups are essentially the same as $G$: $H_j\cong G$ for half-integral~$j$ and $H_j\cong G/\bbZ_2\cong e^{\bbR}\times\SO(3,\bbC)$ for integral~$j$.  This universal instrument is thus a principal instrument.  Differential weak measurements of the three components of angular momentum, performed continuously, become a strong measurement.  For finite times, the POVM moves within a 3-dimensional, hyperbolic base manifold $\sE$ and at late times, approaches the 2-sphere boundary of $\sE$, which is the familiar phase space for spin systems.  In a spin-$j$ representation, the late-time POVM elements are projectors onto spin-coherent states~\cite{CSJackson2021a}.  These universal facts structure every representation~$j$ in a way that is universal and pre-quantum~\cite{STAli2005a}.

Things are quite different for our second example, simultaneous measurements of two components of angular momentum, say $J_z$ and $J_x$.  Such measurements have been analyzed and performed for qubits \cite{Atalaya2018a,Hacohen-Gourgy2016a}; there is a good reason---the point of this paragraph---why only qubits have been considered.  The observable Lie algebra is $\f=\Gamma^{(1)}=\spanvec\{J_z,J_x,-iJ_y\}=\sl(2,\bbR)$.  When one adds the quadratic term $\vec X^2=J_z^2+J_x^2=\vecJsquared-J_y^2$, the universal iterative process doesn't close, as one sees from the fact that nested commutators of $J_y^2$ with either $J_z$ or $J_x$ don't close.  For spin-$\frac12$, the iterative process~(\ref{eq:iteration2}) closes immediately, because $J_z^2+J_x^2=\frac121_{1/2}$, with the result that $\h_{1/2}=\Delta^{(0)}=\sl(2,\bbR)\oplus\bbR=\gl(2,\bbR)$ and thus that $e^{\h_{1/2}}\cong H_{1/2}=\GL(2,\bbR)$.  For $j=1$, one can show that $\h_1=\Delta^{(2)}=\gl(3,\bbR)$ and thus $e^{\h_1}\cong H_1=\GL(3,\bbR)$.  In general, $\h_j\subseteq\gl(2j+1,\bbR)$ and thus $H_j\subgroupeq\GL(2j+1,\bbR)$, and we speculate that the inequalities are actually equalities.  The universal instrumental group is $G=H_\infty$.  The stochastic paths of the universal instrument evolution explore the infinite-dimensional universal instrumental Lie-group manifold. There is no asymptotic approach to a POVM that describes a universal strong simultaneous measurement of $J_x$ and~$J_y$.

Crudely speaking, there is a ``tower'' of chaos in the infinite number of iterations in the universal procedure~(\ref{eq:iteration2}), which leads to higher and higher powers of the observable Lie algebra.  A more developed notion is the ``tower'' of instrumental Lie algebras $\h_j$ and instrumental Lie groups $H_j$.  The universal instrumental Lie group $G$ is the setting of a universal stochastic evolution; the Kraus operators, governed by the outcome paths, wander in this infinite-dimensional Lie-group manifold.  Classical chaos is the study of how this infinite-dimensional stochastic evolution is projected into a classical phase space without loss of information about the paths.   The connection to phase space comes through the Stratonovich-Weyl correspondence~\cite{Weyl1927a,Wigner1932a,Moyal1949a,Brif1999a}, in which powers of observables correspond to scales on phase space.  Quantum chaos is the study of how the paths get projected into the finite-dimensional, quantum instrumental groups $H_j=G/K_j$, with the information about fine-scale phase-space structure disappearing into the kernel~$K_j$.  The right way to think about quantum chaos is that it is about comparing the universal KOD on $G$ with the projected KOD on the quantum $H_j$, with entropies of these KODs perhaps providing the most salient comparison.  Ultimately, this becomes a question of the topologies of the various KODs.  A general feature of the motion of the universal KOD is spiraling~\cite{Frankel2012a}, which when projected onto $H_j$, can appear as motion similar to a solar flare, where the KOD reaches out from the bulk of the support, only to return, producing a handle.  After sufficient time, these handles get filled in, because each $H_j$ has a much simpler fundamental group than $K_j$.

The difference between our measurement setting and conventional studies of quantum chaos is that conventionally one studies nonlinear unitary dynamics, whereas in our measurement setting, the observable Lie algebra $\f$ defines what is linear, and the nonlinearity comes from the unavoidable, completeness-preserving quadratic term.
To study Hamiltonian chaos in a similar way, one would perturb a nonlinear Hamiltonian continuously by the differential stochastic-unitary transformations of Eq.~(\ref{eq:stochasticunitary2}), drawn from the linear Lie algebra of whatever system one is studying.  This is reminiscent of the hypersensitivity to perturbation introduced by Schack and Caves \cite{Schack1996a,Schack1996b} as a way of characterizing both classical and quantum Hamiltonian chaos.  This group-theoretic formulation of classical and quantum chaos is arguably a productive way forward for studying chaos and dynamical complexity.

It is interesting to recall that Poincar\'e~\cite{Poincare2010a} also discovered the fundamental group in the context of his exploration of what we now call chaos; the fundamental group is equal to the kernel of the universal covering group, a concept also accredited to Poincar\'e by Weyl and Bourbaki~\cite{Weyl1913a,Bourbaki1989a,Knapp2002a}.

This paper started as a way to address the question of simultaneous measurements of noncommuting observables: what can be measured simultaneously, and how do we measure it?  The idea was that any set of observables can be measured weakly and simultaneously and that performing these weak measurements continuously would lead to a strong simultaneous measurement of these same observables.  Sections~\ref{sec:weakmeasurements} and~\ref{sec:continuous} developed the general formulation, yet after that development, much remained hidden, hidden behind the very generality of the equations, which treat all sets of measured observables on the same footing, because they only treat them \emph{locally}.  The difference between various sets of measured observables emerged only when, in this section, the \emph{global} treatment of the Instrument Manifold Program was applied to identify instrumental Lie-group manifolds.  The result was to identify and to distinguish quantum instruments and universal instruments, both principal and chaotic.  The universal instrument is detached entirely from Hilbert space and resides in the geometry of the universal covering group.  All instruments for simultaneous measurement have a universal description.  For generic instruments, the universal instrumental group is infinite dimensional, and the instrument moves chaotically in the universal covering group; there is no universal strong measurement, and every quantum instrument does its own thing, in its own Hilbert space, connected to the universal instrument by the projection that maps $G$ to $G/K$.  Only for special instruments, which we call principal instruments, do the quantum instruments resemble the universal instrument, a resemblance so strong that one can speak of a universal strong measurement, independent of Hilbert space.  It is amazing that addressing such a simple question in quantum mechanics---how do we measure noncommuting observables simultaneously?---leads to such a fundamental insight.

This brings us to cases 1-2-3, the primary examples of principal instruments.

\section{Principal instruments.  Cases 1-2-3}
\label{sec:1-2-3}

For the remainder of the paper we set $\hbar=1$.\\

We will now apply the universal instrument program to the three most fundamental principal instruments:
\begin{enumerate}
	\item The measurement of a single observable~$X$.
	\item The simultaneous momentum $P$ and position $Q$ measurement (SPQM), where $P$ and $Q$ have the canonical commutation relation,
	\begin{align}\label{eq:cancomm}
		[Q,P]=i1\,.
	\end{align}
	Generally, $Q$ and $P$ can be thought of as the canonical variables of a bosonic mode.
	\item The 3D isotropic spin measurement (ISM) of the three components of angular momentum, $J_x$, $J_y$, and $J_z$, which have the commutation relations,
	\begin{align}\label{eq:Jcomms}
		[J_\mu,J_\nu]=\sum_\tau\epsilon_{\mu\nu\tau}iJ_\tau\,,
	\end{align}
	where $\epsilon_{\mu\nu\tau}$ is the Levi-Civita symbol.
\end{enumerate}
A notable feature of what happens over the course of each of these measurements is where they end up, at a place that any reasonable physicist might expect:
\begin{enumerate}
	\item Measurement of a single observable $X$ collapses to an \emph{eigenstate\/} of $X$, that is, to a \emph{von Neumann} POVM.
	\item SPQM collapses to the \emph{canonical coherent-state\/} POVM.
	\item ISM collapses to a \emph{spin-coherent-state\/} POVM.
\end{enumerate}
As natural as these collapses appear, the details by which the latter two, the coherent-state cases, occur have only been understood quite recently~\cite{CSJackson2021a,CSJackson2023b}.
Further, ISM appears to have been the first universal method proposed~\cite{Shojaee2018a} for performing the spin-coherent-state POVM~\cite{Klauder1960a,Radcliffe1971a,Massar1995a,DAriano2002}.  Indeed, the proposal by Shojaee~\textit{et al.}~\cite{Shojaee2018a} led to the development of the Instrument Manifold Program and ultimately to the discovery of universal instruments.  Cases~2 and~3 are considered in much greater detail in~\cite{CSJackson2021a,CSJackson2023b}.

\subsection{Getting ready for 1-2-3}
\label{sec:gettingready}

It is important to notice that comparing the measurement of a single observable to SPQM and ISM is like comparing apples and oranges.
Indeed, even the use of the terms ``collapse" and ``state" for SPQM and ISM is a bit problematic because of how these terms are generally thought of as attached to the system's Hilbert space; for SPQM and ISM, these terms are, in fact, \emph{not\/} describing something in Hilbert space, but rather something in the universal measuring instrument that is generated by the simultaneous measuring process.
In simplest terms, the biggest difference is in the nature of the time it takes for the measurement to finish.
For a single observable, the time it takes for the measurement outcomes to come into focus depends on the eigenvalues of the observable, namely the smallest difference among them.  Such a collapse time for a single observable cannot be defined without reference to the Hilbert space in which the observable is represented.
For SPQM and ISM, however, the time it takes for the phase points (which are what the coherent states represent) to come into focus has nothing to do with the specific Hilbert space that is sourcing the outcomes; in these cases, the Hilbert space only adds state-related information such as the size of the quantum uncertainty in phase space and degeneracies.

The reason why this ``collapse time'' for SPQM and ISM is independent of Hilbert space is because these instruments generate universal instrumental Lie groups with far more \emph{structure\/} due to the noncommutativity of the simultaneously measured observables.  Although the identification of the universal instrumental Lie groups for all three cases is easy, we now carry out the procedure outlined in Sec.~\ref{sec:detaching}, partly to illustrate the role of noncommutativity, but also to draw attention to differences among the three cases.  The instrumental Lie groups for all three cases are universal because they are generated solely by the commutator algebra of the measured observables, without reference to any matrix representation; these are principal instruments because the universal instrumental Lie algebra is finite-dimensional.\\

Case~1 is absurdly simple because everything commutes.  The observable Lie algebra is $\f=\spanvec\{X\}$, and after adding the quadratic term $X^2$, the instrumental Lie algebra is $\g=\spanvec\{X,X^2\}$.  The instrumental Lie group $G=e^\g$ is a 2D abelian Lie group that contains only positive transformations.\\

For case~2, SPQM, the observable Lie algebra $\f=\spanvec\{i1,Q,P\}$ comes from one application of the canonical commutation relation~(\ref{eq:cancomm}) to the measured observables $Q$ and $P$.  When the quadratic term,
\begin{align}
Q^2+P^2\equiv2\Ho\,,
\end{align}
is added to $\f$, the first iteration adds the anti-Hermitian generators $-iP=[\Ho,Q]$ and $iQ=[\Ho,P]$, and the second iteration adds $1=[Q,-iP]=[P,iQ]$, after which the Lie algebra closes.  The instrumental Lie algebra,
\begin{align}
\g=\spanvec\{i1,iQ,-iP,1,Q,P,\Ho\}\,,
\end{align}
is 7-dimensional; the 7D universal instrumental Lie group we call the {Instrumental Weyl-Heisenberg Group}, $G=e^\g=\mathrm{IWH}$.  It is worth noting that the quadratic term $\Ho$ plays an essential role, first, in generating the unitary displacement parts of IWH (generators $iQ$ and $-iP$) and, second, in including the real center term $1$ in the Lie algebra.  A productive way to view the structure of IWH is to introduce the 6D Lie algebra $\cwh=\spanvec\{1,i1,iQ,-iP,Q,P\}$, which omits the $\Ho$ that is in $\g$.  The generated Lie group, $e^{\cwh}\equiv\bbC\mathrm{WH}$, called the \emph{Complex Weyl-Heisenberg Group}, is the maximal normal subgroup of \hbox{IWH}; further IWH decomposes as the semidirect product $\mathrm{IWH}\cong\bbC\mathrm{WH}\rtimes e^{\bbR H_o}$.  This semidirect structure means the subgroup $e^{\bbR H_o}$ normalizes $\bbC\mathrm{WH}$ and therefore the co\"ordinate conjugate to $\Ho$, which we call the ruler $r$, has a purely ballistic evolution.  Further discussion of the groups associated with SPQM can be found in~\cite{CSJackson2023b}.

In case~3, ISM, the observable Lie algebra $\f=\spanvec\{-iJ_x,-iJ_y,-iJ_z,J_x,J_y,J_z\}\cong\sl(2,\bbC)$ follows from one application of the commutation relations~(\ref{eq:Jcomms}) to the measured observables $J_x$, $J_y$, and $J_z$; the corresponding Lie group is $F=e^\f\cong\Spin(3,\bbC)\cong\SL(2,\bbC)$.  The quadratic term,
\begin{align}
J_x^2+J_y^2+J_z^2\equiv\vecJsquared\,,
\end{align}
is the Casimir operator: it commutes with all elements of $\f$, and in a spin-$j$ representation is $\vecJsquared=j(j+1)1_{2j+1}$.  The second iterative process therefore goes nowhere, the universal instrumental Lie algebra is
\begin{align}
\g=\spanvec\{-iJ_x,-iJ_y,iJ_z,J_x,J_y,J_z,\vecJsquared\}\cong\sl(2,\bbC)\oplus\bbR\,,
\end{align}
and the universal instrumental Lie group is the 7D group $G=e^\g\equiv\mathrm{ISpin(3)}\cong \mathrm{SL}(2,\bbC)\times e^{\bbR\vecJsquared}$, which we call the \emph{Instrumental Spin Group}.  In this case the quadratic term is essentially trivial, only adding a real center to $\mathrm{SL}(2,\bbC)$.\\

We summarize these group-theoretic considerations as follows:
\begin{enumerate}
	\item Measurement of a single observable $X$ generates a 2D instrument, contained in a 2D \textit{abelian\/} Lie group of positive transformations,
\begin{equation}\label{eq:XG}
	G \equiv \left\{e^{-X^2 r + X a}\right\} \cong \bbR^2\,.
\end{equation}
	\item SPQM generates a 7D instrument, which is contained in the \emph{Instrumental Weyl-Heisenberg Group},
\begin{align}\label{eq:IWH}
\mathrm{IWH} = \bbC\mathrm{WH}\rtimes e^{\bbR H_o}\,,
\end{align}
where $\bbC\mathrm{WH}$ is the complex Weyl-Heisenberg group.
	\item ISM generates a 7D instrument, which is contained in the \emph{Instrumental Spin Group},
\begin{align}\label{eq:ISpin}
\mathrm{ISpin(3)} = \mathrm{Spin}(3,\bbC)\times e^{\bbR\vecJsquared}\,.
\end{align}
\end{enumerate}
The irreducible representations (irreps) of a single observable are all 1D, whereas the irreps of $\mathrm{IWH}$ and $\mathrm{ISpin(3)}$ are multi-dimensional (except the trivial irrep, of course.)
Indeed, the 1D irreps for measuring a single observable are the eigensubspaces of the observable.
Meanwhile the irreps of $\mathrm{IWH}$, and $\mathrm{ISpin(3)}$ include the irreps usually considered for their 3-dimensional unitary subgroups, namely, the unitary Weyl-Heisenberg group, $\mathrm{UWH}$ (this is often thought of as \emph{the\/} Weyl-Heisenberg group), and the universal covering group of 3-dimensional rotations, $\Spin(3,\bbR)\cong\SU(2)$ (often theought  of as \emph{the\/} Spin group).
In summary, the collapse generated by measuring a single observable is entirely about the coherence \emph{between\/} irreps, whereas the collapse generated by SPQM and ISM has a very important \emph{within}-irrep component whose full temporal behavior the Instrument Manifold Program is able to describe.\\

The difference between the group structures of simultaneous measurements for commuting and noncommuting observables is quite dramatic.
Because of this, there appears to be a bit of confusion about the nature of phase space and its relation to quantum measurement, in spite of the fact that the standard and spherical phase spaces appear so plainly in the aforementioned POVMs.
There are at least three major technical obstacles that have to be overcome in order to arrive at the vision of these principal universal instruments: (i) identifying the instrument with the piling up of independent incremental Kraus operators (differential positive transformations); (ii)~placing the instrument's evolution within a finite-dimensional, universal instrumental Lie-group manifold; and (iii) actually integrating the time-ordered exponentials of the instrument's evolution, that is, the piled-up Kraus operators.
These are the essence of the Instrument Manifold Program for principal instruments.
Of course, the first two of these items was the purpose of Sec.~\ref{sec:continuousweak}, a \textit{tour de force\/} in stochastic calculus, the stochastic trinity, and their application to measuring instruments.
The third of these obstacles can be overcome with differential geometry as soon as a co\"ordinate system on the instrumental Lie-group manifold is established, and to this end we have found an invocation of the Cartan decomposition to be just the ticket.
Having established the Cartan decomposition for $\mathrm{IWH}$ and $\mathrm{ISpin(3)}$, both 7D, we will see five of their dimensions are phase space in nature: a 2D phase point in the past, a 2D phase point in the future, and the 1D geodesic curvature of the connection between the two.
The remaining two dimensions are a normalization parameter, which is very different in character between SPQM and ISM, and a ``ruler'' that characterizes the purity of the measuring process (or how close to finished the process is), which is also quite different between SPQM and \hbox{ISM}.

\subsubsection{Recap of the Instrument Manifold Program. Universal notation}\label{sec:StochasticTrinity}

To provide a template for the next three sections on cases 1-2-3, we summarize here the basic elements of the Instrument Manifold Program  as developed in Secs.~\ref{sec:weakmeasurements} and~\ref{sec:continuous}, at the same time taking the opportunity to introduce a notation that is suited to universal instrumental groups. \\

The simultaneous measurement of noncommuting observables corresponds to an instrument consisting of the instrument elements~(\ref{eq:contmeasinstrumentel}) that are associated with each sample path $d\vec W_{[0,T)}$,
\begin{align}
\sD\mu[d\vec{W}_{[0,T)}]\;\Odot\!\left(L[d\vec{W}_{[0,T)}]\right)\,,
\end{align}
where we now introduce the notation for a Kraus-rank-one operation,
\begin{align}
\Odot(A)\equiv A\!\odot\!A^\dagger\,.
\end{align}
The overall Kraus operators are defined by the time-ordered exponentials of Eqs.~(\ref{eq:overallKraus}) and~(\ref{eq:Ldtupdelta}):
\begin{align}\label{eq:Ldtupdelta2}
	L[d\vec{W}_{[0,T)}]&=\sT\,\exp\!\bigg(\int_0^{T_-}\!\!\!\- -\vec{X}^{2}\kappa\,dt+\vec{X}\!\cdot\!\sqrt{\kappa}\,d\vec{W}_t\bigg)\,.
\end{align}
The unconditional quantum operation is woven from the instrument elements,
\begin{align}\label{eq:Wdiffunravel}
		\sZ_T=\int \sD\mu[d\vec{W}_{[0,T)}]\;\Odot\!\left(L[d\vec{W}_{[0,T)}]\right)\,.
\end{align}

The time-ordered exponentials are representations of a universal cover $G$,
\begin{align}
	L[d\vec{W}_{[0,T)}]=R\circ\gamma[d\vec{W}_{[0,T)}]\,,
\end{align}
where $\gamma$ maps Wiener paths to elements of the universal cover $G$, which we generally denote by $x \in G$ to emphasize them as points in a manifold, and $R$ maps elements of $G$ to their operator or matrix representation.  We denote the universal time-ordered exponential by the same notation,
\begin{align}\label{eq:gammaupdelta}
\gamma[d\vec{W}_{[0,T)}]&=\sT\,\exp\int_0^{T_-}\!\!\!\updelta_t\,,
\hspace{20pt}
\text{where}
\hspace{20pt}
\updelta_t\equiv-\vec{X}^{2}\kappa\,dt+\vec{X}\!\cdot\!\sqrt{\kappa}\,d\vec{W}_t\,.
\end{align}
The Wiener differential unraveling~(\ref{eq:Wdiffunravel}) becomes
\begin{align}\label{eq:Wdiffunraveluniversal}
		\sZ_T=\int \sD\mu[d\vec{W}_{[0,T)}]\;\Odot\!\left(R\circ\gamma[d\vec{W}_{[0,T)}]\right)\,,
\end{align}
and this gives rise to the KOD unraveling of Eq.~(\ref{eq:DTLunraveling}),
\begin{align}\label{eq:KODxunravel}
	\sZ_T=\int_G d\mu(x)\,D_T(x)\:\Odot\big(R(x)\big)\,,
\end{align}
where $d\mu(x)$ is the left-invariant Haar measure of $G$, with respect to which is defined the KOD~(\ref{eq:DTL}),
\begin{align}\label{eq:DTxpathintegral}
	D_T(x)\equiv\int\sD\mu[d\vec{W}_{[0,T)}]\;\delta\Big(x,\gamma[d\vec{W}_{[0,T)}]\Big)\,.
\end{align}
This definition can be considered the Feynman-Kac formula of the Fokker-Planck-Kolmogorov diffusion equation defined by Eqs.~(\ref{eq:DTFPKE}), (\ref{eq:Kolmogorovforward}), and~(\ref{eq:Laplacian}),
\begin{align}\label{eq:FPKDelta}
	\frac{1}{\kappa}\frac{\partial D_t(x)}{\partial t}=\Delta[D_t](x)\,,
	\hspace{15pt}
	\text{with}
	\hspace{15pt}
	\Delta\equiv\Rinv{\,\vec X^2}+\frac12\sum_\mu\Rinv{X_\mu}\Rinv{X_\mu}\,,
\end{align}
where the derivatives, with underarrows pointing to the left, are right-invariant derivatives.

\subsubsection{Cartan co\"ordinate systems for principal instruments}\label{sec:IntegratedPile}

The Cartan or ``KAK" decomposition is the universal analog of a singular-value decomposition.
More specifically, every continuous matrix group is a representation of a universal Lie group
in which case the analogy is literally that the singular-value decomposition of a representation is a representation of the Cartan decomposition.
What this means is that the terms of a Cartan decomposition are more about how the dimensions of the Lie group are connected than they are about the Hilbert space that may carry it.  Applied to our three cases, the Cartan decompositions are:
\begin{enumerate}
	\item For measurement of a single observable, the instrumental Lie group is $G\cong\bbR^2$ of Eq.~(\ref{eq:XG}).  The $K$ in the Cartan decomposition is $K=\{1\}$ and
\begin{align}\label{eq:Cartansingle}
x= e^{-X^2 r + X a}\,.
\end{align}
The invariant measure is the familiar Cartesian measure,
\begin{equation}\label{eq:measuresingle}
	d\mu(x) = dr\,da\,.
\end{equation}
	\item For SPQM, the instrumental Lie group is the 7D $\mathrm{IWH}$ of Eq.~(\ref{eq:IWH}).  The $K$ in the Cartan decomposition is $K=\{D_{\beta}e^{i1 \phi}\}$ and
\begin{align}\label{eq:CartanSPQM}
x = \big(D_{\beta}e^{i1 \phi}\big)e^{-\Ho r-1\ell}D_{\alpha}^{-1}\,,
\end{align}
where
\begin{equation}
D_{\alpha} = e^{-iP\alpha_1 + iQ\alpha_2}\,,
\quad
\text{with}
\quad
\alpha \equiv \frac{1}{\sqrt2}(\alpha_1+i\alpha_2)\,,
\end{equation}
is the canonical displacement operator (and similarly for $\beta$).  The Haar measure in Cartan co\"ordinates is
\begin{align}\label{eq:HaarSPQM}
	d^7\!\mu(x)=\,\frac{d^2\beta}{\pi}d\phi\,dr\,\sinh^2\!r\,d\ell\,\frac{d^2\alpha}{\pi}\,,
\end{align}
with $d^2\alpha=\frac12 d\alpha_1\,d\alpha_2$ and similarly for $\beta$.
	\item For ISM, the instrumental Lie group is the 7D $\mathrm{ISpin(3)}$ of Eq.~(\ref{eq:ISpin}).  The $K$ in the Cartan decomposition is $K=\{D_{\hat{m}}e^{-iJ_z \psi}\}$ and
\begin{align}\label{eq:CartanISM}
x = \;(D_{\hat m}e^{-iJ_z\psi})e^{-\vec{J}^{\:2}\ell+J_z a}D_{\hat n}^{-1}\,,
\end{align}
where
\begin{equation}
D_{\hat n} = e^{-iJ_z\phi}e^{-iJ_y\theta}=e^{-i\theta(J_y\cos\phi-J_x\sin\phi)}e^{-iJ_z\phi}\,,
\quad
\text{with}
\quad
\hat{n}\equiv(\sin\theta\cos\phi,\sin\theta\sin\phi,\cos\theta)\,,
\end{equation}
is the spherical displacement operator (and similarly for $\hat m$).  The Haar measure in Cartan co\"ordinates is
\begin{equation}\label{eq:HaarISM}
	d^7\!\mu(x) = d^2\!\mu(\hat m)\,\frac{d\psi}{4\pi}\,d\ell\,da\,\sinh^2\!a\,d^2\!\mu(\hat n)\,,
\end{equation}
where
\begin{equation}
	d^2\!\mu(\hat n) = \frac{d\theta\,\sin\theta}{2}\frac{d\phi}{2\pi}
\end{equation}
is the standard spherical measure normalized to unity (and similarly for $\hat m$).
\end{enumerate}

For a single observable, the parameters $r$ and $a$ do not gain perspective when considered universally.
For $\mathrm{IWH}$ or $\mathrm{ISpin(3)}$, however, one can see the phase point in the past ($\alpha$ or $\hat{n}$), the phase point in the future ($\beta$ or $\hat{m}$), the geodesic curvature of the connection between the two ($\phi$ or $\psi$), the normalization parameter ($\ell$ for both), and the ruler/purity ($r$ or $a$).
Between $\mathrm{IWH}$ and $\mathrm{ISpin(3)}$ the role of the quadratic term ($\Ho$ or $\vec{J}^{\:2}$) flips from ruler to center.

The Cartan co\"ordinates accommodate the POVM because the POVM elements,
\begin{equation}\label{eq:uniproj}
	(1)\Odot\big(R(x)\big) = R(x)^\dag R(x) = R\big(\pi(x)\big)= R\big(``x^\dag x"\big)\,,
\end{equation}
can be lifted to a universal
projection map, $\pi:G\rightarrow\sE$, which maps $G$ to the base manifold symmetric space $\sE \cong \ker\pi\backslash G = K\backslash G$ represented by positive operators.
In other words, Eq.~(\ref{eq:uniproj})
tells us that the POVM for $\mathrm{IWH}$ or $\mathrm{ISpin(3)}$ is 4-dimensional and co\"ordinated by the phase point in the past ($\alpha$ or $\hat{n}$), the normalization parameter ($\ell$ for both), and the ruler ($r$ or $a$).

\subsection{Measuring a single observable continuously}
\label{sec:measX}

We consider first the case of continuous measurement of a single observable $X$.  This case is well understood~\cite{Goetsch1994a,Wiseman1996a,KJacobs1998a,Brun2002b,Silberfarb2005,KJacobs2006a,Barchielli2009,Jacobs2014,LMartin2015a}, making it easy for us to introduce new concepts and techniques in a context where, even though they aren't strictly necessary, one can readily appreciate how to think about and use them.

The continuous measurement of a single observable, $X$, has sample paths generated by the forward generator,
\begin{equation}\label{singleObs}
	\updelta_t = -X^2 \kappa\,dt + X \sqrt{\kappa}\,dW_t\,.
\end{equation}
The Kraus operators generated by $\updelta_t$ represent the 2-dimensional abelian Lie group~(\ref{eq:XG}), which is co\"ordinated in Eq.~(\ref{eq:Cartansingle}) and whose invariant measure is the Cartesian measure~(\ref{eq:measuresingle}).  The time-ordered exponential~(\ref{eq:gammaupdelta}), with forward generator $\updelta_t$ of Eq.~(\ref{singleObs}), doesn't need to be time-ordered because everything commutes, and this means that it can be integrated immediately to
\begin{align}\label{eq:gammasingle}
\gamma[dW_{[0,T)}]=e^{-X^2\kappa T}\exp\!\bigg(X\int_0^{T_-}\sqrt{\kappa}\,dW_t\bigg)\,.
\end{align}
This is equivalent to the co\"ordinate SDEs,
\begin{equation}\label{eq:drtdat}
	dr_t = \kappa\,dt
	\hspace{15pt}
	\text{and}
	\hspace{15pt}
	da_t = \sqrt{\kappa}\,dW_t\,,
\end{equation}
whose solutions,
\begin{equation}\label{eq:rTaT}
	r_T = \kappa T
	\hspace{15pt}
	\text{and}
	\hspace{15pt}
	a_T = a[dW_{[0,T)}] = \int_0^{T_-}\sqrt{\kappa}\,dW_t\,,
\end{equation}
are displayed in $\gamma[dW_{[0,T)}]$.

Meanwhile, the KOD diffuses according to the Kolmogorov forward generator~(\ref{eq:FPKDelta}),
\begin{equation}
	\Delta = \Rinv{X^2} +\frac12\Rinv{X}\Rinv{X}\,,
\end{equation}
where the transformation from the right-invariant frame to the co\"ordinate frame is trivial as the two frames are the same,
\begin{equation}
\Rinv{X^2} = -\partial_r
\hspace{15pt}
\text{and}
\hspace{15pt}
\Rinv{X} = \partial_a\,.
\end{equation}
The solution to the diffusion equation with initial condition $D_0(x) = \delta(x,1)=\delta(r)\delta(a)$ is therefore the familiar
\begin{equation}\label{eq:KODsingle}
	D_T(x) \equiv  e^{\kappa T \Delta}[D_0](x)= \delta(r-\kappa T)\frac{e^{-a^2/2\kappa T}}{\sqrt{2\pi\kappa T}}\,.
\end{equation}
It is useful to note that the KOD also follows directly from applying the stochastic integral~(\ref{eq:gammasingle}) to the KOD path integral~(\ref{eq:DTxpathintegral}),
\begin{align}
\begin{split}
D_T(x)
&=\int\sD\mu[dW_{[0,T)}]\;\delta\big(x,\gamma[dW_{[0,T)}]\big)\\
&=\delta(r-\kappa T)\int\sD\mu[d\vec{W}_{[0,T)}]\;\delta\bigg(a-\int_0^{T_-}\sqrt{\kappa}\,dW_t\bigg)\,.
\end{split}
\end{align}
Since the integral is over the Gaussian Wiener path measure, the distribution for $a$ is a normalized Gaussian whose mean and variance are those determined by the stochastic integral for $a_T$ in Eq.~(\ref{eq:rTaT}); the result is the KOD~(\ref{eq:KODsingle}).

We can now turn to the KOD unraveling~(\ref{eq:KODxunravel}) of the total operation,
\begin{align}
	\sZ_T=\int\sD\sZ[dW_{[0,T)}]
    &=\int_{\bbR^2} d\mu(x)\,D_T(x)\,\Odot\big(R(x)\big)\label{eq:abelianunravel}\\
	&=\int_{\bbR} da\,\frac{e^{-a^2/2\kappa T}}{\sqrt{2\pi\kappa T}}\,\Odot\Big(R\big(e^{-X^2\kappa T+Xa}\,\big)\Big)\label{eq:abelianunravelB}\\
	&=\int_{\bbR} da\;\Odot\Bigg(R\bigg(\frac{e^{-(a-2\kappa T X)^2/4\kappa T}}{(2\pi \kappa T)^{1/4}}\bigg)\Bigg)\,.\label{eq:abelianunravelC}
\end{align}
We say in this case that these KOD unravelings are \textit{abelian unravelings\/} of the unconditional quantum operation~$\sZ_T$ into instrument elements
\begin{align}
d\mu(x)\,D_T(x)\,\Odot\big(R(x)\big)\,.
\end{align}
In these unravelings, one can think that the outcome of the measurement at time $T$ is the Kraus operator $R\big(e^{-X^2\kappa T+X a}\big)$ itself or, equivalently, the parameter $a$.  The KOD $D_T(x)$ measures the weight of the contribution of each $R(x)$ in the abelian unraveling.  Equation~(\ref{eq:abelianunravelC}) incorporates the KOD directly into the Kraus operator, and Eqs.~(\ref{eq:abelianunravelB})~and~(\ref{eq:abelianunravelC}) are therefore considered by Wiseman~\cite{Wiseman1996a} to have different ostensible distributions.  The unraveling~(\ref{eq:abelianunravelC}) is not such a bad idea here, but it will not be such a good idea for SPQM or ISM, where the KOD comes into its own as the way to characterize collapse within an irrep.  In the present case of measuring a single observable, the group being abelian, the irreps are all {1D}.  There being no notion of collapse within an irrep, the KOD is left without much to do.

Up till now, everything has been independent of the spectrum of $X$---Hilbert-space-independent or universal, we would say---but because the irreps are 1D, this universal description is not very enlightening.  To find out more, one does the further integral over~$a$ to get the unconditional quantum operation $\sZ_T$.  In doing so, interpretation is facilitated by introducing the Hilbert-space eigendecomposition,
\begin{align}
X=\sum_j \lambda_jP_j\,,
\end{align}
with $P_j$ being the projector onto the eigensubspace that has eigenvalue $\lambda_j$.  Since we are now manifestly working in a Hilbert space, we drop the map $R$ from the formulas and find for the unconditional quantum operation,
\begin{align}\label{eq:ZTsingle}
	\sZ_T
    =\Odot\big(e^{-X^2\kappa T}\big)\circ\int_{\bbR} da\,\frac{e^{-a^2/2\kappa T}}{\sqrt{2\pi\kappa T}}\,\Odot\big(e^{Xa}\,\big)
    =e^{-(\kappa T/2)(X\odot 1-1\odot X)^2}\,.
\end{align}
This result is the same as that of Eq.~(\ref{eq:contmeasqo2}), giving here the single-observable Lindbladian.  Notice that one is doing here the same Gaussian integral as was done to find the incremental quantum operation~$\ZXdt$ in Eq.~(\ref{eq:unconQOX}); since the incremental Kraus operators commute in this case of measuring a single observable, the same integral appears at finite times.  Plugging in the eigendecomposition of $X$, one finds
\begin{align}
    \sZ_T=\sum_{j,k}P_j\odot P_k\,e^{-(\lambda_j-\lambda_k)^2\kappa T/2}\,,
\end{align}
which shows the well-known effect of the single-observable Lindbladian in the exponential decay of coherence between eigenstates with different eigenvalues, that is, between different irreps.  There is no loss of coherence between equivalent irreps, that is, within degenerate eigensubspaces.   The strong measurement that emerges as $\kappa T$ gets much bigger than 1 is, as one knew from the beginning, a von Neumann measurement of $X$, described by the projectors $P_j$.

It is useful to appreciate that when $X$ has a continuous spectrum and $\delta$-orthogonal eigenvectors, as in the case of the position variable of a particle moving in one dimension, the unconditional quantum operation becomes
\begin{align}\label{eq:ZTX2}
\sZ_T=\int_\bbR dq\int_\bbR dq'\,\proj{q}\odot\proj{q'}\,e^{-(q-q')^2\kappa T/2}\,.
\end{align}
The function $e^{-(q-q')^2\kappa T/2}$ is the simplest and purest expression of the Feynman-Vernon influence functional~\cite{Feynman1963a,Caldeira1983a,Leggett1987a}.

\subsection{Measuring position and momentum continuously}
\label{sec:measQP}

The following is a brief summary of results for \hbox{SPQM}; for further details, please refer to~\cite{CSJackson2023b}.\\

The \emph{Simultaneous $P\& Q$ Measurement} (SPQM) has sample paths generated by
\begin{equation}\label{SPQMexp}
	\updelta_t = -2\Ho\,\kappa\,dt+ P \sqrt{\kappa}\,dW_t^p + Q \sqrt{\kappa}\,dW_t^q\,.
\end{equation}
The universal cover of SPQM is the 7D Lie group that we call the \emph{Instrumental Weyl-Heisenberg Group} $\mathrm{IWH}=\bbC\mathrm{WH}\rtimes e^{\bbR\Ho}$.  The points $x\in\mathrm{IWH}$ can be co\"ordinated by the Cartan-like decomposition~(\ref{eq:CartanSPQM}), and the Haar measure in these Cartan co\"ordinates is given by Eq.~(\ref{eq:HaarSPQM}).

The time-ordered exponential~(\ref{eq:gammaupdelta}), with SPQM foward generator~(\ref{SPQMexp}), is equivalent to the following It\^o-form Cartan-co\"ordinate SDEs:
\begin{align}
	dr_t &= 2\kappa\,dt\,,\\
	d(\beta\sinh r)_t&= \cosh r_t \sqrt{\kappa}\,dw_t\,,\label{eq:betat}\\
	d(\beta\cosh r-\alpha)_t&= \sinh r_t \sqrt{\kappa}\,dw_t\,,\label{eq:betaalphat}\\
	-d\ell_t	&= (\coth r_t- 2|\beta_t|^2)\kappa\,dt+\beta_{t} \sqrt{\kappa}\,dw_t^* +\beta_{t}^* \sqrt{\kappa}\,dw_t\,,\\
\begin{split}
    id\phi_t&=\csch r_t\,(\alpha_t\beta_t^*-\alpha_t^*\beta_t)\kappa\,dt\\
    &\qquad+\smallfrac12(\beta_t\coth r_t-\alpha_t\,\csch r_t)\sqrt\kappa\,dw_t^*
    -\smallfrac12(\beta_t^*\coth r_t-\alpha_t^*\csch r_t)\sqrt\kappa\,dw_t\,,
\end{split}
\end{align}
where
\begin{align}
dw_t=\frac{1}{\sqrt2}(dW_t^q+idW_t^p)
\end{align}
is a complex Wiener increment.  The SDEs for the ruler $r$ and the future and past phase-space co\"ordinates, $\beta$ and $\alpha$, are easy to integrate.  The SDEs for the center co\"ordinates, the normalization $\ell$ and phase $\phi$, look quite complicated, but they can be integrated straightforwardly by a change to what we call Harish-Chandra co\"ordinates; this change of co\"ordinates and the integration of the SDEs are discussed in detail in~\cite{CSJackson2023b}.  Nonetheless, even after integrating the SDEs, interpreting and using the solutions is bedeviled by the normalization center term $e^{-1\ell}$.  Guidance for handling this term comes from how the FPKE is used to solve for the relevant parts of the Kraus-operator distribution function.  This guidance in hand, the SDEs can be used to solve for the KOD to the same level as is provided by the FPKEs.  This approach through the SDEs is, however, nontrivial, and for it the reader is referred to~\cite{CSJackson2023b}.  We turn here to the FPKEs to solve for the relevant part of the \hbox{KOD}.

The SPQM Kolmogorov forward generator is
\begin{equation}
	\Delta = 2\Rinv{\Ho} + \frac12 \Rinv{Q}\Rinv{Q} + \frac12 \Rinv{P}\Rinv{P}\,.
\end{equation}
In the Cartan co\"ordinate basis, these right-invariant derivatives are
\begin{align}
	\Rinv{Q} &=\nabla_1 -\beta_1\partial_\ell +\frac{\beta_2\cosh r-\alpha_2}{2\sinh r}\partial_\phi\,,\\
	\Rinv{P} &=\nabla_2 -\beta_2\partial_\ell -\frac{\beta_1\cosh r-\alpha_1}{2\sinh r}\partial_\phi\,,\\
	-\Rinv{\Ho} &= \partial_{r}-\beta_1\nabla_1-\beta_2\nabla_2+\frac{\beta_1^2+\beta_2^2}{2}\partial_\ell
		+\frac{\beta_1\alpha_2-\beta_2\alpha_1}{2\sinh r}\partial_\phi\,,
\end{align}
where
\begin{equation}\label{nablai}
	\nabla_j \equiv \frac{1}{\sinh r}\left(\partial_{\alpha_j}+\cosh r\,\partial_{\beta_j}\right).
\end{equation}

We do not have the full solution for the KOD~$D_T(x)$, but we can solve a very significant portion of it by binning the Kraus operators that differ only by their normalization and phase dimensions.  To do so, we introduce the center of IWH,
\begin{equation}
	Z \equiv\big\{e^{1 z}\,\big|\, z\in\bbC\big\}\vartriangleleft\mathrm{IWH}\,,
\end{equation}
and the \emph{Reduced Instrumental Weyl-Heisenberg Group\,} $\textrm{RIWH}\cong\textrm{IWH}/Z$.  To denote points in RIWH, we use coset notation $Zx$; readers should interpret $x\in\textrm{IWH}$ as being specified by all seven Cartan co\"ordinates and $Zx\in\textrm{RWIH}$ as being specified by the ruler $r$ and the future and past phase-space co\"ordinates, $\beta$ and $\alpha$.  Now we take the KOD unraveling~(\ref{eq:KODxunravel}) of the unconditional quantum operation and integrate over the center,
\begin{align}\label{eq:ZTunravelC}
\begin{split}
	\sZ_T &= \int_{\lowerintsub{\textrm{IWH}}}\hspace{-3pt}d^7\!\mu(x)\,D_T(x) \Odot\big(R(x)\big)\\
	&= \int_{\lowerintsub{\textrm{IWH}/Z}}\hspace{-6pt}d^5\!\mu(Zx)\left(\int_Z\!d\phi\,d\ell\,D_T(x) e^{-2\ell}\right)
    \Odot\!\left(R\big(D_\beta\,e^{-\Ho r}D_\alpha^{-1}\big)\right)\\
	&=\int_{\lowerintsub{\textrm{RIWH}}}\hspace{-5pt}\!d^5\!\mu(Zx)\,C_T(Zx)\,\Odot\!\left(R\big(D_\beta\,e^{-\Ho r}D_\alpha^{-1}\big)\right)\,.
\end{split}
\end{align}
Here we introduce the invariant measure on RIWH,
\begin{align}\label{eq:d5Zx}
		d^5\!\mu(Zx)=\frac{d^7\!\mu(x)}{d\phi\,d\ell}=\frac{d^2\beta}{\pi}\,dr\,\sinh^2\!r\,\frac{d^2\alpha}{\pi}\,,
\end{align}
and define the \emph{reduced distribution function},
\begin{align}\label{eq:CTZx}
		C_T(Zx)\equiv\int_Z\! d\phi\,d\ell \, D_T(x) e^{-2\ell}\,.
\end{align}
The completeness relation becomes
\begin{align}\label{eq:SPQMcompleteness1}
	1=(1)\sZ_T=\int_{\lowerintsub{\textrm{RIWH}}}\hspace{-5pt}\!d^5\!\mu(Zx)\,C_T(Zx)\,\left(R\big(D_\alpha e^{-\Ho 2r}D_\alpha^\dagger\big)\right)\,.
\end{align}

The reduced distribution function satisfies the partial differential equation,
\begin{align}\label{eq:CTZxPDE}
		\frac{1}{\kappa}\frac{\partial}{\partial t}C_t(Zx) = \Big({-}2\partial_r-2\coth r+ \nabla^*\nabla\Big)[C_t](Zx)\,,
\end{align}
where
\begin{align}\label{eq:nablac}
\nabla\equiv\frac{1}{\sqrt2}(\nabla_1-i\nabla_2)
=\frac{1}{\sinh r}\left(\partial_{\alpha}+\cosh r\,\partial_{\beta}\right)\,.
\end{align}
The appropriate initial condition for solving this equation is a bit tricky.  It is inherited from the initial condition for the KOD, $D_0(x)=\delta(x,1)$, which simply says that the initial Kraus operator is the identity.  When this initial condition is translated through the integration over the center in Eq.~(\ref{eq:ZTunravelC}), one finds that the initial condition for the reduced distribution function~is
\begin{align}
		C_{dt}(Zx)=\frac{2}{r}\delta(r-2\kappa dt)\,\pi\delta^2(\beta-\alpha)\,,
\end{align}
independent of $\beta+\alpha$.  The solution of Eq.~(\ref{eq:CTZxPDE}), given this initial condition, is
\begin{align}
	C_T(Zx) = \frac{2}{\sinh r} \delta(r-2\kappa T)\frac{e^{-|\beta-\alpha|^2/\Sigma_T}}{\Sigma_T}\,,
\end{align}
also independent of $\beta+\alpha$, where
\begin{equation}
\Sigma_T = \kappa T - \tanh\kappa T\,.
\end{equation}
It is important to notice that while $\Sigma_T$ does express a mean-square distance between future and past phase points, the reduced distribution $C_T(Zx)$ is not normalized to unity.  Indeed, the diffusion-like equation~(\ref{eq:CTZxPDE}) for $C_t(Zx)$ does not preserve normalization because of the $-2\coth r$ term.  In fact, the normalization of $C_T(Zx)$ is not well defined.

What is well defined is the completeness relation~(\ref{eq:SPQMcompleteness1}).  Plugging in the solution for $C_T(Zx)$ and assuming we are working in an irrep, this becomes
\begin{align}\label{eq:SPQMcompleteness2}
	1
    = 2\sinh2 \kappa T \int \frac{d^2\alpha}{\pi}D_\alpha\,e^{-\Ho 4\kappa T}D_\alpha^\dagger\int\frac{d^2\beta}{\pi\Sigma_T}e^{-|\beta-\alpha|^2/\Sigma_T}
    = 2\sinh2 \kappa T \int\frac{d^2\alpha}{\pi}D_\alpha\,e^{-\Ho 4\kappa T}D_\alpha^\dagger\,.
\end{align}
The first thing to appreciate is that for late times $T\gg1/\kappa$, $e^{-\Ho 4\kappa T}$ collapses to $e^{-2\kappa T}\proj0$ in the standard quantization and thus $D_\alpha\,e^{-\Ho 4\kappa T}D_\alpha^\dagger$ collapses to $e^{-2\kappa T}\proj\alpha$, where the states $\ket\alpha=D_\alpha\ket0$ are the canonical coherent states.  The completeness relation becomes the coherent-state resolution of the identity,
\begin{align}
1=\int\frac{d^2\alpha}{\pi}\proj\alpha\,.
\end{align}
This completeness is the sense in which the SPQM instrument approaches the coherent-state boundary uniformly at late times.  The SDEs~(\ref{eq:betat}) and~(\ref{eq:betaalphat}) for $\beta$ and $\alpha$ give no hint of this uniform late-time behavior in $\alpha$; the secret to the uniformity lies in the normalization of the Kraus operators coming from the real center term $e^{-1\ell}$, which at late times enhances the weight of Kraus operators with a Gaussian in $|\beta|^2+|\alpha|^2$.

The completeness relation has yet more to say for arbitrary times $T$.  Schur's lemma says that
\begin{align}
\int\frac{d^2\alpha}{\pi}D_\alpha\,e^{-\Ho 4\kappa T}D_\alpha^\dagger
=1\tr\!\big(e^{-\Ho 4\kappa T}\big)\,.
\end{align}
Thus we have evaluated the partition function,
\begin{align}
\tr\big(e^{-\Ho 4\kappa T}\big)
=\frac{1}{2\sinh2\kappa T}\,,
\end{align}
using only the completeness relation for \hbox{SPQM}.  This is an alternative perspective on Planck's energy quantization, based on measuring-instrument considerations instead of on thermal equilibrium.

\subsection{Measuring the three components of angular momentum continuously}
\label{sec:meas3J}

The following is a brief summary of results for \hbox{ISM}; for further details, please refer to \cite{CSJackson2021a}.\\

The \emph{Isotropic Spin Measurement} (ISM) has sample paths generated by
\begin{equation}\label{SpinUpD}
	\updelta_t = -\vec{J}^{\,2} \kappa\,dt + J_x \sqrt{\kappa}\,dW^x + J_y \sqrt{\kappa}\,dW^y + J_z \sqrt{\kappa}\,dW^z\,.
\end{equation}
The universal cover of ISM is the 7-dimensional Lie group we call the \emph{Instrumental Spin Group}~$\mathrm{ISpin(3)}=\Spin(3,\mathbb{C})\times e^{\bbR\vecJsquared}$.
The points $x\in\mathrm{ISpin(3)}$ can be co\"ordinated by the Cartan decomposition~(\ref{eq:CartanISM}), and the Haar measure in Cartan co\"ordinates is given by Eq.~(\ref{eq:HaarISM}).

To find It\^o-form SDEs for the time-ordered exponential~\ref{eq:gammaupdelta}), with ISM forward generator~\ref{SpinUpD}, it is convenient to decompose only partially, writing $x\in\mathrm{ISpin}(3)$ as
\begin{equation}
	x = Ve^{J_z a-\vecJsquared\ell}U\,.
\end{equation}
The co\"ordinate SDEs for the center co\"ordinate $\ell$ and the ruler/purity $a$ are
\begin{align}
	d\ell_t &= \kappa\,dt\,,\\
	da_t &= \kappa\,dt\,\coth a_t + \sqrt{\kappa}\,dY_t^z\,,\label{eq:aSDE}
\end{align}
and the SDEs for the past and future unitaries $U$ and $V$, written as MMCSDs, are
\begin{align}
	dU_t U_t^{-1} - \frac12 (dU_t U_t^{-1})^2 &= \left(-iJ_x\sqrt{\kappa}\,dY^y_t+iJ_y\sqrt{\kappa}\,dY^x_t\right)\csch a_t\,,\\
	dV_t^{-1} V_t - \frac12 (dV_t^{-1} V_t)^2 &= \left(-iJ_x\sqrt{\kappa}\,dY^y_t+iJ_y\sqrt{\kappa}\,dY^x_t\right)\coth a_t\,.
\end{align}
Here the Wiener increments have to be rotated in situ by the future unitary $V$,
\begin{equation}
	dY_t^\mu = {(R_t^{-1})^\mu}_\nu\,dW_t^\nu\,,
\end{equation}
with the rotation matrices defined in the obvious way,
\begin{align}
V_t^{-1} J_\nu V_t = J_\mu{(R_t^{-1})^\mu}_\nu\,.
\end{align}

The diffusion forward generator of ISM is
\begin{equation}
	\Delta = \Rinv{\;\vec{J}^{\,2} } + \frac12 \Rinv{J_x}\Rinv{J_x} + \frac12 \Rinv{J_y}\Rinv{J_y} + \frac12 \Rinv{J_z}\Rinv{J_z}\,.
\end{equation}
In the Cartan partial-co\"ordinate basis, these right-invariant derivatives are
\begin{equation}
	\Rinv{J_\mu} = {(R^{-1})^\nu}_\mu \nabla_\nu\,,
\end{equation}
where
\begin{align}
	\nabla_z &= \partial_a\,,\\
	\nabla_x &= -\frac{1}{\sinh a}\left(\Rinv{L_y^1}-\cosh a\, \Linv{L_y^0}\right)\,,\\
	\nabla_y &= \frac{1}{\sinh a}\left(\Rinv{L_x^1}-\cosh a\, \Linv{L_x^0}\right)\,.
\end{align}
Here we let $L_\mu = -iJ_\mu$ denote the anti-Hermitian generators associated with the components of angular momentum, and we define partial derivatives,
\begin{align}
	\Linv{L_\mu^0}[x] &\equiv \Linv{L_\mu^0}[V]e^{J_z a-\vec{J}^{\:2}\ell}U \equiv V(-iJ_\mu)e^{J_z a-\vec{J}^{\:2}\ell}U\,,\\
	\Rinv{L_\mu^1}[x] &\equiv Ve^{J_z a-\vec{J}^{\:2}\ell}\Rinv{L_\mu^1}[U] \equiv Ve^{J_z a-\vec{J}^{\:2}\ell}(-iJ_\mu)U\,.
\end{align}
In terms of the derivatives $\nabla_\mu$, the Kolmogorov forward generator is
\begin{equation}
	\Delta = \Rinv{\;\vec{J}^{\,2} } + \frac12 \left(\partial_a^2 + 2\coth a\,\partial_a +\nabla_x\nabla_x+\nabla_y\nabla_y\right).
\end{equation}

We do not have the full solution for the KOD~$D_T(x)$, but we can consider the distribution function obtained by integrating over the future unitary,
\begin{align}
	D_T(Kx) &\equiv \int_K d^3\mu(V)\,D_T(x)\,,
\end{align}
where $K$ here stands for unitary group SU(2) of future unitaries $V$.  This distribution function governs completeness, as one sees from the KOD unraveling~(\ref{eq:KODxunravel}),
\begin{align}\label{eq:ISMcompleteness}
\begin{split}
1=(1)\sZ_T&=\int_{\lowerintsub{\textrm{ISpin(3)}}}\hspace{-6pt}d\mu(x)\,D_T(x)\:R(x)^\dagger R(x)\\
&=\int_{\lowerintsub{\textrm{ISpin(3)}}}\hspace{-6pt}d\mu(x)\,D_T(x)\:e^{-\vecJsquared 2\ell}U^\dagger e^{J_z 2a}U\\
&=\int d\ell\,da\sinh^2\!a\,d^2\mu(\hat n)\left(\int_K d^3\mu(V)\,D_T(x)\right)\,e^{-\vecJsquared 2\ell}D_{\hat n}e^{J_z 2a}D_{\hat n}^\dagger\\
&=\int d\ell\,da\sinh^2\!a\,d^2\mu(\hat n)\,D_T(Kx)\,e^{-\vecJsquared 2\ell}D_{\hat n}e^{J_z 2a}D_{\hat n}^\dagger\,.
\end{split}
\end{align}
The solution for $D_T(Kx)$, given the initial condition that sets $x=1$ at $T=0$, is
\begin{align}
	D_T(Kx)=\delta(\ell-\kappa T)\,\frac{P_T(a)}{\sinh^2 a}\,d^2\!\mu(\hat n)\,,
\end{align}
which is symmetric on the sphere of angular variables and thus is specified by a single-variable distribution function $P_t(a)$ for the ruler/purity co\"ordinate.  This distribution is normalized to unity against the measure $da$---that's why we included the $1/\sinh^2\!a$ in this solution---and it diffuses according to the FPKE,
\begin{equation}\label{eq:adiffusion}
\frac{1}{\kappa}\frac{\partial}{\partial t}P_t(a) = \left({-}\frac{\partial}{\partial a}\coth a+\frac{1}{2}\frac{\partial^2}{\partial a^2}\right)\!P_t(a)\,.
\end{equation}
Plugging the solution for $D_T(Kx)$ into the completeness relation~(\ref{eq:ISMcompleteness}) gives
\begin{equation}
	1=e^{-2\kappa T \vec{J}^{\:2}}\!\int_{\lowerintsub{S^2}}d^2\mu(\hat n)\,D_{\hat{n}}\Bigg(\int_{\bbR^+} da\,P_T(a)\,e^{J_z 2a}\Bigg)D_{\hat{n}}^\dagger\,.
\end{equation}
By taking a trace, we can say that for an irreducible spin-$j$ representation,
\begin{equation}
	1=\frac{e^{-2\kappa T j(j+1)}}{2j+1}\!\!\int_0^\infty da\,P_T(a) \tr(e^{J_z 2a})\,.
\end{equation}

More importantly, we can refer to the SDE~(\ref{eq:aSDE}) or the diffusion equation~(\ref{eq:adiffusion}) to find that for late times, $T\gg1/\kappa$, the ruler/purity co\"ordinate $a$ has a mean and variance that are both equal to $\kappa T$.  This means that in a spin~$j$ representation, for late times, $e^{J_z2a}$ is dominated by $e^{2aj}\proj j$.  This in turns means that the ISM POVM at late times approaches the spin-coherent-state boundary, consisting of the states $D_{\hat n}\ket j$, and it does that uniformly over the 2-sphere of spin-coherent states.  A rigorous argument for this conclusion is presented in~\cite{CSJackson2021a}.

\section{Conclusion}
\label{sec:conclusion}

In this article, we formulated the simultaneous measurement of noncommuting observables in terms of a fundamental \emph{differential positive transformation}, defined in Eq.~(\ref{eq:Ldtupdelta}).  These positive transformations we consider fundamental because a large class of independent, sequential weak measurements (Fig.~\ref{fig:sequential}) have transformations that are effectively equivalent to these, similar to how sampling a sum of independent random variables is effectively equivalent to sampling a Gaussian distribution.
Once these fundamental differential positive transformations are recognized, it becomes evident that the simultaneous measurement of noncommuting observables generates over finite amounts of time Kraus operators---or instrument elements---that reside in a Lie group that we call the \emph{instrumental Lie group}.
Recognizing these instrumental Lie groups reveals the existence of a \emph{universal instrument}, which the Hilbert-space-specific quantum instruments can be considered to represent.

These instruments can be analyzed directly in terms of the instrumental Lie groups they reside in, and for this purpose the \emph{Instrument Manifold Program\/} was designed.
The Program interprets the measurement of an observable as a vector field tangent to the instrumental Lie group considered as a manifold.
This, in turn, translates the Kraus operators of an instrument into a distribution function over the instrumental Lie-group manifold.
Once this step is made, it becomes clear that the evolution of the \emph{Kraus-operator distribution function\/} can be analyzed with any of the standard techniques in stochastic calculus, what we call the stochastic trinity: Wiener path integration, stochastic differential equations, or a Fokker-Planck-Kolmogorov equation.

In general, Lie groups are manifolds that are very fundamental, but that have differential geometries that are not most naturally described by the standard definitions of a manifold via co\"ordinate patches.
Rather, the differential geometry of a Lie group is more naturally described by a (co\"ordinate-independent) right-invariant calculus.
The design of a right-invariant stochastic calculus was perhaps the most technically difficult part of the Instrument Manifold Program.

Lifting to the universal instrument, independent of Hilbert space, the Program revealed two quite remarkable insights into the nature of noncommuting observables, insights that come from recognizing the distinction between chaotic and principal universal instruments.
The first of these is that such simultaneous measurements generically produce chaos in the evolution of the instrument.
We believe this could be the beginning of a breakthrough in the study of quantum chaos and dynamical complexity.

The second insight is that the phase-space coherent POVMs are exactly related to those simultaneous measurements that do not produce chaos, the principal instruments.
For principal universal instruments, the quantum instruments strongly resemble each other, and the universal instrument can be analyzed directly with the differential-geometric techniques introduced by Cartan~\cite{Chern1952a, Helgason1978a, Sharpe1997,Borel1998a,Borel2001,Frankel2012a}.
Connecting canonical phase space to the simultaneous measurement of position and momentum, we found that the POVM for finite times offers an alternative perspective to energy quantization, in which the value of the partition function comes directly from the hyperbolic geometry of the instrumental Lie group.
Connecting spherical phase space to the simultaneous measurement of the three components of angular momentum offers a universal way to perform the spin-coherent POVMs.
The \emph{unitary} Weyl-Heisenberg and Spin groups have for some time been understood to describe the virtual motion of canonical phase space and the sphere~\cite{Brif1999a}, which is key to a deeper understanding of quantization and quantum uncertainty;
the \emph{instrumental} Weyl-Heisenberg and Spin Lie groups extend the range of these virtual motions in a profound way, incorporating measurement at the fundamental level and connecting phase space to the identity.

The Instrument Manifold Program emerged from what we call the Principle of Instrument Autonomy, that measuring instruments are fundamentally independent of the states that play them and therefore measurement problems should not be bound to state-based analysis.
There is much more work to be done with the Instrument Manifold Program, on both the principal-instrument and chaotic-instrument fronts.
On the side of principal instruments, there are many principal instruments with features that are still not fully understood: the spin analog of heterodyne detection, heterodyning with small detuning, adaptive phase measurement and other adaptive measurements, fermionic and parafermionic measurements, and measurements of more general semisimple observables, to name a few.
For chaotic instruments, an obvious next step would be to consider the entropies of the Kraus-operator distribution functions and compare those entropies between various quantum instruments and the universal instrument.
In that regard, it should also be noted that the Program for analyzing instrumental chaos is not limited to the measurement chaos encountered in this paper and could as well be applied to unitary chaos~\cite{Schack1996a,Schack1996b}.
It should also be noted that random-matrix theories are also instruments and therefore equivalent to Kraus-operator distribution functions; perhaps the Program could also help to gain insight into the universal properties in random-matrix theory.

In all, a pretty ambitious program of further research, all emerging from taking seriously what it means to measure noncommuting observables simultaneously.

\acknowledgments
CSJ thanks Mohan Sarovar for all of the helpful discussions and financial support through Sandia National Laboratories and thanks CMC for the incredibly fruitful collaboration.
CMC supported himself and is eternally grateful for the opportunity to work with \hbox{CSJ}.

This work was supported in part by the Center for Quantum Information and Control at the University of New Mexico and in part by the U.S. Department of Energy, Office of Science, Office of Advanced Scientific Computing Research, under the Quantum Computing Application Teams (QCAT) program. Sandia National Laboratories is a multimission laboratory managed and operated by NTESS, LLC., a wholly owned subsidiary of Honeywell International, Inc., for the U.S. DOE's NNSA under contract DE-NA-0003525. This paper describes objective technical results and analysis. Any subjective views or opinions that might be expressed in the paper do not necessarily represent the views of the U.S. Department of Energy or the United States Government.


\appendix

\section{Stochastic unitary and jump unravelings}
\label{app:unravelings}

This appendix spells out how the stochastic-unitary and jump unravelings given at the end of Sec.~\ref{sec:multiple} arise from the meter model of Sec.~\ref{sec:single}.  We consider a single observable $X$ and for both unravelings indicate how to generalize to multiple observables.\\

For the stochastic-unitary unraveling, one registers the momentum $p$ of the meter.  The Kraus operator for outcome $p$ within $dp$ is
\begin{align}\label{eq:Kraussinglep}
\begin{split}
\sqrt{dp}\,\langle p|e^{-iH\,dt/\hbar}|0\rangle
&=\sqrt{dp}\,\langle p|e^{-i2\sqrt{\kappa\,dt}\,X\otimes\,\sigma P/\hbar}|0\rangle\\
&=\sqrt{dp}\,e^{-i2\sqrt{\kappa\,dt}\,X\sigma p/\hbar}\langle p|0\rangle\\
&=\sqrt{dp\,\frac{e^{-p^2/2(\hbar^2/4\sigma^2)}}{\sqrt{2\pi\hbar^2/4\sigma^2}}}\,e^{-i2\sqrt{\kappa\,dt}\,X\sigma p/\hbar}\\
&=\sqrt{d(dW)\,\frac{e^{-dW^2/2 dt}}{\sqrt{2\pi dt}}}\,e^{-iX\sqrt\kappa\,dW}\\
&=\sqrt{d\mu(dW)}\,e^{-iX\sqrt\kappa\,dW}\,,
\end{split}
\end{align}
where the outcome $p$ is rescaled to be
\begin{align}
dW=\frac{2\sigma p}{\hbar}\sqrt{dt}
\end{align}
and $d\mu(dW)$ is the standard Wiener measure~(\ref{eq:Wienerdmu}).  As promised, the result for the Kraus operators is the stochastic-unitary transformation~(\ref{eq:stochasticunitary}), specialized to a single observable $X$.  To generalize to multiple observables, one uses the approach that worked for differential weak measurements of noncommuting observables: one can combine the stochastic unitaries for different observables because to order $dt$, the independence of the Wiener increments means that the stochastic unitaries for different observables commute.\\

For the jump unraveling, we introduce the meter's complex-amplitude (annihilation) operator,
\begin{align}
A=\frac{1}{\sqrt2}\left(\frac{1}{\sqrt2\sigma}Q+i\frac{\sqrt2\sigma}{\hbar}P\right)\,,
\end{align}
relative to which the meter state $\ket0$ is the ``vacuum state,'' $A\ket0=0$.  The number states,
\begin{align}
\ket N=\frac{(A^\dagger)^N}{\sqrt{(N+1)!}}\ket0\,,
\end{align}
are eigenstates of $A^\dagger A$, that is, $A^\dagger A\ket N=N\ket N$.  The meter position and momentum can be expressed as
\begin{align}
Q&=\sigma(A+A^\dagger)\,,\\
P&=-i\frac{\hbar}{2\sigma}(A-A^\dagger)\,.
\end{align}
The jump unraveling comes from registering the meter in the number basis.  Expanding the action of the Hamiltonian~(\ref{eq:HdtX}) on the meter vacuum to order $dt$, one finds
\begin{align}
e^{-iH\,dt/\hbar}|0\rangle
&=e^{-i2\sqrt{\kappa\,dt}\,X\otimes\,\sigma P/\hbar}|0\rangle\\
&=\ket0+\sqrt{\kappa\,dt}\,X\otimes\bigg(\!{-}i\frac{2\sigma P}{\hbar}\bigg)\ket0+\frac12\kappa\,dt\,X^2\otimes\bigg(\!{-}i\frac{2\sigma P}{\hbar}\bigg)^2\ket0\\
&=\ket0\bigg(1-\frac12\kappa\,dt X^2\bigg)+\ket1\sqrt{\kappa\,dt}\,X+\ket2\frac{1}{\sqrt2}\kappa\,dt\,X^2\,.
\end{align}
When one projects onto the meter number states, the only states that count are $\ket0$ and $\ket1$---the $\ket2$ projection is order $(dt)^2$ in the instrument element---so there are two Kraus operators,
\begin{align}
K_0&=\langle0|e^{-iH\,dt/\hbar}|0\rangle=1-\frac12\kappa\,dt\,X^2=e^{-(\kappa\,dt/2)X^2}\,,\\
K_1&=\langle1|e^{-iH\,dt/\hbar}|0\rangle=\sqrt{\kappa\,dt}\,X\,.
\end{align}
These match the jump unraveling of Eqs.~(\ref{eq:nojump}) and~(\ref{eq:jump}) for the case of a single observable $X$.  To generalize to multiple observables $X_\mu$, one recognizes that for multiple observables interacting sequentially with meters, the only number-state projectors that survive at order $dt$ are the no-jump projection onto vacuum, which gives the Kraus operator $K_0$, and the single-jump projectors, which give the Kraus operators $K_\mu$.  In the no-jump Kraus operator, the contributions from the multiple observables can be combined because commutators can be ignored at order $dt$.

\section{Chantasri \emph{et al.}\ path integrals}
\label{app:Chantasri}

Chantasri~\textit{et al.}~\cite{Chantasri2013a,Chantasri2015a,Chantasri2018a} formulated a path integral for a distribution function that describes the probability to transition from initial state $\rho_0$ to final state $\rho_F$ at time $T$.  In their paper, Chantasri~\textit{et al.}\ parametrize density operators by ``position'' co\"ordinates ${\bf q}$, considered with respect to a Cartesian measure, so they denote their distribution function as $P({\bf q}_F|{\bf q}_I)$.  Having no need for this parametrization, we prefer to write this distribution function as $P_T(\rho_F|\rho_0)$, with respect to a measure $d\mu(\rho_F)$; this notation also explicitly recognizes the dependence on $T$.  The distribution function $P_T(\rho_F|\rho_0)$ is clearly related to our probability $dp_T(L|\rho_0)=d\mu(L)\,D_T(L)\tr(L^\dagger L\rho_0)$ of Eq.~(\ref{eq:probL}), the probability to transition from $\rho_0$ to $\rho(L|\rho_0)$.  The probability $dp_T(L|\rho_0)$ involves the KOD and has the path-integral expression~(\ref{eq:dpTLrho0paths}).  Chantasri \textit{et al.}\ include an additional path integral for the evolving density operator into their expression for $P_T(\rho_F|\rho_0)$, thus giving a double path integral over both outcome increments and density operators.  This appendix shows how their approach works and how $P_T(\rho_F|\rho_0)$ is related to  $dp_T(L|\rho_0)$.\\

Before getting to that, we note that the Kraus-operator formalism naturally expresses density-operator evolution in a way that generates the desired transition probability.  This comes from the path integral~(\ref{eq:tilderhopathintegral}) for the linear (unnormalized) overall state $\tilde\rho(L|\rho_0)$,
\begin{align}
D_T(L)\,\tilde\rho(L|\rho_0)
&=\int \sD\mu[d\vec{W}_{[0,T)}]\;\tilde\rho\big[d\vec W_{[0,T)}\big|\rho_0\big]\\
&=\int d\sZ_{\vec X}\big(d\vec{W}_{T-dt}\big)\circ \cdots\circ\,d\sZ_{\vec X}\big(d\vec{W}_{1dt}\big)\circ d\sZ_{\vec X}\big(d\vec{W}_{0dt}\big)(\rho_0)\;
\delta\big(L,L[d\vec{W}_{[0,T)}]\big)\,.
\end{align}
Here we write out $\tilde\rho\big[d\vec W_{[0,T)}\big|\rho_0\big]$ in terms of the incremental instrument elements; the integral is over the outcome increments that determine these instrument elements.  Taking the trace of both sides gives the path integral~(\ref{eq:dpTLrho0paths}) for the transition probability distribution,
\begin{align}
\begin{split}
P_T(L|\rho_0)\equiv\frac{dp_T(L|\rho_0)}{d\mu(L)}
&=D_T(L)\,\tr\!\big(\tilde\rho(L|\rho_0)\big)\\
&=\int\tr\!\Big(d\sZ_{\vec X}\big(d\vec{W}_{T-dt}\big)\circ \cdots\circ\,d\sZ_{\vec X}\big(d\vec{W}_{1dt}\big)\circ d\sZ_{\vec X}\big(d\vec{W}_{0dt}\big)(\rho_0)\Big)\,
\delta\big(L,L[d\vec{W}_{[0,T)}]\big)\,,
\end{split}
\end{align}
written here in terms of the incremental instrument elements.

Chantasri~\textit{et al.}\ start with the conditional probability for the density operator $\rho_{t+dt}$ and the outcome increments $d\vec W_t$, given the state $\rho_t$ at time $t$; let us denote this probability as $dp\big(\rho_{t+dt},d\vec W_t\big|\rho_t\big)$.  They factor this probability into further conditionals,
\begin{align}\label{eq:dprhodW}
dp\big(\rho_{t+dt},d\vec W_t\big|\rho_t\big)=dp\big(\rho_{t+dt}\big|d\vec W_t,\rho_t\big)\,dp(d\vec W_t|\rho_{t})\,,
\end{align}
where
\begin{align}\label{eq:dpdWgivenrho2}
dp(d\vec W_t|\rho_t)
=d\mu(d\vec W_t)\tr\!\big(L_{\vec X}(d\vec W_t)\rho_t L_{\vec X}(d\vec W_t)^\dagger\big)
\end{align}
is the Born-rule incremental measure~(\ref{eq:dpdWgivenrho}) and
\begin{align}\label{eq:dprhogivendW}
dp\big(\rho_{t+dt}\big|d\vec W_t,\rho_{dt}\big)
=d\mu(\rho_{t+dt})\:\delta\hspace{-0.5pt}\Bigg(\rho_{t+dt}-\frac{L_{\vec X}(d\vec W_t)\rho_t L_{\vec X}(d\vec W_t)^\dagger}
{\tr\!\big(L_{\vec X}(d\vec W_t)\rho_t L_{\vec X}(d\vec W_t)^\dagger\big)}\Bigg)\,.
\end{align}
The reader should note that the $\delta$-function here is conjugate to the measure $d\mu(\rho)$.  Equally important is that in this approach the density operators $\rho_t$ are independent variables; the $\delta$-function in Eq.~(\ref{eq:dprhogivendW}) constrains $\rho_{t+dt}$ to be the updated state that comes from applying the incremental Kraus operator~$L_{\vec X}(d\vec W_t)$ to $\rho_t$ and normalizing.

The probability for state $\rho_F$ at time $T$ within an infinitesimal volume $d\mu(\rho_F)$, given initial state $\rho_0$, is
\begin{align}
\begin{split}
d\mu(\rho_F)\,P_T(\rho_F|\rho_0)\equiv dp_T(\rho_F|\rho_0)
&=\int\prod_{k=0}^{T/dt-1}dp\big(\rho_{(k+1)dt},d\vec W_{kdt}\big|\rho_{kdt}\big)\,,
\end{split}
\end{align}
where the integral is over the outcome increments $d\vec W_{[0,T)}=\{d\vec W_{T-dt},\cdots,d\vec W_{1dt},d\vec W_{0dt}\}$ and over the density operators $\rho_{(0,T)}\equiv\{\rho_{T-dt},\ldots,\rho_{2dt},\rho_{1dt}\}$.  In the last probability increment, one sets $\rho_T=\rho_F$.  Substituting the further conditionals in the factorization~(\ref{eq:dprhodW}) puts the probability distribution to transition from $\rho_0$ to $\rho_F$ in the form,
\begin{align}\label{eq:Chantasri}
\begin{split}
P_T(\rho_F|\rho_0)
&=\frac{1}{d\mu(\rho_T)}\int\prod_{k=0}^{T/dt-1}d\mu(\rho_{(k+1)dt})\,dp(d\vec W_{kdt}|\rho_{kdt})\,
\delta\Bigg(\rho_{(k+1)dt}-\frac{L_{\vec X}(d\vec W_{kdt})\rho_{kdt}L_{\vec X}(d\vec W_{kdt})^\dagger}
{\tr\!\big(L_{\vec X}(d\vec W_{kdt})\rho_{kdt}L_{\vec X}(d\vec W_{kdt})^\dagger\big)}\Bigg)\\
&=\int\sD\mu\big[\rho_{(0,T)}\big]\,
\prod_{k=0}^{T/dt-1}dp(d\vec W_{kdt}|\rho_{kdt})\;
\delta\Bigg(\rho_{(k+1)dt}-\frac{L_{\vec X}(d\vec W_{kdt})\rho_{kdt}L_{\vec X}(d\vec W_{kdt})^\dagger}
{\tr\!\big(L_{\vec X}(d\vec W_{kdt})\rho_{kdt}L_{\vec X}(d\vec W_{kdt})^\dagger\big)}\Bigg)\,,
\end{split}
\end{align}
where
\begin{align}
\sD\mu\big[\rho_{(0,T)}\big]=d\mu(\rho_{T-dt})\cdots d\mu(\rho_{2dt})\,d\mu(\rho_{1dt})\,.
\end{align}
The double path integral~(\ref{eq:Chantasri}) is the one formulated by Chantasri~\textit{et al.}  Characterizing the functional integral over density operators as a sum over paths is perhaps a bit of a stretch, since the $\delta$-functions restrict the density-operator path to the single path determined by the sequence of outcome increments.  As already mentioned, Chantasri~\textit{et al.}\ parametrize the density operator in terms of ``position'' co\"ordinates; they expand the $\delta$-functions in plane waves, thus introducing ``momentum'' co\"ordinates conjugate to the position co\"ordinates.  This further development---and a subsequent attention to an action-like formulation in terms of the position and momentum co\"ordinates---is not our concern here.

It is evident from the definitions of the transition probabilities that
\begin{align}\label{eq:measureequality}
dp_T(\rho_F|\rho_0)\big|_{\rho_F=\rho(L|\rho_0)}=d\mu(\rho_F)\,P_T(\rho_F|\rho_0)\big|_{\rho_F=\rho(L|\rho_0)}=d\mu(L)\,P_T(L|\rho_0)=dp_T(L|\rho_0)\,.
\end{align}
Clear though this is, it is instructive to see how it follows from the path-integral expressions for the two distribution functions.  Doing the integrals over the density operators in Eq.~(\ref{eq:Chantasri}), starting with $\rho_{1dt}$ and working up to $\rho_{T-dt}$, evaluates these density operators as $\rho_{kdt}=\rho\big[d\vec W_{[0,kdt)}\big|\rho_0\big]$, $k=1,\ldots,T/dt-1$, as one expects.  The result is that
\begin{align}\label{eq:PTrhoFrho0}
P_T(\rho_F|\rho_0)=\int\sD p\big[d\vec W_{[0,T)}\big|\rho_0\big]\,\delta\Big(\rho_F-\rho\big[d\vec W_{[0,T)}\big|\rho_0\big]\Big)\,.
\end{align}
This looks just like the formula~(\ref{eq:dpTLrho0paths}) for $P_T(L|\rho_0)=dp_T(L|\rho_0)/d\mu(L)$ as a path integral over the Born-rule measure,
\begin{align}\label{eq:PTLrho0paths}
P_T(L|\rho_0)
=\int\sD p\big[d\vec W_{[0,T)}\big|\rho_0\big]\;\delta\big(L,L[d\vec{W}_{[0,T)}]\big)\,,
\end{align}
except that the $\delta$-functions in the two formulas are different and conjugate to different measures.  It is easy to verify from these defining path integrals that
\begin{align}
P_T(\rho_F|\rho_0)=\int d\mu(L)\,P_T(L|\rho_0)\,\delta\big(\rho_F-\rho(L|\rho_0)\big)\,;
\end{align}
simply plug Eq.~(\ref{eq:PTLrho0paths}) into this expression and out pops Eq.~(\ref{eq:PTrhoFrho0}).  This ugly duckling of an equation, with its apparent discord between the $\delta$-function and the measure, is actually a swan.  The $\delta$-function enforces that $\rho_F=\rho(L|\rho_0)=L\rho_0L^\dagger/\tr(L\rho_0 L^\dagger)$ and  handles the change in measure between the two probability distributions.  Specifically, one sees this change in measure from the following steps,
\begin{align}
\begin{split}
P_T(\rho_F|\rho_0)\big|_{\rho_F=\rho(L|\rho_0)}
&=\int d\mu(L')\,P_T(L'|\rho_0)\,\delta\big(\rho(L|\rho_0)-\rho(L'|\rho_0)\big)\\
&=\frac{d\mu(L)}{d\mu(\rho_F)}\int d\mu(L')\,P_T(L'|\rho_0)\delta(L,L')\\
&=\frac{d\mu(L)}{d\mu(\rho_F)}\,P_T(L|\rho_0)\,,
\end{split}
\end{align}
which is the relation~(\ref{eq:measureequality}).  Expressing the Chantasri~\textit{et al.} distribution function as
\begin{align}
P_T(\rho_F|\rho_0)\big|_{\rho_F=\rho(L|\rho_0)}
=\frac{d\mu(L)}{d\mu(\rho_F)}\,D_T(L)\,\tr(L\rho_0L^\dagger)
\end{align}
highlights the Kraus-operator density as the state-independent factor in this distribution function.

\end{document}